%% file: main.tex
\documentclass[11pt,dvipsnames,twoside,a4paper]{memoir}
\usepackage[dvipsnames,table]{xcolor}
\usepackage[utf8]{inputenc}
\usepackage[T1]{fontenc}
\usepackage[american]{babel}
\usepackage{amsmath, amssymb, amsthm}
\usepackage{microtype}
\usepackage{todonotes}
\usepackage{graphicx}
\usepackage{tikz}
\usepackage{hhline}
\usepackage{physics}
\usepackage{mathtools}
\usepackage{bm}
\usepackage[linesnumbered,ruled]{algorithm2e}
\usepackage{listings}
\usepackage{booktabs}
\usepackage{longtable}
\usepackage{array}
\usepackage{float}
\usepackage[url=false,isbn=false,doi=false,eprint=true,style=phys,biblabel=brackets,maxbibnames=10,backend=biber,sorting=none,pageranges=false]{biblatex}
\DeclareFieldFormat{journaltitle}{%
  \iffieldundef{doi}
    {#1}
    {\href{https://doi.org/\thefield{doi}}{#1}}%
}
\DeclareFieldFormat[article,periodical]{volume}{\textbf{#1}} 
\usepackage[font={small}]{caption}
\usepackage{subcaption}
\usepackage{xfrac}
\usepackage{multirow}
\usepackage{gensymb}
\usepackage{makecell}
\usepackage[colorlinks=true,citecolor=blue,linkcolor=blue,urlcolor=blue]{hyperref}
\usepackage{csquotes}
\usepackage{rotating}
\usepackage{tabularx}
\usepackage{cleveref}
\usepackage{wrapfig}

\setlrmarginsandblock{4cm}{3cm}{*}
\setulmarginsandblock{3cm}{3cm}{*}
\checkandfixthelayout

\DeclareCaptionLabelFormat{subfigure}{\textbf{#2.}}
\colorlet{theader}{ProcessBlue!30}
\colorlet{tsubheader}{ProcessBlue!20}

\theoremstyle{definition}
\newtheorem{example}{Example}[chapter]

\theoremstyle{definition}
\newtheorem{definition}{Definition}[section]

\newcommand{\Nbar}{\bar{N}}

\definecolor{juliablue}{RGB}{30, 96, 212}
\definecolor{juliagreen}{RGB}{34, 139, 34}
\definecolor{juliapurple}{RGB}{136, 0, 136}
\definecolor{juliagrey}{RGB}{120, 120, 120}
\definecolor{juliabg}{RGB}{248, 248, 248}

\lstdefinelanguage{Julia}{
  morekeywords={
    abstract, break, case, catch, const, continue, do, else, elseif, end,
    export, false, finally, for, function, global, if, import, in, let,
    local, macro, module, mutable, primitive, quote, return, struct,
    true, try, using, where, while
  },
  morekeywords=[2]{
    Int, Float64, Bool, String, Vector, Matrix, Array, Dict, Set,
    println, print, length, push!, pop!, zeros, ones, rand
  },
  sensitive=true,
  morecomment=[l]{\#},
  morecomment=[n]{\#=}{=\#},
  morestring=[b]",
  morestring=[b]',
}

\lstdefinestyle{juliastyle}{
  language=Julia,
  backgroundcolor=\color{juliabg},
  basicstyle=\ttfamily\small,
  keywordstyle=\color{juliablue}\bfseries,
  keywordstyle=[2]\color{juliapurple},
  commentstyle=\color{juliagreen}\itshape,
  stringstyle=\color{juliapurple},
  numbers=left,
  numberstyle=\tiny\color{juliagrey},
  stepnumber=1,
  numbersep=8pt,
  showstringspaces=false,
  breaklines=true,
  breakatwhitespace=true,
  tabsize=4,
  frame=single,
  rulecolor=\color{juliagrey},
  captionpos=b,
  keepspaces=true,
}

\lstdefinestyle{inline}{
  basicstyle=\scriptsize\ttfamily,
}

\title{Neural and Tensor Networks in the Study of Quantum Annealing Processors}
\author{Tomasz {\'S}mierzchalski}
\date{March 2026}

\begin{document}

\begin{titlingpage}
  \begin{center}
    \includegraphics[width=0.4\textwidth]{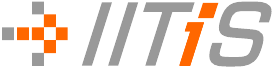}\\
    \vspace{0.5em}
    \textsc{\large Institute of Theoretical and Applied Informatics, Polish Academy of Sciences}
    \vspace*{1in}
    \hrule
    \vspace*{0.5em}
    \textsc{\huge Neural and Tensor Networks in the Study of Quantum Annealing Processors}
    \vspace*{0.5em}
    \hrule
    \vspace*{1em}
    \textsc{\large Doctoral dissertation}
    \par
    \vspace{1.5in}
    {\large mgr Tomasz \textsc{\'{S}mierzchalski}}\\
    \vspace{0.25in}
    Supervisor:\\ dr hab. Bart\l omiej Gardas\\
    \vspace{0.25in}
    Co-supervisor:\\ dr hab. in\.{z}. \L ukasz Pawela\\
    \vfill
    {Gliwice, March 2026}
  \end{center}
\end{titlingpage}

\begin{titlingpage}
  \begin{center}
    \includegraphics[width=0.4\textwidth]{figures/iitis_logo}\\
    \vspace{0.5em}
    \textsc{\large Instytut Informatyki Teoretycznej i Stosowanej Polskiej Akademii Nauk}
    \vspace*{1in}
    \hrule
    \vspace*{0.5em}
    \textsc{\huge Sieci neuronowe i tensorowe w badaniu kwantowych procesor\'{o}w wy\.{z}arzających}
    \vspace*{0.5em}
    \hrule
    \vspace*{1em}
    \textsc{\large Rozprawa doktorska}
    \par
    \vspace{1.5in}
    {\large mgr Tomasz \textsc{\'{S}mierzchalski}}\\
    \vspace{0.25in}
    Promotor:\\ dr hab. Bart\l omiej Gardas\\
    \vspace{0.25in}
    Promotor pomocniczy:\\ dr hab. in\.{z}. \L ukasz Pawela\\
    \vfill
    {Gliwice, Marzec 2026}
  \end{center}
\end{titlingpage}

\frontmatter

\tableofcontents*
\newpage
\input{chapters/acknowledgements.tex}
\input{published_work.tex}
\input{chapters/abstract_en.tex}

\input{chapters/abstract_pl.tex}
\input{chapters/introduction.tex}

\mainmatter

\input{chapters/ising.tex}

\input{chapters/adiabatic.tex}
\input{chapters/software.tex}
\input{chapters/bench.tex}

\input{chapters/thermo.tex}

\input{chapters/error-correction.tex}

\input{chapters/sim.tex}
\input{chapters/last.tex}

{
  \tiny
  \printbibliography
}
\appendix

\input{appendices/appendix-quantum.tex}
\input{appendices/appendix-tn.tex}

\input{appendices/appendix-rl.tex}

\input{appendices/appendix-dwave.tex}

\end{document}

%% file: chapters/acknowledgements.tex
\chapter{Acknowledgements}

I would like to express my sincere gratitude to my supervisor, dr hab. Bart\l omiej Gardas, for his guidance, trust, and continuous support throughout my doctoral studies. His scientific insight, high standards, and ability to identify the essential aspects of a problem have strongly shaped the way I think about research. I am especially grateful for the freedom to explore ambitious ideas, as well as for the constructive criticism that helped turn them into concrete results.

I am also deeply grateful to my co-supervisor, dr hab. in\.{z}. \L ukasz Pawela, for his invaluable advice, technical expertise, and constant readiness to discuss both scientific and practical aspects of my work. His support was particularly important in the development of computational methods and software tools that became a central part of this dissertation.

I would like to thank my collaborators and co-authors, with whom I had the privilege to work during these years. The discussions, joint experiments, code development, manuscript writing, and many difficult debugging sessions were an essential part of this research. I am especially thankful to colleagues from the Institute of Theoretical and Applied Informatics, Polish Academy of Sciences, for creating a stimulating and friendly scientific environment. I also thank all collaborators from other institutions whose expertise contributed to broadening the perspective of this work.

I gratefully acknowledge the administrative and technical staff of the Institute for their help with everyday matters that often remain invisible but are crucial for scientific work. 

Finally, I would like to thank my wife Agnieszka for her patience, understanding, and unconditional support. Her belief in me has been a constant source of motivation throughout my doctoral studies. This dissertation would not have been possible without the people who supported me scientifically, professionally, and personally, and to all of them I am deeply grateful.

This project was partially supported by the National Science Center (NCN), Poland, under Projects: Sonata Bis 10, No. 2020/38/E/ST3/00269 and the Foundation for Polish Science (grant no POIR.04.02.00-00-D014/20-02 co-financed by the European Union under the European Regional Development Fund). I would also like to thank J\"{u}lich Supercomputing Centre for providing me with access to quantum annealers used for some benchmarks presented in this thesis.

%% file: published_work.tex
\chapter{Published work}

\section*{Publications relevant for this dissertation}

\nocite{software, peps, thermo, rl, agv}
\printbibliography[heading=none,keyword=my-used]

\section*{Other publications}
\printbibliography[heading=none,keyword=my-unused]


%% file: chapters/abstract_en.tex
\chapter{Abstract}

Quantum annealing is a hardware realization of adiabatic quantum computation aimed at finding low-energy configurations of Ising/QUBO cost functions and is available in commercial devices such as D-Wave processors. Reliable assessment of these machines requires benchmarking that goes beyond “best solution” and explicitly accounts for strong classical baselines, sampling behavior, and diversity, and the physical cost of computation, including thermodynamic efficiency. This dissertation develops such a framework and applies it to contemporary quantum annealers.

The first contribution is SpinGlassPEPS.jl, a topology-aware tensor-network heuristic designed for optimization and sampling on quasi-two-dimensional graphs aligned with QPU connectivity (e.g., Pegasus/Zephyr). The method reduces problems to local Potts clusters, represents the resulting partition function with PEPS, and performs a branch-and-bound search in probability space, with explicit accuracy cost control via parameters such as inverse temperature and bond dimension. Building on this baseline, the thesis reports head-to-head benchmarks of D-Wave quantum annealers against classical and quantum-inspired solvers on random spin-glass instances defined natively on Pegasus and Zephyr graphs, reaching system sizes of several thousand spins and evaluating both optimization and sampling (including low-energy diversity) using like-for-like metrics. The results establish SpinGlassPEPS.jl as a physically interpretable reference, while also delineating its limitations: for the largest random instances, approximate tensor-network contractions introduce errors that prevent matching highly optimized GPU heuristics at comparable runtimes, positioning tensor-network methods primarily as topology-aware analysis and benchmarking tools for small-to-medium scales.

A second pillar is a thermodynamic characterization of quantum annealers as effective thermal machines, relating computational performance to energy dissipation, entropy production, and effective temperature along programmable schedules. This perspective shows that carefully chosen pauses can simultaneously improve solution quality/success probability and reduce thermodynamic cost relative to simple reverse annealing, while also identifying regimes in which an applied longitudinal field becomes detrimental once pausing is introduced. 

The third pillar is a practical reinforcement-learning post-processing approach that treats returned samples as inputs to an RL agent, which learns spin-flip moves that correct typical error patterns. Once trained, it improves success probability and energy accuracy and can be integrated as an orthogonal mitigation layer.

Finally, exact simulations of small-system adiabatic dynamics support intuition and modeling choices. Comparison to device data suggests that matching observed output statistics may require stronger effective disorder/noise than expected, indicating additional error sources or stronger environmental coupling in the relevant regimes.

Overall, the dissertation argues for benchmarking quantum annealing technologies using coherent protocols that combine rigorous classical comparators with metrics capturing both algorithmic performance and physical expenditure.

%% file: chapters/abstract_pl.tex
\chapter{Streszczenie}

\begin{otherlanguage}{polish}
Wyżarzanie kwantowe stanowi sprzętową realizację adiabatycznych obliczeń kwantowych ukierunkowaną na znajdowanie konfiguracji o niskiej energii dla funkcji kosztu typu Ising/QUBO. Technologia ta została zaimplementowana w komercyjnych urządzeniach, takich jak procesory firmy D-Wave. Rzetelna ocena możliwości tych maszyn wymaga jednak benchmarkingu wykraczającego poza analizę pojedynczego najlepszego rozwiązania. Konieczne jest uwzględnienie silnych klasycznych punktów odniesienia, właściwości próbkowania oraz różnorodności uzyskiwanych konfiguracji, a także fizycznego kosztu obliczeń, w szczególności ich efektywności termodynamicznej. Niniejsza rozprawa proponuje spójne ramy takiej oceny i stosuje je do współczesnych wyżarzaczy kwantowych.

Pierwszym rezultatem pracy jest opracowanie pakietu \texttt{SpinGlassPEPS.jl} - heurystyki opartej na sieciach tensorowych, przystosowanej do grafów o strukturze quasi-dwuwymiarowej zgodnej z łącznością procesorów QPU (np. Pegasus i Zephyr). Metoda wykorzystuje reprezentację funkcji partycji w postaci sieci tensorowej PEPS oraz przeszukiwanie typu branch-and-bound w przestrzeni prawdopodobieństwa, umożliwiając kontrolę dokładności obliczeń poprzez parametry takie jak odwrotna temperatura czy wymiar wiązania. Na tej podstawie przeprowadzono bezpośrednie porównania wyżarzaczy D-Wave z klasycznymi oraz inspirowanymi kwantowo algorytmami optymalizacyjnymi dla losowych instancji szkła spinowego zdefiniowanych natywnie na grafach Pegasus i Zephyr, obejmujących układy liczące kilka tysięcy spinów. Wyniki pokazują, że SpinGlassPEPS.jl stanowi użyteczny i fizycznie interpretowalny punkt odniesienia w benchmarkingu wyżarzaczy kwantowych, choć w przypadku największych instancji jego konkurencyjność ograniczają błędy wynikające z przybliżonej kontrakcji sieci tensorowych.

Drugim filarem pracy jest analiza wyżarzaczy kwantowych z perspektywy termodynamicznej. Urządzenia te traktowane są jako maszyny cieplne, w których wydajność obliczeniowa pozostaje powiązana z dyssypacją energii, produkcją entropii oraz efektywną temperaturą w trakcie realizacji harmonogramu wyżarzania. Przeprowadzone badania pokazują, że odpowiednio dobrane pauzy w trakcie wyżarzania mogą jednocześnie poprawiać jakość uzyskiwanych rozwiązań oraz zmniejszać koszt termodynamiczny procesu. Wskazano również zakresy parametrów, w których obecność pola podłużnego przestaje być korzystna po wprowadzeniu takich pauz.

Trzecim elementem pracy jest metoda postprocessingu oparta na uczeniu ze wzmocnieniem. W zaproponowanym podejściu konfiguracje generowane przez wyżarzacz wykorzystywane są jako dane wejściowe dla agenta RL, który uczy się wykonywać lokalne operacje odwracania spinów korygujące typowe wzorce błędów. Po wytrenowaniu agent zwiększa prawdopodobieństwo uzyskania konfiguracji o niższej energii, poprawiając jakość wyników bez konieczności modyfikacji samego procesu wyżarzania.

Ostatnia część pracy obejmuje dokładne symulacje adiabatycznej dynamiki niewielkich układów kwantowych, które pozwalają lepiej zrozumieć mechanizmy działania wyżarzania kwantowego oraz skonfrontować modele teoretyczne z danymi eksperymentalnymi. Uzyskane wyniki sugerują, że odtworzenie obserwowanych statystyk wyjściowych może wymagać przyjęcia większego poziomu efektywnego szumu, co wskazuje na dodatkowe źródła błędów bądź silniejsze sprzężenie układu ze środowiskiem.

Uzyskane wyniki wskazują, że wiarygodny benchmarking technologii wyżarzania kwantowego powinien opierać się na spójnych protokołach porównawczych, łączących silne klasyczne punkty odniesienia z analizą zarówno wydajności obliczeniowej, jak i fizycznego kosztu obliczeń.
\end{otherlanguage}

%% file: chapters/introduction.tex
\chapter{Introduction}



The concept of quantum computation originated in the early 1980s, when R. P. Feynman and D. Deutsch independently proposed exploiting the principles of quantum mechanics to perform computations beyond the capabilities of classical devices~\cite{Feynman1982,Deutsch1985}. Since then, the field has evolved into a distinct research discipline, encompassing both theoretical formulations of quantum algorithms and the experimental realization of quantum computing by specifically designed controllable quantum systems. These systems, which utilize quantum phenomena such as superposition and entanglement, offer a fundamentally different computational paradigm than classical devices~\cite{Nielsen2011}. Harnessing these properties opens the possibility of significant computational breakthroughs in domains such as combinatorial optimization, quantum simulation, and cryptography~\cite{Shor1997,Kaplan2016,Dridi2017}.

Among the various models of quantum computation, \emph{quantum annealing} constitutes a physically motivated approach based on the adiabatic theorem of quantum mechanics~\cite{Kato1950}. It aims to find the ground state of a problem Hamiltonian that encodes an instance of an optimization problem~\cite{Kadowaki1998, Tanaka2017}. Quantum annealing, as an implementation of adiabatic quantum computation~\cite{Albash2018}, has proven particularly suitable for optimization tasks and has been realized experimentally in commercially available devices, such as those produced by D-Wave Systems~\cite{dwave}.

Parallel to the development of quantum hardware, there has been substantial progress in \emph{quantum-inspired} classical algorithms, \textit{i.e.}, methods that draw inspiration from quantum mechanics to enhance classical optimization and sampling techniques~\cite{Goto2019,Goto2021,Jiang2023}. These approaches are particularly important, as they provide a valuable baseline for benchmarking quantum devices.

Benchmarking constitutes an indispensable component of contemporary quantum computing research~\cite{Munoz-Bauza2025,Tuziemski2025,Pawlowski2025}. It enables systematic comparison between quantum annealers, quantum-inspired classical solvers, and traditional optimization algorithms. Comprehensive benchmarking efforts must consider not only computational performance, such as solution quality, time-to-solution, and success probability, but also the \emph{physical aspects} of these devices, including thermodynamic efficiency. 

Another critical aspect of current research concerns mitigating noise and imperfections inherent to quantum hardware. \emph{Error correction} and \emph{error mitigation} techniques are essential for enhancing computational reliability and accuracy. While fully fault-tolerant quantum computation remains beyond the scope of existing annealing-based architectures, targeted error mitigation strategies can substantially improve their performance.

This thesis is devoted to the assessment of quantum annealers, the development and implementation of a tensor-network-based algorithm, and the investigation of the physical and thermodynamic properties of quantum annealing devices.

\paragraph{Contributions}
The main contributions of this thesis are:
\begin{itemize}
    \item A topology-aware tensor-network software package and heuristic
    (SpinGlassPEPS.jl) for optimization and sampling on quasi-two-dimensional
    graphs aligned with QPU connectivity.
    \item A benchmarking methodology and head-to-head evaluation of quantum
    annealers against classical, quantum-inspired, and tensor-network baselines
    using like-for-like metrics on relevant instance families.
    \item A thermodynamic characterization of quantum annealers as effective
    thermal machines, linking computational performance to physical
    expenditure and enabling schedule assessment via efficiency measures.
    \item A reinforcement learning post-processing approach for improving returned samples and mitigating effective errors without changing the anneal.
    \item Exact unitary simulation of small D-Wave annealing instances used to check quantitative agreement between simulated and experimentally observed output distributions.
\end{itemize}

Overall, the thesis aims to provide a coherent framework for benchmarking
quantum annealing technologies in which \emph{performance claims} are
supported by strong classical baselines and complemented by an explicit
accounting of \emph{physical cost}.

\section{Layout of the thesis}

The thesis begins in Chapter~\ref{chap:ising} by establishing the problem setting: the classical Ising model and its QUBO counterpart are introduced, notation is fixed, and baseline concepts and algorithms used throughout the dissertation are summarized.

Background material on adiabatic quantum computation and quantum annealing is presented in Chapter~\ref{chap:AQC-QA}, with emphasis on the operational characteristics and constraints of D-Wave devices that are most relevant for experimental evaluation and benchmarking.

A baseline is developed in Chapter~\ref{chap:tn} through \texttt{SpinGlassPEPS.jl}, a tensor network based heuristic for optimization and sampling of Ising-like problems. The chapter details the underlying methods, the modular software design, and illustrative examples, motivated by the need for transparent, topology-aware classical comparators.

Benchmarking results are reported in Chapter~\ref{chap:bench}, where quantum annealers are compared with selected classical solvers, including tensor-network and quantum-inspired approaches. The chapter defines the benchmarking protocol, instance families, and evaluation metrics (e.g., best energy/approximation ratio, success probability, time-to-solution, and diversity of low-energy samples) to ensure controlled, like-for-like comparisons.

Thermodynamic aspects are examined in Chapter~\ref{chap:thermo} by modeling a quantum annealer as an effective thermodynamic machine. This analysis connects solution quality and success probability to thermodynamic cost, providing a schedule-evaluation viewpoint that complements purely computational performance measures.

Post-processing as an error-mitigation technique is investigated in Chapter~\ref{chap:rl} using a reinforcement learning agent designed to improve solutions produced by a quantum annealer. The method is interpreted as mitigating effective errors observable in the returned configurations and is evaluated empirically.

To support intuition about annealing dynamics and validate selected modeling choices in classically tractable regimes, exact simulations of adiabatic quantum evolution for small systems are presented in Chapter~\ref{chap:sim}.

The dissertation concludes in Chapter~\ref{chap:last} by summarizing the main findings, discussing limitations, and outlining directions for future work.

%% file: chapters/ising.tex
\chapter{Classical Ising Model and QUBO Problem}
\chaptermark{Ising Model and QUBO}
\label{chap:ising}

The classical Ising model~\cite{Ising1925,Niss2005}  is simultaneously a foundational model of collective behaviour in statistical physics~\cite{Drell1977,Mouritsen1984,Cipra1987,McCabe2006,Dusuel2009,Niss2009,Niss2011,McCoy2012} and a universal intermediate representation for a large class of discrete optimization problems~\cite{Galam2008,Perdomo-Ortiz2012,Lucas2014,Kochenberger2014,Lima2017,Punnen2022,Salehi2022}. This dual role makes it an ideal technical substrate for benchmarking quantum annealing technologies: D-Wave processors natively implement programmable transverse-field Ising Hamiltonians, while many real-world tasks can be reduced to equivalent quadratic binary objectives. Consequently, by expressing problems in the Ising/QUBO language, we can compare different solvers, such as quantum annealers, quantum-inspired Ising machines, and classical algorithms, on exactly the same cost function and with controlled, like-for-like evaluation metrics.

At the same time, ``benchmarking an Ising machine'' is not just benchmarking an optimizer. In practice, these devices are also stochastic samplers whose output statistics reflect a competition between the programmed energy landscape, finite-temperature effects, and implementation imperfections. A careful introduction must therefore cover both viewpoints: (a) ground-state search as an NP-hard optimization problem~\cite{Barahona1982}, where time-to-solution and best-energy metrics are natural, and (b) statistical descriptions via the partition function, where sampling quality, low-energy diversity, and thermodynamic notions become meaningful and later connect to energy-efficiency analysis. This chapter establishes the shared notation and conceptual bridge that the rest of the thesis repeatedly relies on: the same Hamiltonian will be used to define benchmark instance families and to construct strong, topology-aware classical baselines.

Concretely, this chapter: (i) defines the Ising model on general graphs, fixing conventions used throughout the dissertation and introduces the Gibbs distribution and partition function as the statistical-mechanics backbone needed later for thermodynamic characterization (\ref{sec:ising-classical}); (ii) presents the QUBO formulation and the explicit Ising--QUBO mapping, which is the standard interface used by optimization software (\ref{sec:ising-qubo}); and (iii) reviews representative classical methods for low-energy search that serve as baselines in subsequent benchmarking chapters (\ref{sec:ground-state}).

\section{Classical Ising model}
\label{sec:ising-classical}

We will formulate the classical Ising model using graph-theoretic notation. Let $G=(V,E)$ be a simple graph with $|V|=N$ vertices. To each vertex $i\in V$ we associate a binary spin variable $s_i\in\{-1,+1\}$ and a real parameter $h_i\in\mathbb{R}$ representing a local external field (often referred to as a \emph{bias} in the optimization literature). A spin assignment $\bm{s}=(s_1,s_2,\ldots,s_N)$ is called a \emph{state} or \emph{configuration}. To each edge $(i,j)\in E$ we assign a real coupling $J_{ij}\in\mathbb{R}$ that quantifies the interaction (coupling) strength between spins $s_i$ and $s_j$. The energy function (Hamiltonian) $H:\{-1,+1\}^N\to\mathbb{R}$ is then defined by~\cite{Rams2021}:

\begin{equation}
    \label{eq:ising-classical}
    H(\bm{s}) = \sum_{(i, j) \in E} J_{ij} s_i s_j + \sum_{i\in V} h_i s_i.
\end{equation}
	
In the literature, several alternative conventions for the Ising Hamiltonian can be found. These formulations differ only by simple transformations, and they are all equivalent up to rescaling of the parameters $h_i$ and $J_{ij}$. The convention adopted in equation~\eqref{eq:ising-classical} is the one most commonly used in the quantum annealing literature and will be employed throughout this work.

We will consider the general Ising model defined on an arbitrary graph, in which the parameters $h_i$ and $J_{ij}$ can take any real values. Such models are often referred to as spin glasses~\cite{Lucas2014}.

\begin{example}
    \label{example:1}
 Let's consider an Ising model instance defined on the complete graph $K_3$ and given by the Hamiltonian:

 \begin{equation}
    H(\bm{s}) = s_1 - s_2 + 1.5s_3 + s_1s_2 + 0.5 s_1 s_3 - 0.75 s_2 s_3
    \label{eq:example-ising} 
 \end{equation}

\noindent For the state $\bm{s} = (-1, 1, 1)$ the instance~\ref{eq:example-ising} has energy $H((-1, 1, 1)) = -2.75$.
\end{example}

\begin{figure}
    \centering
    \includegraphics[width=0.8\textwidth]{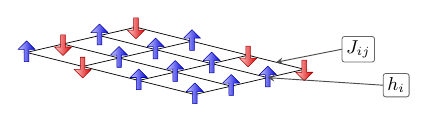}
    \caption[Symbolic representation of an example Ising spin glass system]{Symbolic representation of an example Ising spin glass system. It's defined on a $4 \times 4$ lattice graph with $N=16$ vertices. Here, the configuration of each spin is represented by a blue arrow pointing upwards ($+1$) or a red arrow pointing downwards ($-1$). Additionally, $h_i \in \mathbb{R}$ is the external field acting on the $i-$th spin and $J_{ij}$ denotes coupling strength between spins $i$ and $j$.}
    \label{fig:example-ising}
\end{figure}

Many properties of the classical Ising model follow from its \emph{partition function}. Let $\mathcal{S}=\{-1,1\}^N$ denote the set of all spin configurations and let $H(\bm{s})$ be the Hamiltonian in Eq.~\eqref{eq:ising-classical}. The partition function is defined as:

\begin{equation}
\label{eq:ising-partition}
    Z=\sum_{\bm{s}\in\mathcal{S}} e^{-\beta H(\bm{s})},
\end{equation}

\noindent where $\beta=(k_{\mathrm{B}}T)^{-1}$ is the inverse temperature. Throughout this thesis we set $k_{\mathrm{B}}=1$, so $\beta=1/T$ and $T$ should be interpreted as an \emph{effective} (dimensionless) temperature. The partition function normalizes the corresponding equilibrium distribution, \textit{i.e.}, the probability of observing configuration $\bm{s}$ at inverse temperature $\beta$ is given by the \emph{Boltzmann} (or \emph{Gibbs}) \emph{distribution}~\cite{Chandler1987}:

\begin{equation}
\label{eq:gibbs}
    P_\beta(\bm{s})=\frac{e^{-\beta H(\bm{s})}}{Z}.
\end{equation}

In particular, this distribution allows us to describe thermodynamic properties, which we further investigate in Chapter~\ref{chap:thermo}. Finally, the limiting cases connect statistical physics to optimization: as $\beta\to 0$ the distribution becomes
approximately uniform over $\mathcal{S}$, whereas as $\beta\to\infty$ it concentrates on low-energy configurations of $H$.

A common problem associated with the Ising model is to find a state $\bm{s}^*$ that minimizes the Hamiltonian $H(\bm{s})$ from equation~\eqref{eq:ising-classical}. Such a configuration is called the \emph{ground state}, and is usually very hard to find. This is the bridge that connects the statistical physics model to combinatorial optimisation, which is central to this thesis. The problem of finding the ground state will be analyzed in detail in section~\ref{sec:ground-state}.

\subsection{Connection to QUBO problem}
\label{sec:ising-qubo}

Quadratic Unconstrained Binary Optimization (QUBO) is a mathematical model for combinatorial optimization problems~\cite{Glover2022}. As the name suggests, it concerns quadratic dependencies among variables, no constraints, and binary variables. This description encompasses a surprisingly large class of problems, such as MAX-CUT, the traveling salesman problem, or job shop scheduling~\cite{Kochenberger2014,Punnen2022}.

Formally, given a $N \times N$ matrix of coefficients $Q$, the QUBO objective can be written as:
\begin{equation}
   \bm{x}^\star = \arg \min_{\bm{x}\in\{0,1\}^{N}} E_{\mathrm{QUBO}}(\bm{x}),
\end{equation}

\begin{equation}
E_{\mathrm{QUBO}}(\bm{x}) = \bm{x}Q\bm{x}^T = \sum_{i=1}^{N} Q_{ii} x_i + \sum_{1\le i<j\le N} Q_{ij} x_i x_j,
\label{eq:qubo}
\end{equation}

\noindent where we used the fact that $x_i^2=x_i$ to keep linear terms on the diagonal. If a full symmetric matrix $Q$ is provided, it is common to fold it into an upper-triangular representation by adding symmetric entries, so that each unordered pair $\{i,j\}$ is represented once.

Solving QUBO is equivalent (up to an additive constant) to finding the ground state of a general Ising model. The equivalence follows from the standard change of variables:

\begin{equation}
x_i = \frac{s_i+1}{2}, \qquad s_i = 2x_i - 1.
\label{eq:ising-qubo-subst}
\end{equation}

\paragraph{Ising $\rightarrow$ QUBO.}
Consider the Ising Hamiltonian from Eq.~\eqref{eq:ising-classical}. Substituting $s_i=2x_i-1$ and collecting terms yields a QUBO of the form \eqref{eq:qubo} with coefficients:

\begin{align}
Q_{ij} &= 4J_{ij}, \label{eq:ising-to-qubo-offdiag}\\
Q_{ii} &= 2h_i - 2\sum_{j \neq i} J_{ij}. \label{eq:ising-to-qubo-diag}
\end{align}

\noindent The transformation introduces an additive constant:

\begin{equation}    
C_{I\to Q}=\sum_{i<j} J_{ij} - \sum_{i=1}^N h_i,
\label{eq:ising-to-qubo-const}
\end{equation}

\noindent which does not affect the minimizer but shifts objective values.

\paragraph{QUBO $\rightarrow$ Ising.}
Conversely, given QUBO coefficients $\{Q_{ii},Q_{ij}\}$ (for $i<j$), substituting $x_i=(s_i+1)/2$ gives an Ising Hamiltonian with:

\begin{align}
J_{ij} &= \frac{1}{4}Q_{ij}, \label{eq:qubo-to-ising-offdiag}\\
h_i &= \frac{1}{2}Q_{ii} + \frac{1}{4}\sum_{j\ne i} Q_{ij}. \label{eq:qubo-to-ising-diag}
\end{align}

\noindent Again, an additive constant appears:

\begin{equation}
C_{Q\to I}=\sum_{ i<j}\frac{1}{4}Q_{ij} + \sum_{i=1}^N \frac{1}{2}Q_{ii}.
\label{eq:qubo-to-ising-const}
\end{equation}

In summary, QUBO and Ising are equivalent optimization problems (they share the same $\arg\min$ under \eqref{eq:ising-qubo-subst}), while their objective values differ by a constant shift given in \eqref{eq:ising-to-qubo-const} or \eqref{eq:qubo-to-ising-const}.

\begin{example} Consider an instance from the previous example~(\ref{example:1}). When transformed to QUBO:

    \begin{align}
        \label{eq:example-qubo}
        \begin{split}
        E_{\mathrm{QUBO}}(\bm{x}) &= -x_1 - 2.5 x_2 + 3.5 x_3 + 4x_1 x_2 + 2 x_1 x_3 - 3 x_2 x_3,\\
        C_{I \to Q} &= -0.75.
    \end{split}
\end{align}
    \noindent If we calculate energy for the state $\bm{x} = (0, 1, 1)$, which corresponds to the state $\bm{s} = (-1, 1, 1)$ from the previous example, we get $E((0, 1, 1)) = -2$. Only after adding the offset constant $C_{I \to Q}$ do we get the correct result $E(\bm{x}) + C_{I \to Q} = -2.75$.
\end{example}

\section{Problem of finding the ground state}
\label{sec:ground-state}

A central computational task associated with the Ising model is the determination of its \emph{ground state}, \textit{i.e.}, a configuration $\bm{s}^\star\in\mathcal{S}$ minimizing the energy $H(\bm{s})$ in Eq.~\eqref{eq:ising-classical}. While several restricted cases admit exact solutions, for example, the one-dimensional Ising chain~\cite{Ising1925} and the two-dimensional model with zero external field ($h_i=0$)~\cite{Onsager1944}, the problem is intractable in general. More precisely, finding a ground state of an arbitrary Ising spin glass on a general graph is NP-hard~\cite{Barahona1982}. Consequently, assuming $P\neq NP$, there is no polynomial-time algorithm that can \emph{both} find and certify a ground state for all instances. This motivates the use of heuristic and approximation methods in practical settings, as seen in the benchmarking studies presented in later chapters. By the standard Ising-QUBO correspondence, ground-state search is equivalent to solving the associated QUBO minimization problem.

The importance of this formulation stems from the fact that many discrete optimization problems can be expressed as
QUBO/Ising instances. Canonical examples include Karp's $21$ NP-complete problems~\cite{Karp1972,Lucas2014} as well
as widely studied tasks such as the traveling salesman problem~\cite{Salehi2022}, portfolio optimization~\cite{Lima2017},
and protein folding models~\cite{Perdomo-Ortiz2012}. This expressive power makes the Ising/QUBO formalism a convenient common framework for benchmarking combinatorial-optimization solvers.

A broad spectrum of algorithms exists for Ising/QUBO optimization, ranging from exact methods to scalable heuristics and physics-inspired approaches. In the remainder of this chapter, we will present some of those classical baselines
that will later serve as reference points when evaluating quantum-annealing hardware and a novel tensor-network approach in Chapter~\ref{chap:bench}.

\subsection{Exhaustive search}

The most direct approach to a combinatorial optimization problem is \emph{exhaustive search}: evaluate the objective for every candidate solution and return the best one. For Ising/QUBO instances with $N$ spins (binary variables), the search space has size $|\mathcal{S}|=2^N$. Consequently, the runtime grows exponentially with $N$, and exhaustive enumeration becomes impractical already for moderately sized instances. In practice, even problems with $N\approx 50$--$60$ variables are typically out of reach without highly specialized implementations and
significant computational resources.

Despite this limitation, exhaustive methods remain an active area of development because they provide an unambiguous
reference for optimality. Recent work improves constant factors by: (i) traversing the configuration space using Gray codes to reduce update costs between successive configurations and (ii) exploiting the massive parallelism of GPU architectures to increase the rate of energy evaluations~\cite{Jalowiecki2021,Mucke2023}. These advances do not change the exponential scaling but can extend the range of instances that can be solved exactly.

Exhaustive search is also one of the few approaches that \emph{guarantees} finding a ground state and therefore
supports exact certification. Other exact frameworks include branch-and-bound (with strong bounding rules)~\cite{Hartwig1984},
branch-and-cut~\cite{DeSimone1995}, and selected tensor-network approaches~\cite{Liu2021}. All of these methods, however, still exhibit exponential worst-case complexity in the general case.

Finally, exhaustive enumeration plays a practical role in benchmarking: whenever feasible, it serves as a gold standard for validating solutions returned by heuristic solvers and for quantifying suboptimality in controlled small-scale experiments.

\subsection{Branch and bound}

\emph{Branch and bound} (B\&B) is a general framework for exact discrete optimization based on exploring a search tree of subproblems while pruning regions that can be proven incapable of improving the best solution found so far. In the Ising/QUBO setting, each node of the tree corresponds to a \emph{partial assignment} of spins (binary variables). Branching creates children by fixing an additional variable (or a small block of variables), thereby partitioning the remaining configuration space into smaller subproblems. For each node, one computes a \emph{lower bound} on the minimum energy attainable by any completion of the partial assignment. If this bound is not better than the energy of the current incumbent solution, the entire subtree can be discarded without affecting optimality. When the search terminates, the incumbent is guaranteed to be a global ground state (or global optimum in the QUBO form)~\cite{Lawler1966,Morrison2016}.

The practical performance of B\&B is dominated by the bounding step: tighter bounds yield more pruning but are often
more expensive to compute~\cite{Morrison2016}. For Ising/QUBO problems, bounds can be derived from relaxations (e.g., linear or
semidefinite)~\cite{Rendl1995}, problem decompositions~\cite{Hartwig1984}, or structure-specific arguments exploiting limited connectivity~\cite{Lu2023}. Nevertheless, B\&B retains exponential worst-case complexity on general graphs, and its efficiency depends strongly on the instance family and parameter regime~\cite{Chvatal1980}.

\paragraph{Heuristic B\&B.}
If the requirement of certified bounds is relaxed, the same search-tree viewpoint leads to a practical heuristic (sometimes referred to as \emph{beam search}~\cite{Ow1988}). At each depth $d$ of the tree one keeps only a fixed number $M$ of the most
promising partial assignments. Concretely, each retained node is branched by assigning one additional variable, generating a pool of candidates. These candidates are ranked by a computationally cheap score, for example, the energy of a partial state. Only the top $M$ candidates are kept, and the remainder are discarded. The procedure is iterated until all variables are assigned, yielding up to $M$ complete configurations that approximate low-energy states of the original Hamiltonian~\cite{Rams2021}.

The parameter $M$ controls the trade-off between computational cost and search breadth: increasing $M$ reduces the risk of discarding globally optimal branches but increases runtime and memory. For simple per-extension scoring rules, the dominant cost scales approximately as $\mathcal{O}(MN)$ energy-update operations, multiplied by the cost of the chosen scoring/estimation routine.

This heuristic perspective is useful in this thesis because it provides a common template for combining search with physics-informed estimators. In particular, in Chapter~\ref{chap:tn} we describe \texttt{SpinGlassPEPS.jl}, where tensor-network contractions provide informative estimates for partial assignments, enabling both approximate optimization and sampling of Ising/QUBO models within a heuristic branch and bound framework~\cite{peps,software}.

\subsection{Simulated annealing}

Simulated annealing (SA) is a stochastic optimization heuristic inspired by thermal annealing in statistical physics. In the Ising/QUBO setting, SA can be viewed as a Markov-chain Monte Carlo procedure that performs local updates while gradually decreasing the temperature $T$  (equivalently, increasing the inverse temperature $\beta$). By slowly varying $T$, SA biases the search toward low-energy configurations and, in favorable cases, toward ground states~\cite{Kirkpatric1983,Cerny1985}.

In its basic form, SA maintains a current configuration $\bm{s}$ and repeatedly proposes a neighboring configuration $\bm{s}' \sim \texttt{neighbor}(\bm{s})$. For Ising models, the standard neighborhood is a single-spin flip ($s_i \mapsto -s_i$), and for QUBO, an analogous single-bit flip. Writing $\Delta E = H(\bm{s}')-H(\bm{s})$, the Metropolis acceptance rule is:

\begin{equation}
\label{eq:sa_accept}
P(\beta, \Delta E)=
\begin{cases}
    1, & \mathrm{if} \, \Delta E < 0,\\
    \exp (-\beta \Delta E), & \mathrm{otherwise.}
\end{cases}
\end{equation}

Hence, downhill moves ($\Delta E\le 0$) are accepted deterministically, while uphill moves are accepted with a
probability that decreases with both $\Delta E$ and $\beta$. The behavior of SA is governed by three design choices. First, the \emph{temperature schedule} $\{T_k\}$ determines how quickly the algorithm transitions from exploratory to exploitative dynamics. Common choices include geometric schedules ($T_{k+1}=\alpha T_k$) and other monotone schedules~\cite{Yao1995}. Second, one must
choose the number of Metropolis updates performed at each temperature. Third, a stopping criterion specifies when the run terminates, e.g., after a fixed computational budget, after a target $T_{\max}$ is reached, or after the acceptance rate (or
best energy) stabilizes.

Although SA is primarily used as a heuristic in practice, it is worth noting that sufficiently slow cooling schedules can yield theoretical convergence guarantees under standard assumptions (at the cost of impractically long runtimes). In this thesis, we employ the basic single-variable Metropolis SA as a classical baseline for Ising/QUBO ground-state search, with the schedule and neighborhood specified by the functions \texttt{schedule} and \texttt{neighbor}.

SA is also straightforward to parallelize. A simple strategy is to run multiple independent replicas (independent Markov chains) in parallel, each initialized from a different random configuration. This increases the probability of finding low-energy states under a fixed number of annealing steps.

\begin{algorithm}[H]
\label{sa}
\caption{Simulated annealing for Ising model}
\KwData{Initial configuration $S_{init}$, initial temperature $T_{0}$, number of steps $K$, number of repetitions $M$}
\KwResult{Approximate solution to the ground state problem}

\SetKwFunction{random}{random}
\SetKwFunction{neighbor}{neighbor}
\SetKwFunction{schedule}{schedule}

\tcp{Initialization}

$T \gets T_0$\;
$S \gets S_{init}$\;

\For{k = 0 \KwTo $K$}{

    \For{m = 0 \KwTo $M$}{
    	$S' \gets$ \neighbor{$S$}\; 
    	$E' \gets E(S')$\;
        $\Delta E \gets E(S') - E(S)$\;
    	
    	\eIf{$\Delta E < 0$}{
    		$S \gets S'$\;

    	}{
    		$r \gets$ \random{$0, 1$}\;
    		$\beta \gets \frac{1}{T}$\;
    		\If{$r < \exp\left(-\beta\Delta E\right)$}{
    			$S \gets S'$\;
    		}
	   }
       }
       
	$T \gets$ \schedule{$T$} \tcp{Update temperature}
}
\Return $S$ \tcp{Return the best found solution}
\end{algorithm}

\subsection{Parallel annealing}

Parallel Annealing (PA) is a quantum-inspired classical heuristic that is highly parallelizable. It was originally demonstrated on an analog memristor crossbar architecture~\cite{Yesheng2021,Jiang2023}, but its update rule consists of vector--matrix operations and elementwise nonlinearities, making it also well suited for efficient GPU implementations.

PA constructs a time-dependent auxiliary objective that interpolates between an easy energy landscape and the target
Ising Hamiltonian. Following~\cite{Jiang2023}, one introduces continuous variables $\bm{x}\in\mathbb{R}^N$ and defines
the binary spin configuration by projecting those variables into the binary ($\pm 1$) domain. The easiest, but not the only, method is to use the $\text{sign}()$ function. The central object is the scheduled system Hamiltonian:

\begin{equation}
\label{eq:pa}
    H_{\text{system}} = H_{\text{Ising}} + \lambda(t)\,H_{\text{initial}},
\end{equation}

\noindent where $\lambda(t)$ is decreased from a large value to $0$ during the run. The initialization term is typically chosen as:

\begin{equation}
    H_{\text{initial}}(\bm{x}) = \frac{1}{2}\sum_{i=1}^{N} x_i^2,
\end{equation}

\noindent so that for large $\lambda(t)$ the dynamics are dominated by a simple convex objective. As $\lambda(t)\to 0$, the optimization progressively focuses on the target Ising energy. This construction mirrors the \emph{interpolation idea} of adiabatic evolution (Chapter~\ref{chap:AQC-QA}), but in PA the trajectory toward low energy is enforced algorithmically by gradient descent~\cite{Boyd2004} rather than arising from physical adiabatic dynamics (cf.~\cite{Kato1950}).

A direct gradient method is obstructed by the discontinuity of the spin variables. PA therefore employs a straight-through estimator (STE)~\cite{Bengio2013,Hubara2016}. A core idea of the STE is to maintain a full-precision “latent” variable that serves as a proxy for a binary variable. The latent value is binarized in both the forward and backward passes to compute the gradient, which is then used directly to update the underlying full-precision latent variable. The analog spin variable $x_i$ serves as a proxy for the spin.

To implement the STE algorithm for the system described by~\eqref{eq:pa}, we first projected the analog spin variable $\bm{x}$ onto the binary domain using a $\mathrm{sign}()$ function to obtain the spin configuration. We then used $\bm{s}$ to compute the gradient of the system Hamiltonian. Under this method, the gradient used for updates can be written as~\cite{Petersen2012,Jiang2023}:

\begin{equation}
    \label{eq:pa-gradient}
   \nabla H_{\text{system}} = \nabla H_{\text{Ising}} + \lambda(t) \nabla H_{\text{initial}} = -\bm{J}\bm{s}  \bm{h} + \lambda(t) \bm{x},
\end{equation}

\noindent in the common case of symmetric couplings (Hermitian $J$) with minus sign convention. In general setting, the gradient $\nabla H_{\text{Ising}} = (\bm{J}+\bm{J}^T)\bm{s} + \bm{h}$.

\begin{figure}[H]
    \centering
    \includegraphics[width=\textwidth]{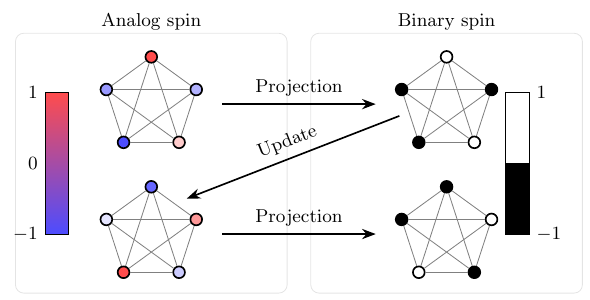}
    \caption[Schematic representation of parallel annealing algorithm]{Schematic representation of parallel annealing algorithm. The analog spin variables $\bm{x}$ are projected onto the binary domain using the $\mathrm{sign}$ function. Thus, the obtained real spin configuration $\bm{s}$ is used to calculate the gradient and update the analog variables. This process continues until a specified number of steps is reached.}
    \label{fig:pa-schema}
\end{figure}

To stabilize and accelerate convergence, PA adopts standard stochastic optimization techniques, including clipping and momentum~\cite{Qian1999,Pascanu2013,Jiang2023}. Clipping constrains the analog variables to a bounded domain:
\begin{equation}
    \mathrm{clip}(x,-1,1)=\max\!\left(-1,\min(1,x)\right),
\end{equation}

\noindent preventing numerical divergence. With momentum, one updates:

\begin{align}
\label{eq:pa_momentum_rewrite}
    \bm{m}(t+1) &= \beta\,\bm{m}(t) - \alpha\,\nabla_{\bm{x}} H_{\text{system}},\\
    \bm{x}(t+1) &= \bm{x}(t) + \bm{m}(t+1),
\end{align}

\noindent where $\alpha$ is the step size and $\beta$ is the momentum parameter (often chosen to be $1-\alpha$). In practice, $\bm{x}$ and $\bm{m}$ are clipped after each update to avoid explosion.

Below, we present pseudocode for the parallel annealing algorithm. The \texttt{schedule} function is a function that governs the values and interpolation of $\lambda(t)$.

\begin{algorithm}
\label{alg:pa}
\caption{Parallel annealing for Ising model}
\KwData{Step size $\alpha$, Momentum parameter $\beta$, Number of steps $K$, Number of trajectories $M$}
\KwResult{Approximate solution to the ground state problem}

\SetKwFunction{random}{random\_state}
\SetKwFunction{grad}{calculate\_gradient}
\SetKwFunction{clip}{clip}
\SetKwFunction{projection}{projection}
\SetKwFunction{schedule}{schedule}

$\bm{x} \gets \mathbf{0}$ \tcp*[l]{Initialize spin variable in ground state of $H_{\text{initial}}$}
$\bm{m} \gets \mathbf{0}$ \tcp*[l]{Initialize momentum at 0}
$\bm{s} \gets$ \random{} \tcp*[l]{Initialize random state}

\For{$t = 0$ \KwTo $K$}{
    $\lambda \gets$ \schedule{$t$}\;
    $\nabla \gets$ \grad{$\bm{s}, \bm{x}$, $\lambda$}\; \tcp{as in equation (\ref{eq:pa-gradient})}
    $\bm{m} \gets \beta * \bm{m} - \alpha * \nabla$\; 
    $\bm{m} \gets$ \clip{$\bm{m}, -1, 1$}\;
    $\bm{x} \gets \bm{x} + \bm{m}$\;
    $\bm{x} \gets$ \clip{$\bm{x}, -1, 1$}\;
    $\bm{s} \gets$ \projection{$\bm{x}$}\;
}
\Return $\bm{s}$ \tcp{Return the solution}
\end{algorithm}

\subsection{Simulated bifurcation}

Simulated bifurcation (SB) is a physics-inspired heuristic for Ising/QUBO optimization based on the dynamics of a network of parametrically driven nonlinear oscillators. The method was proposed as a classical analog of adiabatic optimization, where the target Ising instance is encoded in coupling terms of an effective Hamiltonian and a slowly varying parameter induces a bifurcation that selects one of $2^N$ stable configurations~\cite{Goto2019}. In this thesis, we focus on the practically useful discretized variant, \emph{Discrete Simulated Bifurcation} (dSB), which introduces explicit discretization steps~\cite{Goto2021}.

Full derivation of relevant equations can be found in~\cite{Goto2019}. Here we will focus more on the overall picture and relevant intuitions. One can think of this algorithm as encoding the given Ising optimization problem as a dynamical system of oscillating particles, where each one is described by position $x_i$ and momentum $y_i$. This system is described by the classical Hamiltonian:

\begin{equation}
    \mathcal{H}_{\mathrm{SBM}} = \frac{a_0}{2} \sum_i y_i^2 + \frac{a_0 - a(t)}{2} \sum_i x_i^2 + c_0 H_{I}(\bm{x}), 
\end{equation}

\noindent  $a_0>0$ and $c_0>0$ are hyperparameters, $a(t)$ is a continuous function going from 0 at $t=0$ to $a_0$ at $t=\tau$, where $\tau$ is total time. Function $H_{I}(\bm{x})$ is reformulation of Ising model as:

\begin{equation}
    H_{I}(\bm{x}) = -\frac{1}{2} \sum_{i,j} J_{i,j} x_i x_j - \sum_i^N h_i x_i.
\end{equation}

The evolution of $\mathcal{H}_{\mathrm{SBM}}$ described by the following differential equations:

\begin{equation}
    \label{eq:sbm-x}
    \dot{x}_i = \frac{\partial \mathcal{H}_{\mathrm{SBM}}}{\partial y_i} = a_0 y_i,
\end{equation}

\begin{equation}
    \label{eq:sbm-y}
    \dot{y}_i = - \frac{\partial \mathcal{H}_{\mathrm{SBM}}}{\partial x_i} = -[a_0 - a(t)]x_i + c_0 \left(\sum_{j=1}^N J_{ij} x_j + h_i \right).
\end{equation}

The system's nonlinearity is introduced in two places. First, we add a sign function to the Ising gradient, replacing $x_j$ with $\mathrm{sign}(x_j)$. The new derivative has the following form:

\begin{equation}
    \label{eq:sbm-y-sign}
    \dot{y}_i^{\text{nonlinear}}  = -[a_0 - a(t)]x_i + c_0 \left(\sum_{j=1}^N J_{ij} \mathrm{sign}(x_j) + h_i \right).
\end{equation}

\noindent We will drop the superscript in the rest of the text and use the nonlinear form of the $y$ derivative.  

Another source of nonlinearity arises from a perfectly inelastic wall at $\vert x_i \vert = 1$. That is, if in any moment of the system's evolution $\vert x_i \vert > 1$, we replace $x_i$ with its sign, $\mathrm{sign}(x_i) = \pm 1$, and set $y_i=0$. It can be thought of as a particle hitting a wall and sticking to it. It can be realized by the function:

\begin{equation}
\label{eq:sbm-wall}
    \text{wall}(x_i, y_i) = \left\{
                            \begin{array}{ll}
                            (\mathrm{sign}(x_i), 0), & \text{if} \, \, \vert x \vert > 1 ,\\
                            (x_i, y_i), & \text{otherwise.}\\
                            \end{array}
                            \right.
\end{equation}

The mechanism of SBM is as follows. When $a(t)$ is gradually increased from $0$ to a sufficiently large value $a_f$, each oscillator exhibits a bifurcation with two stable branches whose positions are given by~\cite{Goto2019} $\pm \sqrt{(a_f - a_o)/K}$, where $K$ is a positive coefficient. Consequently, the final potential energy has $2^N$ minima corresponding to spin configurations. If $a(t)$ varies slowly enough, the state will follow one of the potential energy minima. Assuming that minima associated with lower final energies emerge earlier in the evolution and remain energetically below those corresponding to higher final energies, we expect the system to settle in a minimum with a comparatively low final energy. Thus, we will find a low-energy approximate solution.

To implement the SBM algorithm, we must solve the aforementioned differential equations~\eqref{eq:sbm-x} and~\eqref{eq:sbm-y}. This can be done numerically using the symplectic Euler method. This method works here because SB is based on Hamiltonian dynamics, and the symplectic Euler integrator is the simplest explicit, structure-preserving, and hardware-friendly integrator that maintains the stability of these dynamics~\cite{Leimkuhler2005}. The updating rule is as follows:

\begin{equation}
\label{eq:dsbm-euler-y}
    y_i(t_{n+1}) = y_i(t_n) + \left( -[a_0 - a(t_k)] x_i(t_k) + c_0 G(\bm{x}(t_{n+1}))\right)\Delta_t,
\end{equation}

\begin{equation}
\label{eq:dsbm-euler-x}
    x_i(t_{n+1}) = x_i(t_n) + a_0 y_i(t_n) \Delta_t.
\end{equation}

\noindent Here, $G(\bm{x}(t_{n+1}))$ is gradient of the Ising Hamiltonian defined as:

\begin{equation}
    G(\bm{x}(t_{n+1})) = \sum_{j=1}^N J_{ij}\mathrm{sign}(x_j(t_{n+1})) + h_i.
\end{equation}

\noindent $\Delta_t$ is the time step and $t_k$ is the discretized time satisfying $t_0 = 0$ and $t_{k+1} = t_k + \Delta_t$. The hyperparameters are typically set as $a_0 = 1$ and $c_0 = \frac{0.7 a_0}{\sigma \sqrt{N}}$, with $\sigma$ the standard deviation of the off-diagonal elements of $\bm{J}$. Value of $c_0$ is based on a random-matrix estimate for the largest eigenvalue of $\bm{J}$~\cite{Kanao2022} and its role is to guide the strength of the gradient upgrades in relation to the energy scale of matrix $\bm{J}$.

The dSB updates are dominated by dense/sparse matrix--vector products of the form $\sum_j J_{ij}\,\mathrm{sign}(x_j)$ and by elementwise operations (sign and wall), which map efficiently to GPU kernels. Moreover, the dynamics can be chaotic and sensitive to initial conditions, so practical deployments typically run many independent replicas with different initializations in parallel and return the best (lowest-energy) solution observed~\cite{Goto2021}.

\begin{algorithm}
\label{alg:sbm}
\caption{Simulated bifurcation for Ising model}
\KwData{Parameters: $a_0$, $c_0$, $\Delta t$, Function $a(t_k)$, Total time $T$}
\KwResult{Approximate solution to the ground state problem}

\SetKwFunction{random}{random\_vector}
\SetKwFunction{grad}{calculate\_gradient}
\SetKwFunction{wall}{wall}
\SetKwFunction{binarize}{binarize}

$\bm{x} \gets \mathbf{0}$\;
$\bm{y} \gets$ \random{-0.1, 0.1} \tcp*[l]{Initialize uniform random vector}

\For{$k = 0$ \KwTo $T$}{
    calculate $y(t_{k+1})$\;
    calculate $x(t_{k+1})$\;
     $y(t_{k+1}), x(t_{k+1}) \gets$ \wall{$x(t_{k+1}), y(t_{k+1})$} \;
}

$\bm{s} \gets$ \binarize{$x(t_T)$}\;
\Return $\bm{s}$ \tcp{Return the solution}
\end{algorithm}

\begin{example}
Let us conduct a simple benchmarking example to compare the methods described above. We have randomly generated 10 small instances with $N=30$ spins. We will compare the time needed to obtain an optimal solution for each instance (which we can certify due to the small instance size). More formally, we used the time-to-solution metric (Eq.~\ref{eq:tts}) described in detail in Chapter~\ref{chap:bench}. We must stress that this isn't a formal benchmark but a demonstrative example of the algorithms presented in this chapter. 

\begin{figure}
    \centering
    \includegraphics[width=0.9\textwidth]{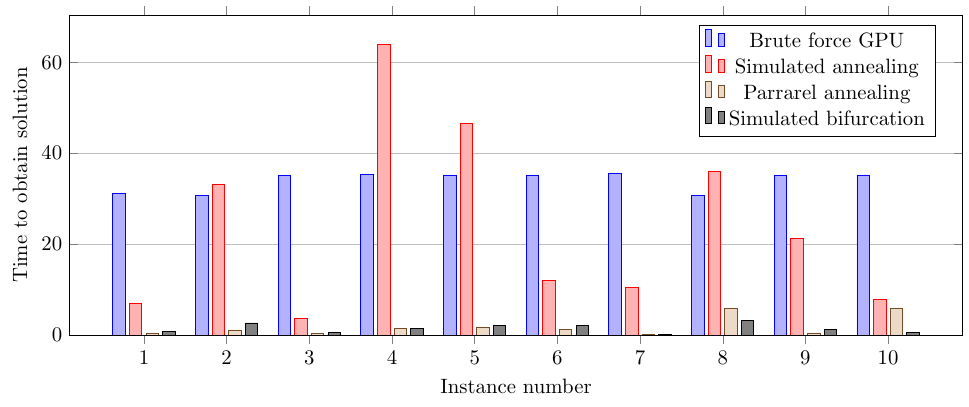}
    \caption[Solvers example]{Time to obtain ground state for each instance. Due to the small size of the instances, we were able to certify the ground state using exhaustive search. For brute-force, we used GPU acceleration~\cite {Jalowiecki2021}, as CPU-only time was two orders of magnitude higher (around 2 hours). With the rest of the solver, we used CPU-only versions. We used the time-to-solution metric (Eq.~\ref{eq:tts}), which accounts for the probability of obtaining the ground state in the given sample. Each heuristic solver was run with the same number of steps and sample size. }
\end{figure}

This example shows the different behavior of the presented algorithms. The simulated annealing results show significant differences across examples due to the relatively low probability of finding the ground state in a given sample. 

\end{example}

\section{Summary}

The classical Ising model offers a unified framework for both statistical physics and combinatorial optimization. In this chapter, we defined the model on arbitrary graphs, introduced its thermodynamic formulation via the partition function and the Boltzmann distribution, and established its equivalence to a QUBO via a simple variable transformation. This equivalence makes the ground-state problem a natural target for algorithmic and hardware approaches.

We then reviewed algorithmic baselines for ground-state search. Beyond exhaustive enumeration, heuristic branch-and-bound variants offer structured exploration with pruning; simulated annealing provides a temperature-driven stochastic search; simulated bifurcation leverages the dynamics of nonlinear oscillators, and parallel (quantum-inspired) annealing exploits gradient-like updates with high parallelism. These methods provide practical references and complementary strengths. They sit alongside other established solvers—e.g., tabu search and related local-improvement metaheuristics, digital MIP/CP-based solvers (as in modern B\&B/branch-and-cut frameworks), coherent Ising machines, and message-passing or tempering strategies—which together form the classical and physics-inspired landscape against which specialized hardware should be compared.

Finally, this discussion motivates an alternative approach to solving Ising instances: adiabatic quantum computation, implemented as quantum annealing. The next chapter introduces AQC/QA and hardware embodiments (notably D-Wave systems), positioning them within the same Ising/QUBO framework to enable apples-to-apples benchmarking against the classical baselines summarized here.

%% file: chapters/adiabatic.tex
\chapter{Adiabatic Quantum Computation and Quantum Annealing}
\chaptermark{AQC and QA}
\label{chap:AQC-QA}

In the previous chapter, we formulated the task of finding the ground state of an Ising Hamiltonian, an NP-hard combinatorial optimization problem, and reviewed classical heuristics commonly used in practice. Despite substantial progress, no classical method provides a decisive solution to this problem. This motivates the exploration of \emph{quantum computation}, a distinct computational paradigm that exploits quantum-mechanical dynamics to process information and, in some settings, to offer advantages for optimization, sampling, and simulation.

Quantum computation is commonly divided into two paradigms: \emph{digital} and \emph{analog}. Digital quantum computation, also known as the \emph{gate model}, represents algorithms as sequences of discrete unitary gates acting on qubits, closely mirroring classical circuit-based computation~\cite{Nielsen2011}. In contrast, analog quantum computation leverages the continuous-time evolution of a controllable quantum system whose Hamiltonian is engineered so that its dynamics encode the problem of interest~\cite{Das2008}. A prominent model within the analog paradigm is \emph{adiabatic quantum computing} (AQC), which is the focus of this chapter.

This chapter introduces the AQC framework and its practical realization in \emph{quantum annealing} (QA) hardware, with emphasis on aspects critical for benchmarking. In particular, we discuss  the mapping between Ising/QUBO formulations and hardware-implementable Hamiltonians,  annealing schedules and control parameters, and  major non-idealities such as finite-temperature and open-system effects, integrated control errors, and the embedding overhead imposed by limited device connectivity. These considerations are essential for interpreting empirical performance and for designing fair comparisons between quantum annealers and classical baselines.

\section{Adiabatic quantum computation}
\label{sec:aqc}

Building on previous work~\cite{Kadowaki1998}, Farhi \emph{et al.}~\cite{Farhi2001} proposed a continuous-time quantum algorithm for solving classical optimization problems based on the \emph{adiabatic theorem}~\cite{Kato1950,Messiah2014}. This proposal initiated the model of \emph{adiabatic quantum computation} (AQC)~\cite{vanDam2001}, in which a quantum system is steered from an easily prepared ground state to (approximately) the ground state of a problem Hamiltonian that encodes the solution. AQC is computationally equivalent to the standard gate-based (circuit) model up to polynomial overhead in resources~\cite{Aharonov2007}, establishing it as a universal model of quantum computation.

In AQC, the computational output is obtained by measuring the final state of an evolution generated by a slowly varying Hamiltonian. One begins with an \emph{initial (driver) Hamiltonian} $\mathcal{H}_{\mathrm{init}}$ whose ground state $\ket{\psi_0}$ can be prepared efficiently, and defines a \emph{problem Hamiltonian} $\mathcal{H}_{\mathrm{problem}}$ whose ground state encodes the optimal solution. The system Hamiltonian is then deformed continuously from $\mathcal{H}_{\mathrm{init}}$ to $\mathcal{H}_{\mathrm{problem}}$ over a total runtime $\tau$. Introducing the reduced time $s=t/\tau\in[0,1]$, a common choice is the linear interpolation~\cite{Aharonov2007,Albash2018}

\begin{equation}
\label{eq:aqc}
\mathcal{H}(s) = (1-s)\,\mathcal{H}_{\mathrm{init}} + s\,\mathcal{H}_{\mathrm{problem}},
\end{equation}

\noindent although more general schedules are often employed in practice.

Let $\Delta(s)=E_1(s)-E_0(s)$ denote the instantaneous energy gap between the ground state and the first excited state of $\mathcal{H}(s)$, and define $\Delta_{\min}=\min_{s\in[0,1]}\Delta(s)$. In the idealized closed-system setting, the adiabatic theorem implies that if $\tau$ is sufficiently large relative to problem-dependent scales determined by $\Delta_{\min}$ and the rate of change $\partial_s \mathcal{H}(s)$, the state remains close to the instantaneous ground state throughout the evolution. Consequently, the final state $\ket{\psi(\tau)}$ approximates the ground state of $\mathcal{H}_{\mathrm{problem}}$, and a measurement in the computational basis yields the desired solution with high probability~\cite{Farhi2001,Jansen2007,Albash2018}.

\begin{figure}
\centering
\includegraphics[width=0.7\textwidth]{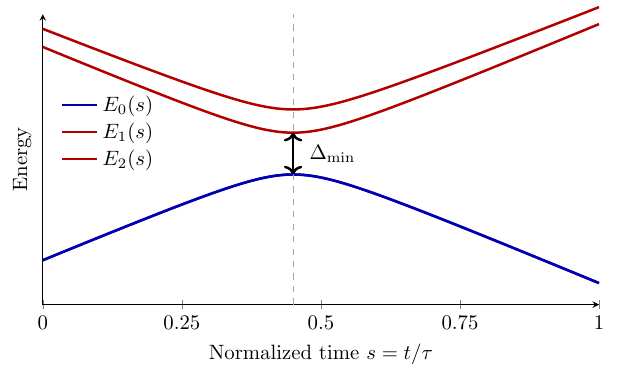}
\caption[Example energy paths of adiabatic evolution.]{Example energy paths of adiabatic evolution. The expression $E_{n}(s)$ represents the instantaneous energy of $n$-th excited state with $E_0(s)$ denoting the ground state. }
\label{fig:adiabatic-example}
\end{figure}

AQC is motivated by the quantum adiabatic theorem. We do not state the theorem in its most formal form here, precise statements with explicit error bounds and regularity assumptions can be found in~\cite{Kato1950,Jansen2007,Messiah2014,Tanaka2017,Albash2018}. Instead, we present a standard informal version sufficient for intuition.

Let $\mathcal{H}(s)$ be a smooth interpolation as in Eq.~\eqref{eq:aqc}, with reduced time $s=t/\tau\in[0,1]$, and let $\ket{\psi_0(s)}$ denote the instantaneous ground state of $\mathcal{H}(s)$ with eigenvalue $E_0(s)$. Assume that the ground state is non-degenerate for all $s\in[0,1]$, and that it is separated from the rest of the spectrum by a strictly positive gap:

\begin{equation}
  \Delta(s) = E_1(s) - E_0(s) > 0, \qquad s\in[0,1],
\end{equation}

\noindent where $E_1(s)$ is the first excited energy. If the system is initialized in $\ket{\psi_0(0)}$ and evolves according to the Schr\"odinger equation

\begin{equation}
  i\,\frac{d}{dt}\ket{\psi(t)}=\mathcal{H}(t)\ket{\psi(t)},
\end{equation}

\noindent then, in the closed-system setting, sufficiently slow evolution guarantees that the final state remains close to the instantaneous ground state at $s=1$. A commonly quoted \emph{sufficient} scaling condition~\cite{Jansen2007} is

\begin{equation}
\label{eq:adiabatic-theorem}
  \tau \gg
  \frac{\max_{s\in[0,1]} \bigl\|\partial_s \mathcal{H}(s)\bigr\|}{\Delta_{\min}^{\,2}},
  \qquad
  \Delta_{\min} := \min_{s\in[0,1]} \Delta(s),
\end{equation}

\noindent where $\|\cdot\|$ denotes the operator norm.\footnote{Rigorous bounds include prefactors and, depending on the chosen formulation, may also involve higher derivatives of $\mathcal{H}(s)$ and stronger gap dependence. Eq.~\eqref{eq:adiabatic-theorem} should therefore be interpreted as an order-of-magnitude guide rather than a tight runtime predictor~\cite{Jansen2007,Tanaka2017,Albash2018}.}
In other words, under suitable smoothness and gap conditions, a system starting in the ground state of $\mathcal{H}(0)$ approximately follows the ground-state branch of $\mathcal{H}(s)$ during the interpolation and ends near $\ket{\psi_0(1)}$.

\section{Quantum annealing}
\label{sec:qa}

Quantum annealing (QA) is a realization of adiabatic quantum computation, tailored specifically to finding low-energy states of Ising/QUBO cost functions. In contrast to AQC as a universal model of computation, QA is primarily used as an \emph{optimization and sampling heuristic} whose output is a classical spin configuration obtained by measurement at the end of the evolution~\cite{Kadowaki1998}.

The standard QA model is based on a transverse-field Ising Hamiltonian:

\begin{equation}
    \mathcal{H}(s) = \Gamma(s) \sum_{i=1}^N \sigma^x_i + \sum_{<i,j> \in E} J_{i,j} \sigma^z_i \sigma^z_j + \sum_{i \in V} h_i \sigma^z_i,
\end{equation}

\noindent where $\sigma^z_i$ and $\sigma^x_i$ denote Pauli operators acting on qubit $i$. $\sigma_i^{z}$ defines the computational basis in which the Ising problem Hamiltonian is diagonal, with eigenvalues \(\pm 1\) corresponding to classical spin values, while \(\sigma_i^{x}\) is the transverse-field (driver) operator that induces quantum fluctuations and mediates transitions between \(\sigma^{z}\)-basis states. More details can be found in Appendix~\ref{apx:qconventions}. $\Gamma(s) \in \mathbb{R}$ is a parameter that controls the strength of the transverse field (intuitively, it can be understood as a "quantumness" parameter~\cite{Tanaka2017}). It is assumed that $\Gamma(0)$ has a high value and $\Gamma(1)\approx 0$. Additionally, $i$ and $j$ indicate the indices of sites, $N$ is the number of spins, $J_{ij}$ is an interaction between neighboring sites, $h_i$ is the external field, and $s = t/\tau$ denotes normalized time.

At the beginning of the anneal ($s\approx 0$), the driver term $\sum_i\sigma_i^x$ dominates, and the ground state is the product state:

\begin{equation}
\ket{G} := \ket{+}^{\otimes N},
\qquad
\ket{+} = \frac{1}{\sqrt{2}}\left(\ket{\uparrow}+\ket{\downarrow}\right),
\end{equation}
\noindent \textit{i.e.}, an equal-weight superposition over all $\sigma^z$-basis configurations. During the anneal, $\Gamma(s)$ is decreased, so that the final Hamiltonian ($s\approx 1$) is approximately classical and diagonal in the computational basis. In the idealized closed-system and adiabatic limit, sufficiently slow evolution would keep the state near the instantaneous ground state, implying that the final measurement returns a ground state (or low-energy state) of the target Ising model with high probability.




\section{D-Wave's quantum annealers}
\label{sec:dwave}

D-Wave quantum annealers are purpose-built analog devices that implement a programmable transverse-field Ising model for optimization and sampling tasks. Compared to the idealized AQC setting, their dynamics is realized in superconducting hardware, on a fixed sparse interaction topology, and under open-system conditions at millikelvin temperatures. Nevertheless, the central principle remains the same: a strong transverse-field ``driver'' is gradually reduced while the programmed Ising terms are turned on.

\begin{figure}
    \begin{subfigure}[b]{0.3\textwidth}
        \centering
        \includegraphics[width=\textwidth]{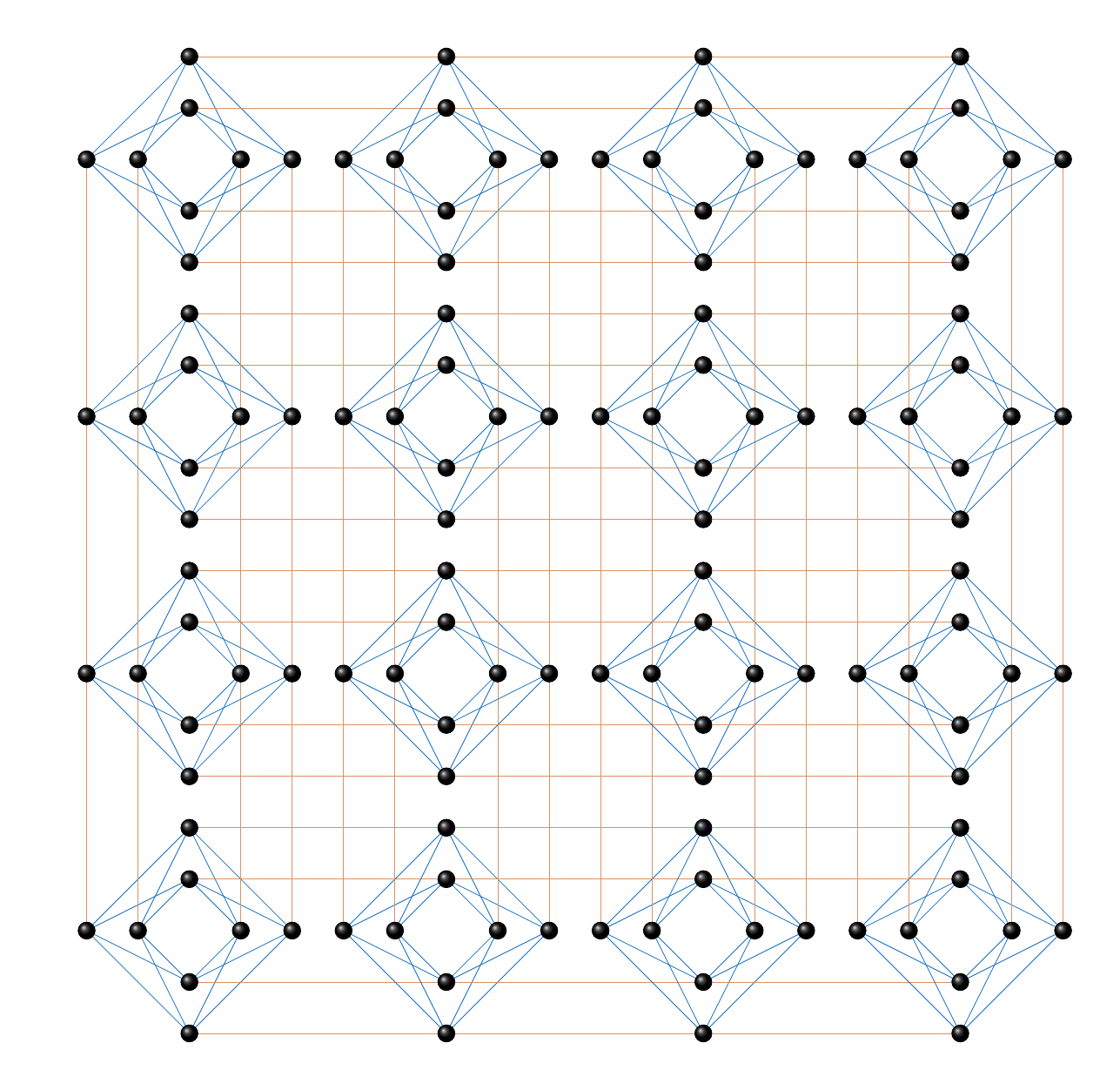}
         \caption{Chimera}
         \label{fig:chimera-qpu}
    \end{subfigure}
    \hfill
    \begin{subfigure}[b]{0.3\textwidth}
        \centering
        \includegraphics[width=\textwidth]{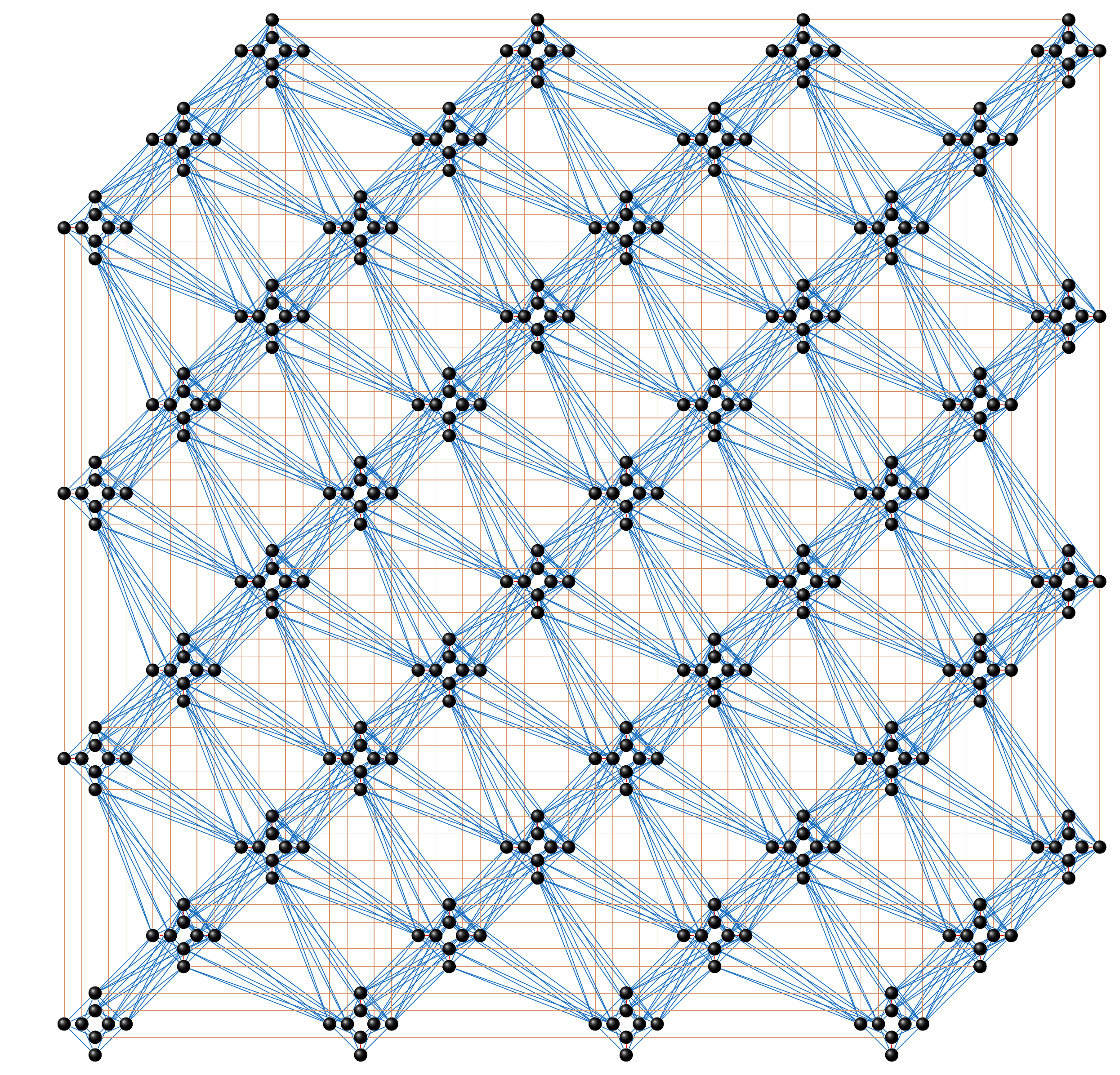}
         \caption{Pegasus}
         \label{fig:pegasus-qpu}
     \end{subfigure}
     \hfill
     \begin{subfigure}[b]{0.3\textwidth}
        \centering
        \includegraphics[width=\textwidth]{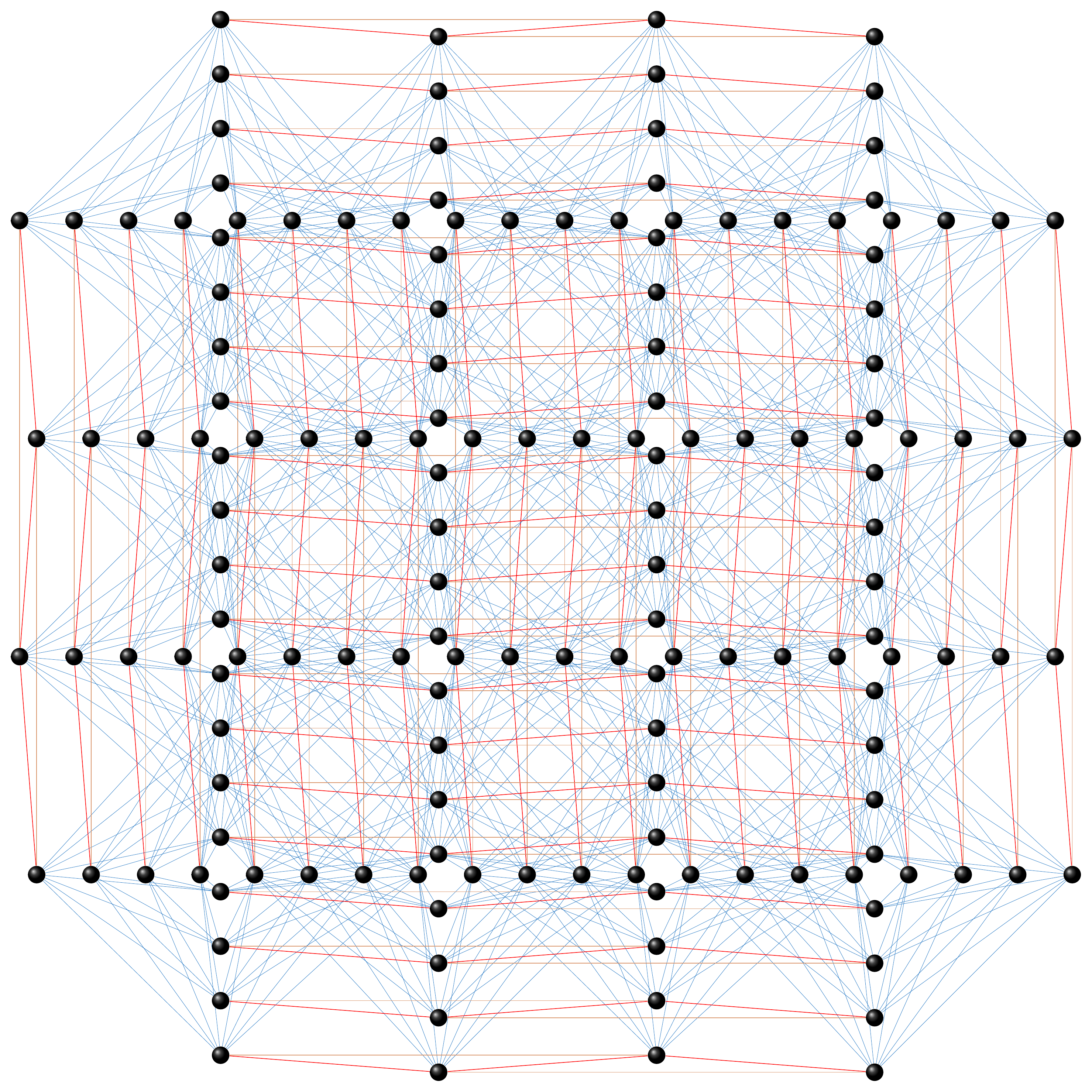}
         \caption{Zephyr}
         \label{fig:zephyr-qpu}
     \end{subfigure}
     \hfill
    \caption[Connectivity structures of D-Wave Quantum Processing Units (QPUs)]{Connectivity structures of D-Wave Quantum Processing Units (QPUs), also called Topology~\cite{Boothby2021,Boothby2020,Dattani2019,Lanting2014}.}
    \label{fig:dwave-topologies}
\end{figure}

At the level of an effective model, a D-Wave QPU implements a Hamiltonian of the form~\cite{dwave}
\begin{equation}
\label{eq:ising-dwave}
\mathcal{H}(s)
=
-\frac{A(s)}{2}\sum_{i=1}^{N}\sigma^{x}_{i}
+\frac{B(s)}{2}\left(
\sum_{(i,j)\in E_{\mathrm{QPU}}} J_{ij}\,\sigma^{z}_{i}\sigma^{z}_{j}
+\sum_{i=1}^{N} h_{i}\,\sigma^{z}_{i}
\right),
\end{equation}

\noindent where $s=t/\tau\in[0,1]$ is the normalized annealing time, $\tau$ is the total annealing time, and $A(s)$, $B(s)$ are device-specific schedule functions. The graph $G_{\mathrm{QPU}}=(V_{\mathrm{QPU}},E_{\mathrm{QPU}})$ specifies the programmable connectivity: vertices correspond to physical qubits and edges to tunable couplers. Due to fabrication defects and calibration constraints, a subset of qubits/couplers may be disabled. The user-visible \emph{working graph} is therefore the operational subgraph of the nominal topology (see Fig.~\ref{fig:dwave-topologies}).

\begin{figure}
    \centering
    \includegraphics[width=0.65\textwidth]{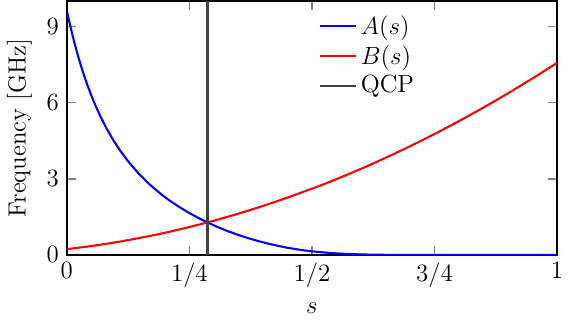}
    \caption[Example of schedule functions $A(s)$ and $B(s)$]{Example of schedule functions $A(s)$ and $B(s)$ for a representative D-Wave QPU (values shown as frequencies in the manufacturer's convention). The shaded mid-anneal window indicates the region where the driver and problem energy scales compete most strongly ($A(s)\approx B(s)$), often coinciding with an increased probability of diabatic excitations~\cite{dwave}.}
    \label{fig:example-schedule}
\end{figure}

The schedules satisfy boundary conditions $A(0)\gg B(0)$, $B(1)\gg A(1)$, and $A(1)\approx 0$, so that the system is initialized close to the ground state of a strong transverse field and (ideally) ends in a classical state governed by the programmed Ising Hamiltonian. The precise shapes of $A(s)$ and $B(s)$ depend on the processor generation and calibration. An example schedule for a representative processor is shown in Fig.~\ref{fig:example-schedule}. The region where $A(s)$ and $B(s)$ become comparable is often associated with a heightened susceptibility to excitations.

D-Wave QPUs are based on superconducting flux qubits: each qubit is a superconducting loop interrupted by Josephson junctions, and couplings are implemented via tunable inductive elements that realize effective Ising interactions. The qubits, couplers, and on-chip control circuitry form a planar superconducting integrated circuit operated in a dilution refrigerator at millikelvin temperatures to reduce thermal excitations and decoherence~\cite{Harris2010a,Harris2010b,Johnson2010,Bunyk2014}.

Finally, the fixed sparsity of \(G_{\mathrm{QPU}}\) has algorithmic consequences. Many practical Ising/QUBO instances induce logical graphs that are denser than the hardware connectivity, so an embedding step (typically minor embedding using ferromagnetic chains~\cite{dwave}) is required to map the logical instance onto \(G_{\mathrm{QPU}}\) before execution on the QPU. We discuss this mapping and its overhead in the subsequent section.

Successive generations of D-Wave processors employ different topologies, with the general trend towards larger system sizes and higher average degree. The earliest commercial systems used the Chimera architecture~\cite{Boothby2015,Lanting2014}, in which the QPU is arranged as an $L \times L$ grid of identical \emph{unit cells}. Each unit cell contains eight qubits connected as a complete bipartite graph $K_{4,4}$, with four ``vertical'' and four ``horizontal'' qubits. Additional couplers connect corresponding qubits in neighboring unit cells, forming a sparse quasi-two-dimensional lattice (see Fig.~\ref{fig:chimera-qpu}).

More recent D-Wave systems employ the Pegasus and Zephyr families of topologies. These are presented in Fig.~\ref{fig:pegasus-qpu} and~\ref{fig:zephyr-qpu}. Pegasus, introduced with the \texttt{Advantage} QPU, increases the nominal degree to fifteen by augmenting the Chimera-like layout with additional inter-cell couplers and longer-range connections~\cite{Boothby2020}. The unit cell in Pegasus consists of three Chimera-like graphs, giving $24$ qubits.  The Zephyr topology, used in the \texttt{Advantage2} processors, further increases the nominal degree to twenty~\cite{Boothby2021}. In each case, the working graph presented to the user is large and sparse, with a highly regular local structure but non-trivial global connectivity.

D-Wave topologies are typically labeled by a letter indicating the family and an integer size parameter. For example, a Chimera graph C$6$ consists of $6\times 6$ array of \(K_{4,4}\) unit cells, while P$N$ (Pegasus) and Z$N$ (Zephyr) denote the corresponding lattices with size parameter $N$, which fixes the nominal qubit count and connectivity~\cite{dwave}.

\begin{table}[h!]
\centering
\begin{tabular}{|lccc|}
    \hline
   & $2000$Q & Advantage & Advantage2 \\
    \hline
    \hline
    Graph topology & Chimera & Pegasus & Zephyr \\
    Graph size & C16 & P16 & Z12\\
    Number of qubits & $> 2000$ & $> 5000$ & $>4000$ \\
    Number of couplers & $> 6 000$ & $> 35 000$ & $> 40 000$\\
    Couplers per qubit & $6$ & $15$ & $20$\\
    \hline
\end{tabular}
\caption{Typical characteristics of subsequent generations of QPUs.}
\label{tab:dwave}
\end{table}

\subsection{Embedding}
\label{subsec:embedding}

As the connectivity structure of the quantum processing unit (QPU) is fixed by the underlying hardware graph, arbitrary Ising or QUBO problem instances cannot, in general, be implemented directly on the device. The native Hamiltonian realized by the QPU is defined on a sparse quasi-two-dimensional graph $G_{\mathrm{QPU}} = (V_{\mathrm{QPU}}, E_{\mathrm{QPU}})$, such as the Chimera, Pegasus, or Zephyr topology used in successive generations of D-Wave processors. In contrast, typical optimization problems give rise to a logical interaction graph $G_{\mathrm{logical}} = (V, E)$ whose connectivity is denser and structurally different from $G_{\mathrm{QPU}}$. Consequently, a dedicated embedding step is required before the annealing process can be executed on hardware.

The standard approach is \emph{minor embedding}~\cite{Choi2008,Choi2011}. In this procedure, each logical variable $i\in V$ is mapped to a connected set of physical qubits (a \emph{chain}) $K_i\subseteq V_{\mathrm{QPU}}$, and logical couplings are realized by at least one physical coupler between the corresponding chains. Formally, a minor embedding is a mapping:

\begin{equation}
\phi: V \to 2^{V_{\mathrm{QPU}}},\qquad i \mapsto K_i,
\end{equation}

\noindent such that every $K_i$ induces a connected subgraph of $G_{\mathrm{QPU}}$ and for every logical edge $(i,j) \in E$ there exists at least one physical edge $(u,v) \in E_{\mathrm{QPU}}$ with $u \in K_i$ and $v \in K_j$. Under this mapping, the logical Ising problem can be implemented as a physical Ising Hamiltonian on $G_{\mathrm{QPU}}$ by appropriately redistributing the logical couplings $J_{ij}$ and fields $h_i$ over the chains and their inter-chain couplers. Finding an efficient embedding that minimizes chain length is a non-trivial combinatorial problem~\cite{Lobe2024}, which motivates the use of heuristic embedding algorithms in practice~\cite{Cai2014}.

\begin{figure}
    \begin{subfigure}[c]{0.3\textwidth}
        \centering
        \includegraphics[width=\textwidth]{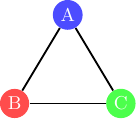}
        \captionsetup{labelformat=empty}
        \caption{Logical graph}
        \label{fig:eb-triangle}
    \end{subfigure}
    \hfill
    \begin{subfigure}[c]{0.3\textwidth}
        \centering
        \includegraphics[width=\textwidth]{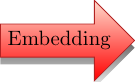}
        \captionsetup{labelformat=empty}
        \caption{}
        \label{fig:eb-arrow}
     \end{subfigure}
     \hfill
     \begin{subfigure}[c]{0.3\textwidth}
        \centering
        \includegraphics[width=\textwidth]{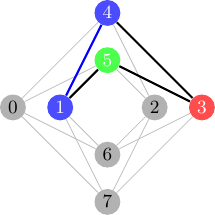}
        \captionsetup{labelformat=empty}
        \caption{Embedded instance}
        \label{fig:eb-chimera}
     \end{subfigure}
    \caption[Example minor embedding]{Example minor embedding of the complete graph $K_3$ onto a Chimera unit cell. Here, vertex $A$ is mapped to the chain $K_A$ consisting of qubits $1$ and $4$, while $B$ and $C$ are mapped to single qubits (3 and 5). Logical couplings are implemented by physical couplers between the corresponding chains.}
    \label{fig:example-embedding}
\end{figure}

\subsubsection{Annealing schedules}
\label{subsubsec:anneal-schedules}

An annealing schedule on D-Wave processors specifies how the normalized annealing parameter $s\in[0,1]$ is traversed in physical time $t\in[0,\tau]$. Concretely, the device implements the Hamiltonian $\mathcal{H}(s)$ from Eq.~\eqref{eq:ising-dwave}, and a user-defined schedule $s(t)$ induces a time-dependent evolution

\begin{equation}
  \mathcal{H}(t)=\mathcal{H}(s(t)), \qquad s:[0,\tau]\to[0,1].
\end{equation}

Modern D-Wave processors allow the user to override the default monotone schedule by specifying custom piecewise-linear functions $s(t)$, enabling reverse annealing, mid-anneal pauses, quenches, and fast-anneal protocols~\cite{dwave}. Below, we briefly summarize the schedule types used in this thesis.

\paragraph{Forward annealing.}
In the standard mode (forward annealing), the schedule is monotone, increasing from $s=0$ to $s=1$ over a user-chosen total anneal time $\tau$. This corresponds to gradually reducing the driver term while increasing the problem term.

\paragraph{Reverse annealing.}
In reverse annealing, the system is initialized at $s=1$ in a user-provided classical configuration. The schedule then reverses to a turning point $s_a<1$, reintroducing a transverse field and allowing quantum (and thermal) transitions among nearby configurations, after which it proceeds forward to $s=1$. Operationally, this implements a form of local search around the initial state and is particularly useful when good approximate solutions are available or when probing the local energy landscape near selected configurations~\cite{Mehta2025}. Additionally, this schedule allows us to pose thermodynamic inquiries (see Chapter~\ref{chap:thermo}).

\paragraph{Mid-anneal pause.}
A mid-anneal pause is implemented by holding the schedule at a fixed value $s=s_a$ for a programmable dwell time $t_{\mathrm{pause}}$. During the pause, the Hamiltonian is static, and the system evolves only through its interaction with the environment. Empirically, pausing in a narrow window in the mid-anneal region, typically shortly after the minimum-gap region, can increase ground-state success probabilities by enabling relaxation of diabatic and thermal excitations~\cite{Marshall2019}.

\paragraph{Fast annealing.}
Fast-anneal protocols reduce the total anneal time $\tau$ by orders of magnitude relative to standard settings (to the range of few nanoseconds) while preserving the qualitative shape of the underlying schedules $A(s)$ and $B(s)$. This drives the system far from the adiabatic regime and reduces the time available for thermalization, thereby emphasizing coherent and diabatic contributions to the dynamics~\cite{dwave_fast_anneals_2024,King2022,King2023}. In current D-Wave implementations, this mode may be restricted to instances without longitudinal fields (\textit{i.e.}, with $h_i=0$ in Eq.~\eqref{eq:ising-dwave})~\cite{dwave}.

\subsubsection{Errors in quantum annealers}
\label{subsubsec:qa-errors}

The performance of quantum annealers is limited by several error sources that can degrade solution quality and sampling statistics. It is convenient to group the dominant effects into two broad classes: (i) \emph{integrated control errors} (ICE), which quantify inaccuracies in the analog realization of the programmed Hamiltonian parameters, and (ii) \emph{open-system effects}, which arise from the unavoidable coupling of the QPU to its environment (finite temperature, dissipation, and decoherence)~\cite{dwave}.

\paragraph{Integrated control errors (ICE).}
Although the user specifies $h_i$ and $J_{ij}$ as real numbers, the analog control circuitry implements only an approximation due to calibration uncertainty, control noise, crosstalk, and limited precision. A standard phenomenological model is that the device realizes a \emph{perturbed} Ising objective,

\begin{equation}
\label{eq:ising-ice}
H_{\mathrm{Ising}}^{\delta}(\mathbf{s}) = \sum_{(i,j)\in E_{\mathrm{QPU}}} \bigl(J_{ij}+\delta J_{ij}\bigr)\, s_i s_j
+ \sum_{i\in V_{\mathrm{QPU}}} \bigl(h_i+\delta h_i\bigr)\, s_i,
\end{equation}

\noindent where $\delta J_{ij}$ and $\delta h_i$ represent implementation errors. Even when these perturbations are small in absolute magnitude, they can change the identity and ordering of low-energy states, especially for instances with many near-degenerate configurations or small energy gaps between competing solutions~\cite{Pearson2019}. From a benchmarking perspective, ICE introduces an instance-dependent \emph{effective problem drift} between the programmed and realized Hamiltonians, which can increase run-to-run variability.

\paragraph{Open-system effects.}
D-Wave QPUs operate at millikelvin temperatures but are not closed quantum systems. Interaction with the environment leads to thermal excitations, relaxation, and dephasing during the anneal. The impact of these processes depends strongly on the annealing schedule and runtime $\tau$: very short anneals enhance diabatic transitions due to rapid driving, whereas longer anneals provide more opportunity for thermalization and noise-induced transitions. As a result, there is generally no monotone relationship between $\tau$ and success probability. Performance can be limited either by non-adiabatic excitations at short $\tau$ or by thermally assisted transitions and "freeze-out" phenomena at longer $\tau$~\cite{Boixo2014,Marshall2017,Albash2018}.

\paragraph{Manifestation in samples and mitigation.}
Algorithmically, these non-idealities appear in the returned samples as deviations from the intended low-energy configurations. Depending on the setting, this can include apparent spin flips relative to an ideal ground state, biased sampling of low-lying excited states, and (for embedded problems) chain inconsistencies that require decoding. Post-processing methods can improve solution quality by exploiting structure in the returned samples. In this thesis, we employ a reinforcement-learning-based post-processing procedure that learns to identify and correct error manifestations in the raw annealer outputs without modifying the hardware or annealing protocol. This approach is described in Chapter~\ref{chap:rl}.

A broad range of additional error suppression, correction, and mitigation strategies has been proposed for quantum annealers, including classical post-processing, encoding, and repetition/aggregation techniques~\cite{Pudenz2014,Pudenz2015,Vinci2016,Shingu2024,Raymond2025}. A detailed comparison of these methods is beyond the scope of this work.

\section{Summary}
\label{sec:ch2-summary}

This chapter introduced \emph{adiabatic quantum computation} (AQC) as an analog model of quantum computation and discussed \emph{quantum annealing} (QA) as its optimization-oriented realization. Starting from the quantum adiabatic theorem, we outlined how a computational task can be encoded into a \emph{problem Hamiltonian} and addressed by interpolating from an easily prepared \emph{driver} ground state to the final Hamiltonian whose ground state represents the desired solution. Using the transverse-field Ising model as the canonical QA setting, we showed how the Ising/QUBO formulations from Chapter~\ref{chap:ising} naturally fit into the AQC framework and provide a direct route from combinatorial optimization to programmable quantum dynamics.

We then focused on D-Wave quantum annealers as a concrete hardware instantiation of QA. These devices implement a time-dependent transverse-field Ising Hamiltonian on sparse hardware topologies (Chimera, Pegasus, and Zephyr) with device-specific schedule functions. The restricted connectivity of the QPU working graph necessitates \emph{minor embedding}, introducing chain variables and associated overheads that must be accounted for when interpreting performance and scaling. We also discussed how user-accessible schedule controls, including custom anneal paths, reverse annealing, and mid-anneal pauses, provide additional degrees of freedom for shaping the dynamics, while integrated control errors and open-system effects place fundamental constraints on attainable solution quality and sampling behavior. Together, these considerations motivate rigorous experimental methodology: transparent reporting of embedding and schedule choices, careful selection of classical baselines, and systematic benchmarking protocols.

In the next chapter, we introduce \texttt{SpinGlassPEPS.jl}, a tensor-network-based solver tailored to QPU-relevant graph families, which provides a complementary classical baseline for benchmarking and expands the set of tools available for evaluating quantum annealing technologies.

%% file: chapters/software.tex
\chapter{SpinGlassPEPS.jl -- Tensor-network Package for Optimization and Sampling}
\chaptermark{SpinGlassPEPS.jl}
\label{chap:tn}

To computationally benchmark quantum annealers, it is essential to establish classical baselines that quantify the solution qualities and time-to-solution achievable without quantum hardware. Although many general-purpose approaches exist (Chapter~\ref{chap:ising}), quantum annealing processors exhibit strong structural constraints. Most importantly, they feature sparse and quasi-two-dimensional interaction graphs with fixed geometry (Chapter~\ref{chap:AQC-QA}). For this reason, it is valuable to study \emph{structure-exploiting} classical heuristics that are tailored to native QPU topologies and can leverage their locality rather than treating the input as an arbitrary graph. Tensor-network (TN) methods provide a natural language for such algorithms: by representing Gibbs probabilities or low-energy amplitudes as contracted networks, locality can be used to control computational cost via systematic truncations. A brief introduction to tensor networks is provided in the appendix~\ref{apx:tn}.

This chapter presents \texttt{SpinGlassPEPS.jl}~\cite{software}, a collection of Julia~\cite{Bezanson2017} packages implementing the TN-based heuristic introduced in~\cite{peps}. The method targets sampling and low-energy search for Ising-type models on graphs relevant to contemporary quantum annealers (e.g., Pegasus- and Zephyr-like quasi-2D connectivities). It uses a topology-aware reduction to local Potts clusters,  a projected entangled-pair state (PEPS) representation of the resulting partition function, and a branch-and-bound search in configuration space. In practice, solution quality and runtime are controlled by a set of parameters (notably the inverse temperature $\beta$, the maximal bond dimension $\chi$, cut-off probability $p$, and the number of retained branches $M$), which makes the approach well-suited for benchmarking studies where accuracy--cost trade-offs must be explicit.

The algorithm and the software were developed concurrently and are tightly coupled: implementation choices (data structures, contraction backends, GPU acceleration) directly shape the accessible parameter regimes and, consequently, the solver's empirical performance. The author's primary contribution is the software development and engineering effort, which is therefore the main emphasis of this chapter.

We first outline the algorithmic workflow and its underlying intuitions. We then describe the software architecture, key components, and user-facing interfaces. Detailed benchmarking experiments and comparisons against other baselines are deferred to the next chapter for clarity of presentation.

\paragraph{Author contributions} 
The algorithm and its software implementation were developed concurrently and are therefore tightly coupled. My primary contribution is the software development and engineering effort, which is therefore the main emphasis of this chapter. As I have contributed to various parts of the software, it is impractical to list them all individually. The project was carried out as a team effort, with shared contributions to software architecture, validation, and scientific interpretation.

\paragraph{Relation to published work}
This chapter is based on the author's prior work~\cite{software,peps} and primarily provides a self-contained description of the underlying algorithm and its implementation details. To keep the exposition coherent, the chapter focuses on methodology rather than performance. The corresponding experimental evaluation, benchmarking protocol, and result analysis are presented in Chapter~\ref{chap:bench}.

\begin{table}[t]
\begin{tabular}{|>{\raggedright\arraybackslash}p{0.46\textwidth}|>{\raggedright\arraybackslash}p{0.46\textwidth}|}
\hline
Current code version & v1.5.0\\
\hline
Permanent link to repository  & \url{https://github.com/euro-hpc-pl/SpinGlassPEPS.jl} \\
\hline
Legal Code License   & Apache 2.0 \\
\hline
 Software code languages, tools, and services used & Julia \\
\hline
Compilation requirements, operating environments \& dependencies & \texttt{Julia 1.11, CUDA.jl v5, TensorOperations.jl, Memoize.jl} \\
\hline
Documentation & \url{https://euro-hpc-pl.github.io/SpinGlassPEPS.jl/dev/}\\
\hline

\end{tabular}
\caption{Software metadata.}
\label{tab:code-metadata} 
\end{table}


\section{Algorithm overview}
\label{sec:overview}

Building on the framework of~\cite{Rams2021}, we recast the Ising optimization problem (Eq.~\eqref{eq:ising-classical}) as an inference task in a Boltzmann distribution. Instead of searching directly for the lowest-energy configurations, we identify the configurations with the highest probability in a thermal state at inverse temperature $\beta>0$. Concretely, the probability of observing a configuration $\bm{s}$ is

\begin{equation}
p(\bm{s}) = \frac{1}{Z} \exp{(-\beta\, H_{\text{Ising}}(\bm{s}))},
\label{eq:boltzmann-ising}
\end{equation}

\noindent where $Z$ is the partition function (Eq.~\eqref{eq:ising-partition}), $\bm{s}=[s_1,\dots,s_N]$ is a spin configuration, and $H_{\mathrm{Ising}}(\bm{s})$ is given by Eq.~\eqref{eq:ising-classical}. Since $p(\bm{s})$ is a strictly decreasing function of $H_{\mathrm{Ising}}(\bm{s})$ for $\beta>0$, the maximizers of $p(\bm{s})$ coincide with the ground-state configurations (up to degeneracy). In practice, we treat $\beta$ as an algorithmic control parameter: increasing $\beta$ concentrates probability mass around low-energy states, while moderate $\beta$ yields broader distributions useful for sampling and for improved numerical stability.

\begin{figure}[!h]
    \centering
    \includegraphics[width=\textwidth]{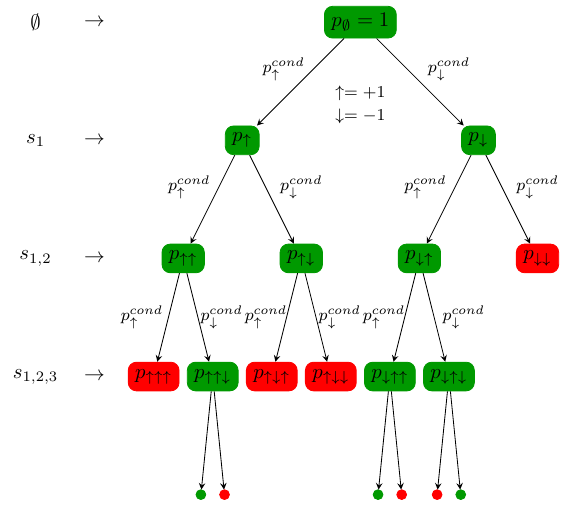}
    \caption[Branch-and-bound (tree) search in the probability space.]{Branch-and-bound search in probability space. The solution is built sequentially by assigning spins. For each partial assignment $\bm{s}_{1:k}$ (green), the algorithm branches by appending $s_{k+1}\in\{-1,+1\}$ and evaluating the corresponding conditional probabilities $p(s_{k+1}\!\mid\!\bm{s}_{1:k})$ (equivalently, the marginals $p(\bm{s}_{1:k+1})$). The bounding step retains at most $M$ partial assignments with the largest probabilities (green) and discards the remainder (red). The figure shows $M=3$ for clarity. In \texttt{SpinGlassPEPS.jl}, the required marginals are obtained from approximate tensor-network contractions.}
    \label{fig:branch-and-bound-ising-prob}
\end{figure}

The set of most probable configurations can be constructed incrementally by exploring a search tree over partial assignments. Let $\bm{s}_{1:k}=(s_1,\dots,s_k)$ denote a partial configuration. Its probability

\begin{equation}
p(\bm{s}_{1:k})=\sum_{\bm{s}_{k+1:N}} p(\bm{s}),
\end{equation}

\noindent is a marginal of~\eqref{eq:boltzmann-ising}, and the branching probabilities are the corresponding conditionals
$p(s_{k+1}\mid \bm{s}_{1:k})$. Figure~\ref{fig:branch-and-bound-ising-prob} illustrates the resulting branch-and-bound~\cite{Morrison2016} procedure used in this work. At depth $k$ we maintain a set of at most $M$ partial assignments with the largest marginal probabilities (a \emph{beam} of width $M$). Each partial assignment is extended by $s_{k+1}\in\{-1,+1\}$, producing up to $2M$ candidates. We then retain only the $M$ most probable ones. For the parameter regimes relevant here, both the marginal evaluation and the truncation to finite $M$ make the method heuristic. 

If we could perform the exact calculations of the relevant marginal probabilities, then this approach would guarantee finding the ground state~\cite{Kardar2007}. However, to our knowledge, computing the required marginals exactly is generally intractable. In our approach, they are obtained from a projected entangled-pair states (PEPS) tensor network representation of the partition function~\cite{Nishino2001,Verstraete2006}. Computing marginal probabilities, then transform to contracting networks with appropriate boundary conditions. Exact contraction is, in general, a $\#\mathrm{P}$-hard task~\cite{Schuch2007}, which motivates approximate contraction schemes.

A second challenge is geometric: many efficient PEPS contraction schemes assume a regular two-dimensional lattice~\cite{Ran2020}, while quantum annealers implement sparse, structured, but non-planar interaction graphs. To bridge this gap, we map the original Ising instance to a Potts model defined on a king's graph, which admits a PEPS representation on a square lattice. Extending this construction beyond earlier settings to modern QPU topologies (notably Pegasus and Zephyr) requires handling large unit cells and the resulting large local dimensions, which lead to large structured tensors and nontrivial memory/computation patterns. The next sections describe how these issues are addressed algorithmically and in software.

\begin{figure}[!t]
    \centering
    \includegraphics[width=\textwidth]{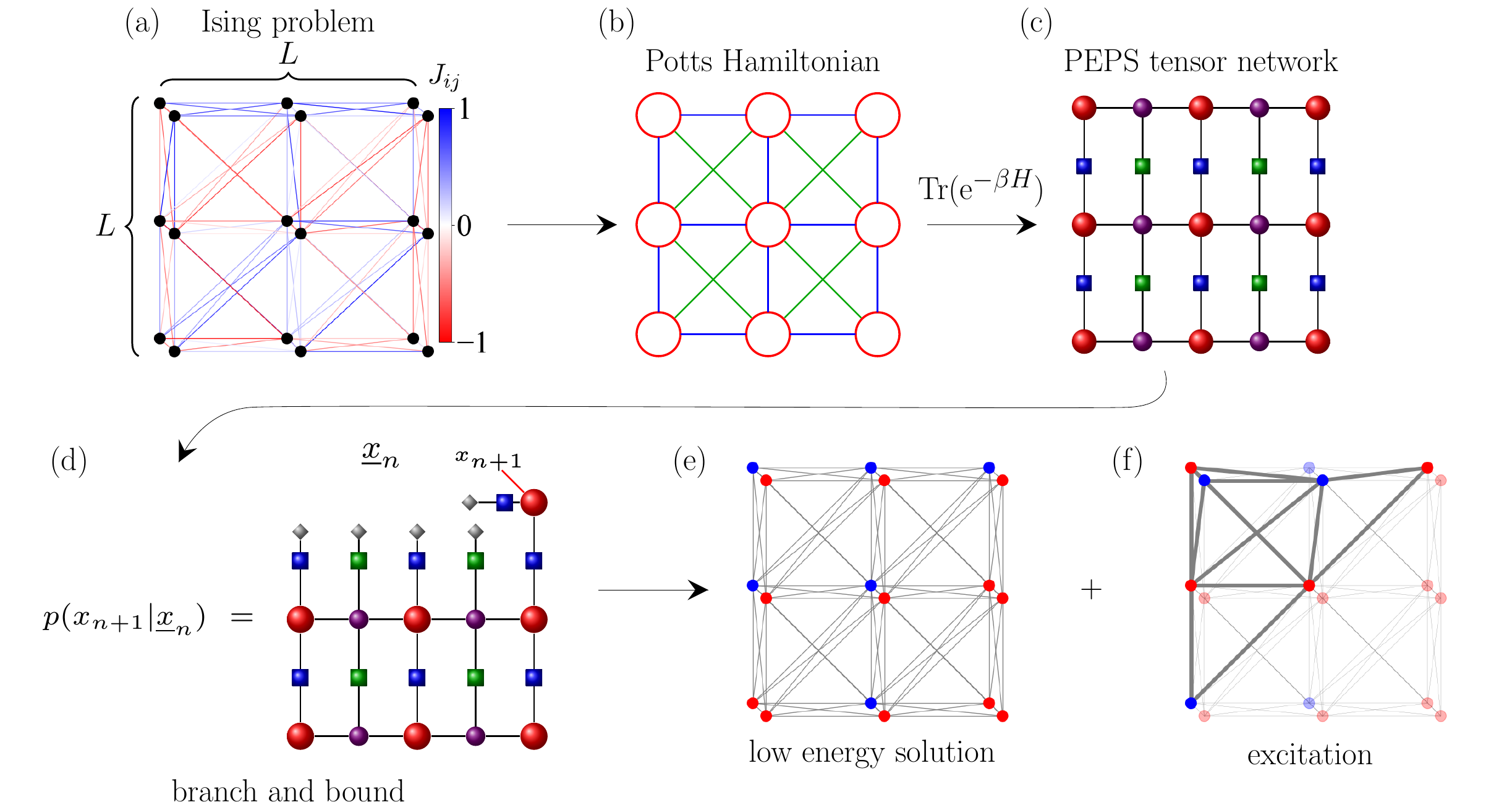}
    \caption[Execution flow of \texttt{SpinGlassPEPS.jl}.]%
    {Execution flow of \texttt{SpinGlassPEPS.jl}. (a) An Ising instance on the native interaction graph is transformed into (b) a local Potts Hamiltonian on a king's graph via clustering. (c) The resulting (Potts) partition function is represented as a PEPS tensor network on a square lattice. (d) Conditional marginals required by the search are computed by approximately contracting the PEPS. (e) A branch-and-bound sweep yields a candidate configuration, which corresponds to a ground-state configuration of the original Ising model. (f) The same machinery can be used to find localized low-energy excitations around the returned configuration.}
    \label{fig:algorithm-overview}
\end{figure}

Figure~\ref{fig:algorithm-overview} summarizes the workflow of \texttt{SpinGlassPEPS.jl}. The software is deliberately modular: the mapping, tensor construction, contraction backends, and the search routine are separated into composable components. This design enables interchangeable contraction schemes, exploitation of internal tensor structure, and hardware acceleration (CPU/GPU) via Julia's multiple dispatch.

\subsection{The Potts model}
\label{subsec:potts}

Recent quantum annealing architectures (notably Pegasus and Zephyr, see Fig.~\ref{fig:dwave-topologies}) have large, repeating unit cells and a structured pattern of short-range couplings. To exploit this locality with tensor networks, we transform the original Ising Hamiltonian (Eq.~\eqref{eq:ising-classical}) into a generalized Potts model with fewer variables but larger local state spaces. 

Let $\mathcal{F}$ denote the resulting two-dimensional ``King's graph'' layout of clusters (square lattice with nearest and diagonal neighbors), with $\Nbar$ nodes and configuration vector $\bm{x}=[x_1,\dots,x_{\Nbar}]$. Each Potts variable $x_n$ labels one of the $d_n=2^{k_n}$ binary configurations of cluster $n$, where $k_n$ is the number of spins in that cluster (e.g., $k_n\le 24$ for Pegasus and $k_n\le 16$ for Zephyr in the maximal, fully-populated case).\footnote{In practice, $k_n$ may be smaller due to inactive qubits or because only a subset of qubits participates in the embedded Ising instance. We therefore allow cluster-dependent local dimensions $d_n=2^{k_n}$.} The energy can then be written as a Potts Hamiltonian:

\begin{equation}
    \label{eq:potts-peps}
    H(\bm{x}) = \sum_{(m,n)\in\mathcal{F}} E_{x_m x_n} + \sum_{n=1}^{\Nbar} E_{x_n},
\end{equation}

\noindent where $E_{x_n}$ is the intra-cluster energy of state $x_n$, and $E_{x_m x_n}$ is the interaction energy between clusters $m$ and $n$ induced by the original Ising couplers crossing the cluster boundary. Equation~\eqref{eq:potts-peps} defines a factor graph with unary and pairwise factors~\cite{Mezard2009}, which is convenient for subsequent tensor-network construction and for probabilistic inference.

All subsequent steps of the solver operate on the Potts variables. In particular, the branch-and-bound search returns a high-probability Potts configuration $\bm{x}$, which is then decoded back into the original spin assignment $\bm{s}$ by expanding each cluster state.

\begin{figure}[!t]
    \centering
    \begin{subfigure}{0.9\textwidth}
        \centering
        \includegraphics[width=\linewidth]{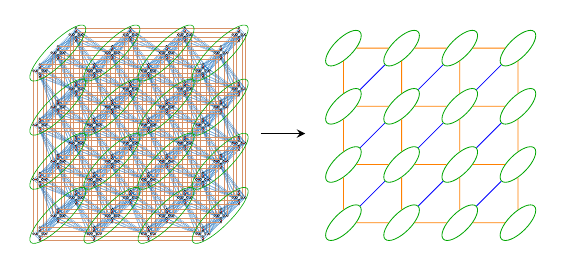}
        \caption{Pegasus: clusters of up to $k=24$ spins ($d=2^{24}$ cluster states in the maximal case).}
        \label{fig:pegasus-to-potts}
    \end{subfigure}
    \par\vspace{1em}
    \begin{subfigure}{0.9\textwidth}
        \centering
        \includegraphics[width=\linewidth]{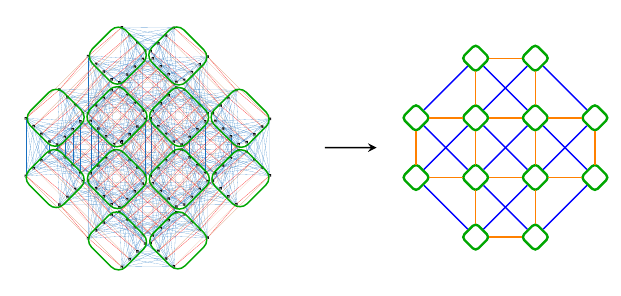}
        \caption{Zephyr: clusters of up to $k=16$ spins ($d=2^{16}$ cluster states in the maximal case).}
        \label{fig:zephyr-to-pots}
    \end{subfigure}
    \caption[Transforming hardware graphs into Potts clusters.]%
    {Transforming hardware graphs into Potts clusters. Black dots denote physical spins, and colored edges denote couplers: orange for intra-row/column links and blue for diagonal and inter-cell links (schematic). Each circled block indicates a cluster, which is replaced by a single Potts variable $x_n$ whose state enumerates the cluster's binary microstates.}
    \label{fig:dwave-to-potts}
\end{figure}

\paragraph{Projected variables and compressed pair factors.}
A direct representation of $E_{x_m x_n}$ would scale as $d_m\times d_n$, which is prohibitive when $d_n=2^{k_n}$ is large. However, inspection of the unit-cell structure in Fig.~\ref{fig:dwave-to-potts} shows that only a \emph{subset} of spins in cluster $m$ actually couples to cluster $n$. Therefore, the pair interaction depends only on the spins participating in couplers between these two clusters. We exploit this by introducing an edge-dependent projection

\begin{equation}
    \bar{x}_m = P^{mn}(x_m), \qquad \bar{x}_n = P^{nm}(x_n),
    \label{eq:projector}
\end{equation}

\noindent where $P^{mn}$ maps the full cluster state index $x_m\in\{1,\dots,d_m\}$ to a reduced index $\bar{x}_m\in\{1,\dots,d_m^{mn}\}$ that enumerates only the spin subconfigurations relevant for the $(m,n)$ coupling. The pair energy can then be represented by a much smaller matrix,

\begin{equation}
    E_{x_m x_n} = \bar{E}_{\bar{x}_m \bar{x}_n},
    \label{eq:proj_energy}
\end{equation}

\noindent where $\bar{E}_{\bar{x}_m \bar{x}_n}$ is defined over $d_m^{mn}\times d_n^{nm}$ states and is evaluated by summing the original Ising couplings crossing the cluster boundary for the corresponding boundary-spin assignments. These projections are an important step toward reducing complexity: they preserve the exact energy contributions while drastically reducing the size of the pair factors used by the tensor network and inference routines.

\begin{figure}[!t]
    \centering
    \begin{subfigure}{\textwidth}
        \centering
        \caption{}
        \includegraphics{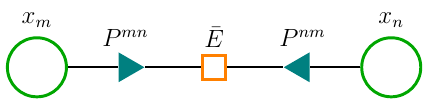}
    \end{subfigure}
    \vspace{2pt}
    \begin{subfigure}{0.4\textwidth}
        \centering
        \includegraphics[width=\linewidth]{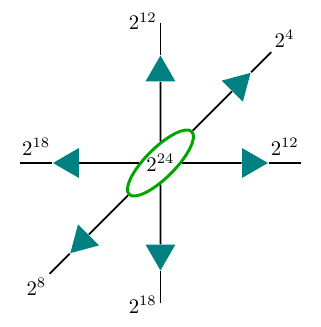}
        \caption{Example reduced dimensions for Pegasus.}
    \end{subfigure}
\hfill
    \begin{subfigure}{0.4\textwidth}
        \centering
        \includegraphics[width=\linewidth]{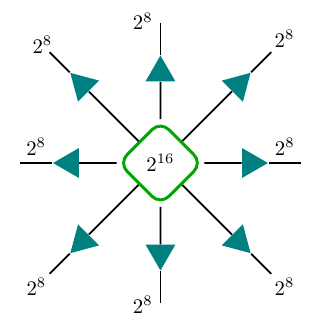}
        \caption{Example reduced dimensions for Zephyr.}
    \end{subfigure}
    \caption[Edge-specific projections for compressing Potts pair energies.]%
    {Edge-specific projections for compressing Potts pair energies. A naive representation of $E_{x_m x_n}$ scales as $d_m\times d_n$ with $d_n=2^{k_n}$ (up to $2^{24}$ for Pegasus and $2^{16}$ for Zephyr in the maximal case). Since only boundary spins couple across a given edge $(m,n)$, we project each cluster microstate $x_m$ to a reduced index $\bar{x}_m=P^{mn}(x_m)$ that retains only the boundary-spin subconfiguration relevant to that edge. The pair term is then stored as a compressed matrix $\bar{E}_{\bar{x}_m\bar{x}_n}$ (Eq.~\eqref{eq:proj_energy}). Panel (a) shows the schematic construction of edge-specific projected indices and the resulting compressed pair matrix $\bar{E}_{\bar{x}_m\bar{x}_n}$. Panels (b) and (c) illustrate typical reductions in local dimensions obtained in practice.}
    \label{fig:projectors}
\end{figure}

\paragraph{Branch-and-bound over Potts variables.}

The solver performs a sweep through the tensor network nodes, incrementally constructing a set of high-probability partial configurations. Let $\bm{x}_{1:n}$ denote assignments of the first $n$ Potts variables in the chosen traversal order. At step $n+1$, each of the retained $M$ partial assignments is branched over the $d_{n+1}$ possible states of $x_{n+1}$, producing up to $d_{n+1} \times M$ candidates. To keep the computation tractable, we retain only the $M$ candidates with the largest marginal probabilities. Using the chain rule,

\begin{equation}
    p(\bm{x}_{1:n+1}) = p(x_{n+1}\mid \bm{x}_{1:n})\, p(\bm{x}_{1:n}),
    \label{eq:chain_rule}
\end{equation}

\noindent where $p(x_{n+1}\mid \bm{x}_{1:n})$ is obtained from tensor-network contractions of the corresponding partial model. The traversal order can always be enforced by reindexing. 

\begin{algorithm}[t]
\label{alg:peps}
\caption{SpinGlassPEPS}
\KwData{Instance $I$, Lattice structure $L$, Inverse temperature $\beta$, Maximum bond dimension $\chi$, cut-off probability $p$, number of retained branches $M$, other parameters $*\mathrm{params}$}
\KwResult{Approximate solutions to the ground state problem}

\SetKwFunction{potts}{potts\_hamiltonian}
\SetKwFunction{pepsf}{PEPSNetwork}
\SetKwFunction{branch}{branch\_solutions}
\SetKwFunction{probability}{calculate\_marginal\_probabilities}
\SetKwFunction{bound}{bound\_solutions}

\SetKwData{ph}{potts}
\SetKwData{node}{node}
\SetKwData{peps}{peps}
\SetKwData{sol}{solution}
\SetKwData{tempsol}{candidate\_solutions}
\SetKwData{probs}{probs}

\tcp{Initialization}

\ph $\gets$ \potts{$I, L$}\;
\peps $\gets$ \pepsf{\ph, $*\mathrm{params}$}\;
\sol $\gets$ $\{\}$\ \tcp*{empty solution struct}

\tcp{Main loop}
\ForEach{\node $\in$ \peps} {
    \tempsol $\gets$ \branch{\sol, \node}\;
    \probs $\gets$ \probability{\tempsol, \peps, $\beta$, $\chi$, $p$}\;
    \sol $\gets$ \bound{\tempsol, \probs, $M$}\;
}

\Return \sol
\end{algorithm}


\section{Software description}

We have implemented the algorithm described in the previous section into the \texttt{SpinGlassPEPS.jl} Julia~\cite{Bezanson2017} package. The goal of this software is to lower the barrier to using complex tensor network methods for optimization and sampling. It offers clean, efficient (GPU-accelerated) and modular software written in Julia. This package serves as both a standalone Ising solver and a source of independent tools for developing physics-inspired algorithms.

\subsection{Software architecture}

\begin{figure}[!t]
    \centering
    \includegraphics[width=\textwidth]{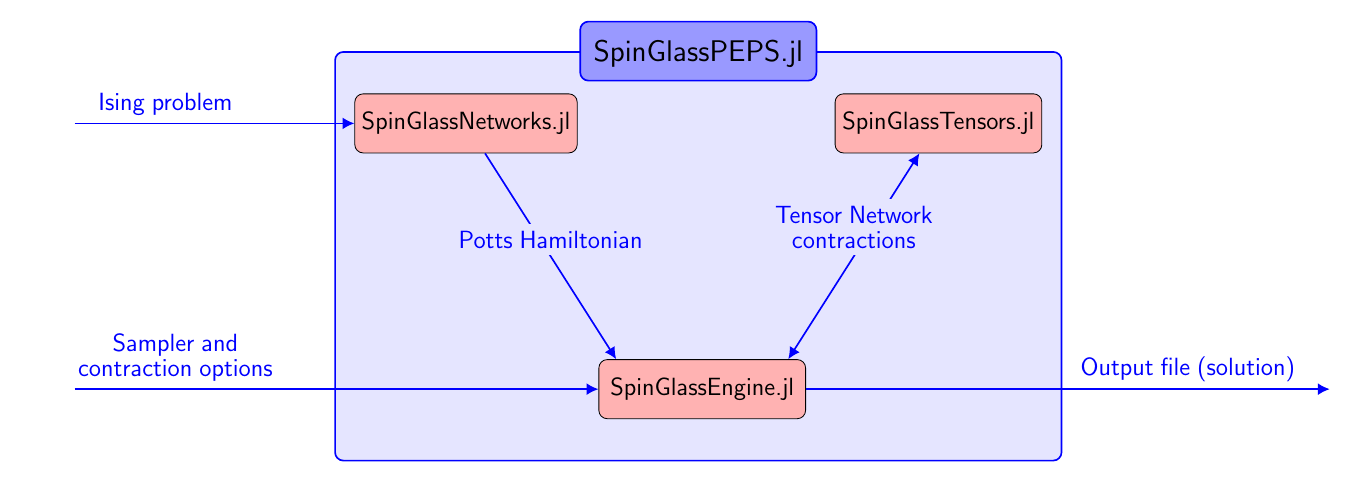}
    \caption[Data flow between \texttt{SpinGlassPEPS.jl} sub-packages.]%
    {Data flow between \texttt{SpinGlassPEPS.jl} sub-packages. An input Ising instance is parsed by \texttt{SpinGlassNetworks.jl}, which constructs the interaction graph, performs clustering, and outputs a Potts Hamiltonian. The Potts instance and solver parameters (e.g., $\beta$, search width $M$, and contraction settings such as maximal bond dimension $\chi$) are passed to \texttt{SpinGlassEngine.jl}, which executes the branch-and-bound search in probability space. Whenever conditional marginals are required, \texttt{SpinGlassEngine.jl} delegates their computation to \texttt{SpinGlassTensors.jl}, which builds the corresponding PEPS and approximately contracts it. The final output consists of a decoded Ising configuration (optionally), together with energies and, when requested, a set of localized low-energy excitations (droplets).}
    \label{fig:software-architecture}
\end{figure}

The structure of \texttt{SpinGlassPEPS.jl} mirrors the computational pipeline required to perform branch-and-bound search in probability space: (i) parse an Ising instance and construct the corresponding interaction graph, (ii) cluster the graph into Potts variables and build the unary/pair energies, (iii) construct the PEPS representation of the partition function, and (iv) repeatedly evaluate conditional marginals via approximate contractions during the search. To keep these concerns separated, the code base is organized into three sub-packages (Fig.~\ref{fig:software-architecture}):

\begin{itemize}
    \item \texttt{SpinGlassNetworks.jl} handles instance ingestion and graph preprocessing. It constructs Ising graphs from standard input formats (including those used in the D-Wave Ocean ecosystem), performs clustering into unit cells, and produces the corresponding Potts Hamiltonian, along with the auxiliary functions needed to decode solutions back to spins.

    \item \texttt{SpinGlassEngine.jl} implements the main solver logic. It performs a branch-and-bound search over Potts variables, manages branching and pruning, and provides routines for post-processing, including the reconstruction of low-energy spectra and the identification of diverse localized excitations above the ground state (droplets). 

    \item \texttt{SpinGlassTensors.jl} provides the core tools for constructing and manipulating the tensors that form the PEPS network, with support for both CPU and GPU execution. It implements the fundamental tensor-network operations required by the solver, most importantly, approximate PEPS contraction via the boundary matrix-product-state (boundary-MPS) scheme~\cite{Nishino2001}. In the default workflow, this package acts as a backend: it is invoked by higher-level components (e.g., the solver in \texttt{SpinGlassEngine.jl}), and most users do not interact with it directly.
.
\end{itemize}

Each sub-package can be used independently, but the default workflow combines them to provide an end-to-end solver.

\subsubsection{Work division between CPU and GPU}

The package supports both CPU-only execution and configurations in which selected kernels are offloaded to a single GPU. In the workflow of Fig.~\ref{fig:algorithm-overview}, the dominant cost is the repeated approximate contraction of PEPS networks required to evaluate conditional marginals during the search. We therefore focus acceleration efforts on tensor contractions.


Contractions are implemented via \texttt{TensorOperations.jl}~\cite{Devos2023}, which provides a unified interface for expressing tensor networks and dispatching to different numerical backends. In CPU mode, contraction kernels are lowered to Julia code calling the standard BLAS/LAPACK stack (e.g., OpenBLAS or MKL). In GPU mode, the same high-level tensor expressions are mapped to CUDA kernels via the Julia wrapper of cuTENSOR~\cite{cutensor}. This design keeps the algorithmic code path identical across CPU and GPU configurations while enabling substantial speed-ups where contraction cost dominates.

From a resource perspective, the method is primarily compute-bound rather than memory-bound in the parameter ranges considered in this thesis. All experiments reported later were executed on a single consumer-grade GPU with 24~GB of device memory, which was sufficient for the selected values of $\chi$ and for the instance sizes considered.


\subsection{Software functionalities}

For a given input instance, the solver returns a \emph{set of low-energy configurations} together with their energies, \textit{i.e}., a practical approximation to the low-energy spectrum. The output also includes \emph{approximate probabilities} of the reported states, obtained from the same tensor-network contractions used to compute conditional marginals during the search. 

The package's modular structure provides a range of options for controlling various aspects of the main algorithm. These include the details and control parameters of the tensor network contraction schemes, the ability to use either CPU or GPU for low-level operations, and the option to construct the tensor network in dense or sparse format. The sparse format~\cite{peps} leverages the internal structures of individual tensors and is particularly essential for handling large clusters, such as those relevant to Pegasus and Zephyr graphs. Detailed information on all required and optional parameters is available in the documentation~\cite{documentation}.

The conceptual workflow in Fig.~\ref{fig:algorithm-overview} is reflected directly in the program interface: an instance is loaded and clustered into a Potts model, contraction and search parameters are specified, and the branch-and-bound routine is executed to produce solutions and post-processed excitations. The following code snippet illustrates this typical usage pattern. The subsequent subsections describe each stage in detail.



\subsubsection{Lattice geometry}

First, one must define and load an instance of the Ising model to solve. 

\begin{lstlisting}
m, n, t = columns, rows, cell_size
instance = "path/to/instance"
lattice = super_square_lattice(topology)
potts_h = potts_hamiltonian(ising_graph(instance),                                spectrum=full_spectrum,                               cluster_assignment_rule=lattice)
\end{lstlisting}

The parameters \lstinline|n, m, t| define the size of the used lattice. Their precise meaning depends on the shape of the underlying lattice. For example, when using \lstinline|super_square_lattice|, which represents a pseudo-king's graph, $n$ and $m$ are rows and columns of the square lattice of unit cells. Parameter $t$ tells how many spins are in each unit cell. Other types of supported lattices are \lstinline|pegasus_lattice| and \lstinline|zephyr_lattice|.

Function \lstinline|potts_hamiltonian| creates a Potts model Hamiltonian (energy function) as described in~\ref{subsec:potts}. The arguments are:

\begin{itemize}
    \item \lstinline|ising_graph|: Loads the instance from \texttt{.txt} file into a graph structure representing spin interactions. The text file should be in the COO convention \textit{i.e.}, the spins are numbered $1$ to $N$ and arranged in rows as $i$ $j$ $v$, where $v$ represents the coupling value between vertices $i$ and $j$, or the local magnetic field when $i = j$.

    \item \lstinline|spectrum|: Specifies to use the complete energy spectrum (rather than a truncated version). The spectrum is calculated independently for each Potts variable. 

    \item \lstinline|cluster_assignment_rule|: Determines how spins are grouped into Potts variables and how those variables are connected. Here we specify the lattice shape that we want to use.
\end{itemize}

\begin{figure}
    \begin{subfigure}{0.4\textwidth}
        \centering
        \includegraphics[width=\linewidth]{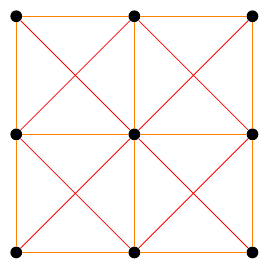}
        \caption{The king's graph with \\$3 \times 3 \times 1$ spins.}
        \label{fig:lattice-1}
    \end{subfigure}
    \hfill
    \begin{subfigure}{0.4\textwidth}
        \centering
        \includegraphics[width=\linewidth]{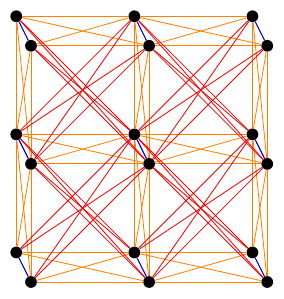}
        \caption{The king's graph with \\$4 \times 4 \times 2$ spins.}
        \label{fig:lattice-2}
    \end{subfigure}
    \caption{Example of lattice graphs represented by \lstinline|super_square_lattice| with $n=m=4$ and $t=1$ (panel~\ref{fig:lattice-1}) or $t=2$ (panel~\ref{fig:lattice-2}). The unit-cells are nodes of a king's graph. Each spin in the unit cell is connected to all other spins on neighboring sites. The couplings within the unit cell are marked blue, horizontal and vertical are colored orange, and diagonal connections are red.}
    \label{fig:lattices}
\end{figure}

\subsubsection{PEPS tensor network construction}

The PEPS tensor network is governed by two structures, \lstinline|PEPSNetwork()| and \lstinline|MpsContractor()|. The first one holds details about how the tensor network is constructed. The second one determines how the tensor network will be contracted.

\vspace{1cm}

\begin{lstlisting}
Node =  KingSingleNode
Layout = GaugesEnergy
net = PEPSNetwork{Node{Layout}, Sparsity, T}(m, n,                                                  potts_h,                                               transform)

\end{lstlisting}

The control parameter \lstinline|transform| manages lattice transformations (rotations) applied to the graph. The branch-and-bound search employed here operates with a fixed order of sweeping through local variables, corresponding to the fixed order of tensor network contraction. This introduces bias as, in general, approximate contractions of PEPS are not transformation-invariant~\cite{Vanderstraeten2022}. To combat this inherent bias, we use \emph{lattice transformations} in our algorithm. They consist of $90\degree$ rotations and reflections along horizontal, vertical, and diagonal axes. It means that, to obtain the best possible solutions, we perform calculations several different times, each time applying a different transformation to create the PEPS network. This approach enhances the stability of the results~\cite{peps,software}. The control parameter \lstinline|transform| manages which (if any) lattice transformations are applied.

For ease of iteration, all possible transformations are wrapped into a single iterator \lstinline|all_lattice_transformations|. 

\begin{lstlisting}
const all_lattice_transformations =                              (rotation.([0, 90, 180, 270])...,                       reflection.([:x, :y, :diag, :antydiag])...)
\end{lstlisting}

Next is the parameter \lstinline|Sparsity| which determines whether a sparse or dense tensor structure should be used. The sparsity used here is distinct from that of traditional sparse matrices~\cite{Saad2003}. The sparse tensor structure circumvents the need for direct construction of individual tensors by performing optimal contractions on small tensor diagrams utilizing the internal structure of individual tensors~\cite{peps}. These diagrams are then combined to efficiently contract the entire created network. A dense tensor structure means that every tensor is explicitly constructed.

The next useful structure is \lstinline|Node|. It specifies the type of node used within the tensor networks. It determines expected structure of clustered Hamiltonian node, if it is a single spin (like in the $n \times m \times 1$ king's graph) or if it is group of spins connected in certain way. Details can be found in the software's documentation~\cite{documentation}. The \lstinline|Layout| parameter decides the division of the PEPS network into boundary Matrix Product States~\cite{Verstraete2008,Orus2014} used to contract the network.

\begin{figure}
    \centering
    \includegraphics[width=0.9\textwidth]{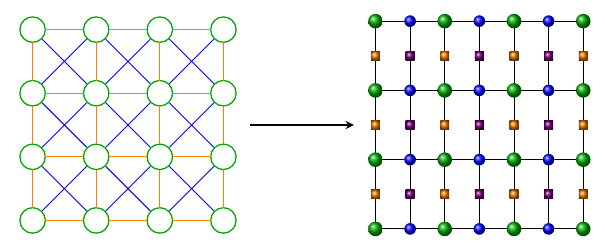}
    \caption{Construction of PEPS tensor network from clustered Hamiltonian (Potts model) defined in Eq.~\ref{eq:potts-peps}. The partition function of a generalized Potts Hamiltonian on a square lattice with nearest-neighbor and diagonal interactions,  can be expressed as a contraction of a tensor network depicted above.}
    \label{fig:potts-to-peps}
\end{figure}

\texttt{SpinGlassPEPS.jl} provides users with a high degree of control over the construction and contraction of the PEPS tensor network. To keep it all manageable, the structures below are used to store and manipulate control parameters. The \lstinline|MpsParameters| \lstinline|struct| encapsulates various control parameters that influence the behavior and accuracy of the MPO-MPS contraction scheme used in PEPS network calculations. This allows fine-tuning of tolerances, iteration limits, and methods for efficient and accurate tensor network contractions. Next useful structure is \lstinline|MpsContractor|. It is a mutable \lstinline|struct| that represents the contractor responsible for contracting a PEPS network.

\begin{lstlisting}
params = MpsParameters{Float64}(; bond_dim=16)
ctr = MpsContractor(Strategy,                                              net,                                                   params;                                                onGPU=true,                                           beta=2)
\end{lstlisting}

Example control parameters are shown in the code above. The \lstinline|bond_dim| specifies the maximum bond dimension between tensors used in the contraction scheme. The user can choose the contraction strategy used in calculations by setting \lstinline|Strategy| parameter. The \lstinline|beta| here is the inverse temperature $\beta$ used for calculating conditional probabilities in the main branch and bound search.

\subsubsection{Tensors}

\begin{figure}[h!]
    \centering
    \includegraphics[width=\textwidth]{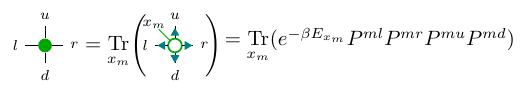}
    \caption{Site tensor. Four virtual legs mediate the interaction structure and trace out Hamiltonian degrees of freedom. Square markers denote the exponents of the corresponding interaction matrices.}
    \label{fig:site-tensor}
\end{figure}

\begin{figure}[h!]
    \begin{subfigure}[t]{0.48\textwidth}
        \centering
        \caption{}
        \includegraphics[width=\linewidth]{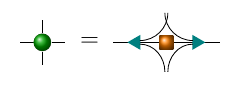}
    \end{subfigure}
    \hfill
    \begin{subfigure}[t][][c]{0.48\textwidth}
        \centering
        \caption{}
        \includegraphics[width=\linewidth]{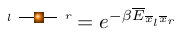}
        
    \end{subfigure}
    \caption{Additional structures. A diagonal tensor (a) carries diagonal interactions. Square markers (b) denote the exponents of the corresponding interaction matrices.}
    \label{fig:diagonal-tensor}
\end{figure}

We implement several variants of the generic \lstinline|Tensor| data structure, each associated with a distinct role in the PEPS network. \lstinline|SiteTensor| objects encode the local energy contributions of a cluster together with the structure of its couplings to neighbouring sites. \lstinline|DiagonalTensor| objects represent diagonal interactions between neighbouring clusters, allowing us to rewrite quasi-two-dimensional interaction graphs as a PEPS defined on a square lattice, which can then be treated using standard approximate tensor-network contraction techniques. The structure of these tensors is shown in Figs.~\ref{fig:site-tensor} and~\ref{fig:diagonal-tensor}.

\subsubsection{Sampling and droplets}

This is the final stage of the SpinGlassPEPS.jl workflow. The \lstinline|SearchParameters| encapsulates parameters that control the behavior of low-energy spectrum search algorithms. The main ones are the maximum number of states to consider during the search and the cutoff probability for terminating the search.
\vspace{1.6cm}

\begin{lstlisting}
search_params = SearchParameters(; 
                                 max_states=2^8,                                        cut_off_prob=1E-4)

droplets = SingleLayerDroplets(max_energy=10,                                         min_size=54,                                           metric=:hamming)

merge_strategy = merge_branches(ctr;                                                  merge_prob=:none,                                      droplets_encoding=droplets)

solution, schmidt = low_energy_spectrum(ctr,                                                   search_params,                                         merge_strategy)
\end{lstlisting}

The call to \lstinline|low_energy_spectrum| produces a \lstinline|Solution| object summarizing the low-energy spectrum identified during the search and a dictionary containing Schmidt spectra~\cite{Verstraete2008} for each row of the PEPS network.

\begin{lstlisting}
struct Solution
    energies::Vector{<:Real}
    states::Vector{Vector{Int}}
    probabilities::Vector{<:Real}
    degeneracy::Vector{Int}
    largest_discarded_probability::Real
    droplets::Vector{Droplets}
    spins::Vector{Vector{Int}}
end
\end{lstlisting}

This structure aggregates all information produced by the branch-and-bound exploration: the energy spectrum, the corresponding spin configurations, their estimated Boltzmann weights, degeneracies, and an explicit list of droplets.

Low-energy excitations (i.e., droplets)~\cite{Newman2024,Shen2024} above the best solution can be identified during the optional \lstinline|merge_branches| step. This option can be provided as an argument to the function that executes the branch-and-bound algorithm \lstinline|low_energy_spectrum|. The algorithm searches for diverse excitations within a specified energy range above the ground state. An excitation is accepted only if its Hamming distance from any previously identified excitation exceeds a predefined threshold. This is governed by the parameters \lstinline|energy_cutoff|, which sets the maximum allowed energy above the ground state, and \lstinline|hamming_cutoff|, which determines the minimum Hamming distance required between excitations for them to be considered distinct. To view all droplets found, one may invoke \lstinline|unpack_droplets(solution)|.

Inside the \lstinline|low_energy_spectrum| function, we perform the search as depicted in Fig.~\ref{fig:branch-and-bound-ising-prob}. It is done by calling \lstinline|branch_solution| and \lstinline|bound_solution| for each node in the created clustered Hamiltonian.

\begin{lstlisting}
sol = empty_solution(S)
old_row = ctr.nodes_search_order[1][1]

for node in ctr.nodes_search_order
    ctr.current_node = node
    current_row = node[1]
    if current_row > old_row
        old_row = current_row
        clear_memoize_cache_after_row()
    end
    sol = branch_solution(sol, ctr)
    sol = bound_solution(sol,                                                   sparams.max_states,                                    sparams.cutoff_prob,                                   merge_strategy)
    Memoization.empty_cache!(precompute_conditional)
    if no_cache
       Memoization.empty_all_caches!()
    end
end

clear_memoize_cache_after_row()
empty!(ctr.peps.lp, :GPU)
\end{lstlisting}

We have omitted some parts of the code to focus on the main ideas. The algorithm traverses Potts clusters in a fixed \lstinline|nodes_search_order| and incrementally builds a set of candidate partial configurations \lstinline|sol|. At each node, \lstinline|branch_solution| extends every retained partial assignment by all admissible local states, while \lstinline|bound_solution| prunes the resulting pool back to at most \lstinline|sparams.max_states| candidates using their (approximate) marginal probabilities, optionally merging near-duplicate branches via \lstinline|merge_strategy| and discarding states below \lstinline|sparams.cutoff_prob|. These marginals are evaluated by the \lstinline|MpsContractor| (\lstinline|ctr|) through repeated contractions of the underlying PEPS, which dominate the runtime. To keep this step feasible, contraction results are memoized and selectively cleared: caches are reset when the sweep advances to a new lattice row (to limit memory growth and avoid stale intermediates), and are fully emptied when \lstinline|no_cache| is enabled. After the sweep, remaining caches and GPU work buffers are released, yielding the final low-energy configurations and their associated energies,  contraction-derived probability estimates, degeneracies, etc.

\begin{figure}[h]
    \centering
    \includegraphics[width=\textwidth]{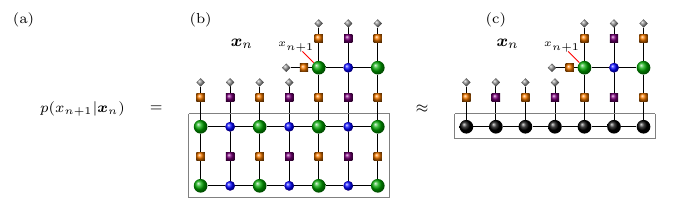}
\caption{Computation of conditional probabilities via approximate PEPS contraction.
    The marginal update in the branch-and-bound sweep reduces to evaluating
    conditional probabilities \(p(x_{n+1}\mid \partial\bm{x}_n)\), which (up to an overall normalization)
    are represented by the tensor network shown in panel~(a), closely related to the PEPS construction in
    Fig.~\ref{fig:potts-to-peps}. Gray diamonds fix (project) the virtual indices on the boundary
    \(\partial\bm{x}_n\) associated with a given partial configuration \(\bm{x}_n\), while the \((n+1)\)th site tensor
    carries an additional open leg corresponding to the untraced physical degree of freedom \(x_{n+1}\).
    In panel~(b), the lower part of the network is compressed using the boundary-MPS method (highlighted by gray boxes),
    producing an approximate effective environment with bond dimension \(\chi\).
    This preprocessing is performed once and is independent of the particular \(\bm{x}_n\).
    The resulting reduced network is then contracted exactly, and the outcome is normalized to obtain
    \(p(x_{n+1}\mid \partial\bm{x}_n)\).
    }

    \label{fig:peps-prob}
\end{figure}

\section{Illustrative examples}

This section demonstrates the main capabilities of \texttt{SpinGlassPEPS.jl} and how its modular components can be combined in practice. We consider two representative workloads. First, we solve an Ising instance defined on a king's graph, corresponding to the square-lattice PEPS setting of Fig.~\ref{fig:algorithm-overview}(a). While this example uses a regular topology for clarity, the same workflow applies after clustering to the structured hardware graphs of D-Wave processors (Pegasus and Zephyr). Second, we illustrate the generality of the Potts-layer interface by solving an image inpainting problem formulated as a Potts model~\cite{Lellmann2011,Kappes2013,openGM}.

\subsection{Solving the Ising Model}

The example here presents a complete Julia script that defines and solves an Ising problem on a king's graph. We first load a small instance and construct the corresponding Potts Hamiltonian using a prescribed cluster assignment. In this example we use a $3\times 3\times 2$ king's-graph instance:

\begin{lstlisting}
T = Float64
m, n, t = 3, 3, 2
instance = get_instance(m, n, t)
lattice = super_square_lattice(m, n, 3)
potts_h = potts_hamiltonian(
    ising_graph(instance);
    spectrum=full_spectrum,
    cluster_assignment_rule=lattice
)
\end{lstlisting}

We then specify the contraction parameters (bond dimension and sweep count) and the search parameters controlling the search width and probability cutoff:

\begin{lstlisting}
params = MpsParameters{T}(; bond_dim=16, num_sweeps=1)
search_params = SearchParameters(; max_states=2^8, cut_off_prob=1e-4)
hamming_dist = 5
eng = 10
\end{lstlisting}

Finally, we run the solver. To reduce orientation-dependent artifacts introduced by approximate contraction and truncation, we sweep over all lattice symmetries via \lstinline|all_lattice_transformations| and keep the best energy encountered. For each transformation, we construct the PEPS network, initialize the \lstinline|MpsContractor| (here in CPU mode), configure the droplet-based excitation handling and branch-merging strategy, and call \lstinline|low_energy_spectrum|:

\begin{lstlisting}
best_energies = T[]
for transform in all_lattice_transformations
    net = PEPSNetwork{KingSingleNode{GaugesEnergy}, Dense, T}(
        m, n, potts_h, transform
    )
    ctr = MpsContractor(
        SVDTruncate, net, params;
        onGPU=false, beta=T(2), graduate_truncation=true
    )
    single = SingleLayerDroplets(eng, hamming_dist, :hamming)
    merge_strategy = merge_branches(
        ctr; merge_type=:nofit, update_droplets=single
    )
    sol, _ = low_energy_spectrum(ctr, search_params, merge_strategy)
    push!(best_energies, sol.energies[1])
    clear_memoize_cache()
end
\end{lstlisting}

\subsection{Solving the Potts model: inpainting}
\label{MRFS}

The Potts Hamiltonian in Eq.~\eqref{eq:potts-peps} is not restricted to Ising models. It is a standard representation of discrete Markov random fields (MRFs) used in computer vision. A representative example is \emph{image inpainting}, where the goal is to fill missing or corrupted pixels in a visually plausible way~\cite{Zeng2020}. In this setting, each variable $x_n$ corresponds to a pixel label (e.g., color), unary terms encode agreement with observed data, and pairwise terms penalize discontinuities to encourage piecewise-smooth reconstructions with sharp edges.

\begin{figure}[!t]
    \centering
    \begin{subfigure}[b]{0.31\textwidth}
        \centering
        \includegraphics[width=\textwidth]{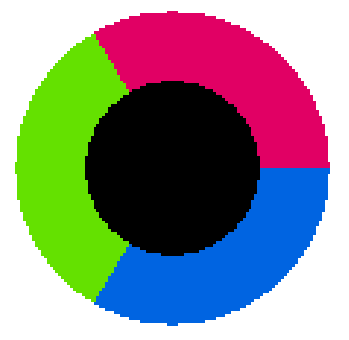}
        \caption{Input data.}
    \end{subfigure}
    \hfill
    \begin{subfigure}[b]{0.33\textwidth}
        \centering
        \includegraphics[width=0.97\textwidth]{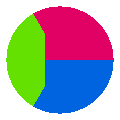}
        \caption{Best Found}
    \end{subfigure}
    \hfill
    \begin{subfigure}[b]{0.31\textwidth}
        \centering
        \includegraphics[width=0.98\textwidth]{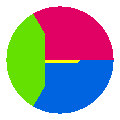}
        \caption{Droplet.}
    \end{subfigure}
    \caption[Triple-junction inpainting as a Potts-model benchmark.]%
    {Triple-junction inpainting benchmark~\cite{Lellmann2011,openGM}. Panel (a) shows the input: pixel labels are fixed on a circular boundary, while the interior (black region) is unknown. Panel (b) shows a lowest-energy configuration found by \texttt{SpinGlassPEPS.jl}. Due to the grid-based discretization, the energy function is mildly anisotropic, which can bias reconstructed interfaces toward axis-parallel directions~\cite{Lellmann2011}. Panel (c) highlights a droplet, \textit{i.e.}, a localized set of Potts variables that can be collectively relabeled to produce an alternative low-energy solution. Such droplets provide a compact description of near-degeneracies and are useful for characterizing solution diversity in subsequent benchmarking.}
    \label{fig:inpainting}
\end{figure}

Here we consider a small benchmark instance from~\cite{Lellmann2011,openGM}, provided in the OpenGM2 dataset format. The instance corresponds to a discretized \emph{triple-junction} inpainting problem (Fig.~\ref{fig:inpainting}): boundary data are specified on a ring, and the interior region must be reconstructed. \texttt{SpinGlassPEPS.jl} can load OpenGM-style instances with nearest-neighbor and diagonal interactions, which match the King's graph geometry used by the PEPS construction in this chapter.

\begin{lstlisting}
instance = "$(@__DIR__)/instances/triplepoint4-plain-ring.h5"
# Image size in pixels
potts_h = potts_hamiltonian(instance, 120, 120)
\end{lstlisting}

As in the Ising example, the solver can return a best-found configuration together with a set of localized low-energy excitations (droplets). For inpainting/MRF instances, it is recommended to use the \lstinline|:RMF| droplet mode, which is tailored to multi-label Potts variables rather than binary spins:

\begin{lstlisting}
droplets = SingleLayerDroplets(;
    max_energy=100,
    min_size=100,
    metric=:hamming,
    mode=:RMF
)
\end{lstlisting}

The remaining setup (choice of contraction accuracy $\chi$, inverse temperature $\beta$, beam width $M$, and CPU/GPU backend) is analogous to the King's graph case described in the previous subsection. In practice, since the inpainting landscape is typically highly degenerate, the solver may find multiple visually plausible reconstructions.

\section{Conclusion}

In this chapter, we have introduced \texttt{SpinGlassPEPS.jl}, a tensor-network-based solver tailored to Ising-like optimization problems defined on quasi-two-dimensional graphs, with a particular focus on geometries relevant to current quantum annealing processors such as Pegasus and Zephyr. The central algorithm reformulates the ground-state search as a branch-and-bound procedure in probability space: low-energy configurations are identified as the most probable states of an effective Potts model at finite inverse temperature, with the corresponding Boltzmann marginals approximated via PEPS contractions on a square lattice. This construction allows one to handle large unit cells by clustering physical spins into higher-dimensional Potts variables and exploiting locality.

On structured, grid-like instance families (square and king's graphs), the PEPS-based tensor-network solver \texttt{SpinGlassPEPS.jl} is the most accurate method in our benchmark, for the median instance it reaches $E_{\mathrm{approx}}=0$, \emph{i.e.}, it attains the best-reference energy among all compared solvers. This advantage, however, comes at a higher computational budget. The regime in which the TN solver closes the remaining optimality gap is the long-runtime tail (typically $\sim 10^{1}$--$10^{3}\,$s), whereas faster baselines achieve near-optimal energies quickly but plateau at a small residual error that grows mildly with system size.

From an implementation perspective, \texttt{SpinGlassPEPS.jl} provides a modular Julia ecosystem that mirrors the algorithm's structure. The package \texttt{SpinGlassNetworks.jl} maps Ising instances to clustered Potts Hamiltonians, \texttt{SpinGlassEngine.jl} implements the deterministic branch-and-bound search and reconstruction of the low-energy spectrum, and \texttt{SpinGlassTensors.jl} provides an approximate contraction back-end with seamless CPU/GPU integration built on well-established \texttt{TensorOperations.jl}. This design lowers the barrier to using tensor-network techniques for classical optimization~\cite{Cichocki2014} by combining sparse tensor structures~\cite{peps}, GPU acceleration, and a clean, composable API, while still allowing individual components to be reused in other physics-inspired algorithms.

Within the broader context of this thesis, SpinGlassPEPS.jl plays a dual role. First, it serves as a QPU topology-aware classical baseline, enabling fair, like-for-like comparisons to quantum annealers on native hardware graphs in the benchmarking study of Chapter~\ref{chap:bench}. Second, it provides a flexible platform for further methodological developments, such as alternative contraction schemes (e.g., Corner Transfer Matrix~\cite{Lukin2024, Mangazeev2023}), extended Potts geometries (e.g., LHZ-type architecture~\cite{Lechner2015}), multi-GPU execution, and approximate thermodynamic observables~\cite{Nishimori2025}. Together, the algorithmic and software contributions of this chapter establish a practically useful tensor-network solver that both clarifies the strengths and limitations of TN-based optimization~\cite{peps} for QPU-relevant graphs and supports the subsequent analysis of quantum annealers in the rest of the thesis.

%% file: chapters/bench.tex
\chapter{Benchmarking Quantum Annealers Against Selected Classical Solvers}
\chaptermark{Benchmarking QA}
\label{chap:bench}

Benchmarking quantum optimization devices is essential for separating genuine computational advantage from improvements that can be explained by instance selection or evaluation artifacts~\cite{Tasseff2024,Pawlowski2025,Munoz-Bauza2025,Schulz2025}. At the same time, it is technically demanding: quantum annealers are analog, open, and hardware-constrained systems whose performance depends not only on problem structure, but also on control errors, finite temperature, limited connectivity, parameter setting, and sampling noise (see Chapter~\ref{chap:AQC-QA}). These features blur the boundary between algorithmic performance and hardware idiosyncrasies, making “fair” comparisons to classical solvers nontrivial. Meaningful benchmarking, therefore, requires carefully designed instance families that probe relevant structure while avoiding triviality,  explicit runtime and success criteria, and metrics that capture both optimization quality and the ability to generate diverse low-energy solutions rather than a single best outcome~\cite{Ronnow2014}.

In this chapter, we benchmark quantum annealing devices against a set of competitive classical baselines under a consistent experimental protocol. Performance is evaluated using complementary indicators: the best energy observed, time-to-approximation-ratio, and diversity of near-optimal solutions. Together, these metrics provide a multi-faceted picture of annealer behavior across instance classes and problem sizes and enable a principled assessment of where quantum annealing is competitive, where it saturates, and how its performance compares to classical heuristics. An additional goal of this chapter is to benchmark \texttt{SpinGlassPEPS.jl} as a transparent, topology-aware tensor-network baseline for Ising optimization and low-energy sampling.

\paragraph{Author contributions} 
My contribution in this chapter is the development and execution of a controlled benchmarking framework for quantum annealers versus classical baselines, including the construction of instance families and the systematic collection and interpretation of experimental results under a unified protocol.

\paragraph{Relation to published work}
This chapter synthesizes and reinterprets results previously reported in my publications~\cite{software,peps}, organizing them into a unified benchmarking narrative and a consistent set of metrics and comparisons used throughout the thesis. While the majority of experimental results and baselines follow the published studies, the chapter also includes new material that was not part of the original publications: in particular, an additional analysis of D-Wave performance on lattice-structured instances and the resulting insights into the practical impact of minor-embedding.


\section{Methodology}

Establishing a fair benchmarking methodology for comparing quantum annealers with classical solvers is nontrivial. The outcome depends not only on the quantum device, but also on the strength of the chosen classical baselines, implementation details, and hyperparameter tuning (e.g., temperature schedules, stopping criteria, and parallelization strategy)~\cite{Ronnow2014}. Instance selection is equally important: different instance families can probe different algorithmic mechanisms and may favor particular solver architectures. In this chapter, we therefore distinguish two broad classes of problem instances: \emph{topology-native} instances (defined directly on the hardware connectivity graph) and \emph{lattice-based} instances (defined on regular lattices and, when needed, embedded into the target topology). Both classes are described in detail in the following sections.

A second methodological issue concerns the definition of runtime. For deterministic solvers (e.g., \texttt{SpinGlassPEPS.jl}), it is natural to report the elapsed wall-clock time of the computation, together with the computational setting (CPU/GPU model, number of threads, and solver parameters). For stochastic solvers, including quantum annealers and randomized classical heuristics such as simulated bifurcation, a single run does not certify optimality, and performance is typically summarized by a \emph{time-to-solution} (TTS) metric, defined as the expected time required to obtain at least one successful sample with a prescribed confidence level.

Let $p_s$ denote the probability that a single solver run returns a \emph{successful} sample (according to a specified success criterion), and let $t_{\mathrm{run}}$ denote the duration of one run. The time-to-solution for target confidence $p_{t}$ is then

\begin{equation}
    \label{eq:tts}
\mathrm{TTS}(p_t) =  t_{\mathrm{run}} \, \frac{\ln (1 - p_t)}{\ln(1 - p_s)}.
\end{equation}

In our experiments, for D-Wave processors, we set $t_{\mathrm{run}}$ to the total QPU use time as reported by the Ocean toolkit, while for classical stochastic solvers, we use the measured wall-clock compute time per run on the corresponding hardware. 

Since identifying the true optimum energy $E^\star$ is often infeasible at the scales considered here, we also report a \emph{time-to-target} metric based on approximate solutions. Concretely, we use time-to-approximation-ratio (also referred to as time-to-$\varepsilon$), where a run is deemed successful if it returns a configuration with energy within a prescribed approximation ratio using a relative tolerance. Defining

\begin{equation}
    \label{eq:tt-epsilon}
p_{\varepsilon} = P\left(E(x)\le E^\star+\varepsilon\right),
\end{equation}

\noindent the corresponding runtime estimator is

\begin{equation}
\mathrm{TT}_{\mathcal{R}}(p_t) = t_{\mathrm{run}}\, \frac{\ln\!\bigl(1-p_t\bigr)}{\ln\!\bigl(1-p_{\varepsilon}\bigr)} .
\end{equation}

When using approximation ratios, $\varepsilon$ is replaced by a relative threshold (defined in Section~\ref{sec:metrics}), while the probabilistic structure of the metric remains unchanged.

Finally, empirical TTS distributions are often heavy-tailed and strongly instance-dependent~\cite{Steigner2015}. To reduce sensitivity to outliers and improve interpretability across heterogeneous instance families, we report robust summary statistics (in particular medians and selected quantiles) rather than relying exclusively on means. All classical computations were performed on the hardware reported in Table~\ref{tab:benchmark-hardware}.


\begin{table}[!t]
    \centering
    \caption[Benchmark hardware]{Hardware configuration used for the benchmarks reported in this chapter.}
    \label{tab:benchmark-hardware}
    \begin{tabular}{ll}
        \toprule
        Component & Specification \\
        \midrule
        CPU   & Intel Core i9-7980XE, 2.60\,GHz \\
        System memory & 128\,GB RAM \\
        GPU   & NVIDIA TITAN RTX \\
        GPU memory & 24\,GB VRAM \\
        \bottomrule
    \end{tabular}
\end{table}

\subsection{Problem instances}

We consider two broad families of benchmark instances. The first family consists of instances native to the connectivity graphs of contemporary quantum processing units (QPUs), such as the Pegasus and Zephyr topologies realized in D-Wave devices; see section~\ref{sec:dwave} and Figure~\ref{fig:dwave-topologies}. In this setting, the logical interaction graph of the Ising model coincides with the hardware graph, so that no additional embedding overhead is incurred. We focus on several chosen instance sizes that correspond to theoretical hardware graph topologies. For Pegasus, these sizes are $216$ spins (P4), $1176$ spins (P8), and $5375$ spins (full working graph). For Zephyr, due to the small size of the prototype machine, we use instance sizes of $332$ spins (Z3) and $563$ spins (full working graph).
 
The second family comprises lattice instances defined on simple two-dimensional graphs with small repeating unit cells (see Fig.~\ref{fig:lattices}). The unit cell consists of one or two spins. These lattices interpolate between purely nearest-neighbor square geometries and more densely connected King's graphs, while retaining a high degree of spatial regularity. We consider $50 \times 50$ site lattices. 
 
We further subdivide the instances according to the way in which the coupling strengths $J_{ij}$ and local fields $h_i$ are generated. In particular, we distinguish three principal classes, denoted RCO (Random Coupling Only), RAU (Random Uniform), and CBFM-P (Corrupted Biased Ferromagnet for Pegasus)~\cite{Tasseff2024}. Their characteristics are summarized in Table~\ref{tab:instances}. For Pegasus-native instances, we used all three classes; Zephyr-native used two classes (RAU and RCO), and all lattices are of the RCO class.

\begin{itemize}
    \item RAU - Random Uniform. The couplings are taken uniformly at random from the [-1, 1] interval, $J_{i,j} \in \mathcal{U}(-1,1)$, and the external fields are taken from $[-0.1, 0.1]$ interval, $h_i \in \mathcal{U}(-0.1, 0.1)$.

    \item RCO - Random Coupling Only. The external fields are all set to $0$ and couplings are generated from $[-1, 1]$ interval uniformly at random,  $J_{i,j} \in \mathcal{U}(-1,1)$.  

    \item CBFM-P - Corrupted Biased Ferromagnet instances for Pegasus~\cite{Tasseff2024}. It uses the following distribution of parameters for the connectivity graph $G$:

        \begin{equation*}
            \begin{split}
            \forall_{(i,j) \in E(G)} \,P(J_{ij} = 0) = 0.35, \, P(J_{ij} = -1) = 0.10, \, P(J_{ij} = 1) = 0.55 \\
            \forall_{i \in V(G)} \, P(h_i = 0) = 0.15, \, P(h_i = -1) = 0.85, P(h_i = 1) = 0.
            \end{split}
        \end{equation*}
\end{itemize}

\begin{table}[h]
    \centering
    \begin{tabularx}{\textwidth}{|c|X|}
        \hline
        Class & Description  \\ \hline \hline
        RCO &  Random couplings in $[-1, 1]$ \\ \hline
        RAU &   RCO $+$ random local fields in  $[-0.1, 0.1]$ \\ \hline
        CBFM-P & Instances of Ref.~\cite{Tasseff2024} for Pegasus graph.\\ \hline
    \end{tabularx}
    \caption{Classes of problem instances on native QPU hardware graph that are considered in this benchmark}
    \label{tab:instances}
\end{table}

It is worth noting that RCO instances exhibit a global reflection symmetry. That means that for any given solution $\bm{s}$, the solution $-\bm{s}$, where every spin is flipped, gives exactly the same energy. This symmetry is broken in the RAU instances. The instances of the class CBFM-P are taken from~\cite{Tasseff2024} and serve as a standard benchmarking class.

For each of the aforementioned categories and given instance size, we generated $20$ test instances. The number of instances allows us to test a variety of instances while keeping the time cost reasonably low.

\subsection{Performance metrics}
\label{sec:metrics}

To enable a meaningful comparison between quantum annealers and classical solvers, we employ several complementary performance metrics. The two fundamental metrics are \emph{quality of solutions} denoted by $E_{\mathrm{approx}}$ and \emph{diversity of solutions} $\mathcal{D}$. Both will be considered within the time-to-solution (TTS) framework.

The first metric, $E_{\mathrm{approx}}$, is defined as the relative difference between the solution found by the given method and the best one found across all methods. For a given instance, let $E$ be the energy of the best solution found by a particular method, and let $E_{\text{best}}$ denote the lowest energy obtained for that instance across \emph{all} methods and all runs considered in the benchmark. Since, in many cases, certifying the true ground-state energy is practically infeasible, $E_{\text{best}}$ serves as a pragmatic surrogate.

\begin{equation}
    E_{\mathrm{approx}}
    = \frac{E - E_{\text{best}}}{2\lvert E_{\text{best}} \rvert}.
\end{equation}

The diversity of solutions~\cite{Zuca2021} metric serves as a way to probe the quality of sampling of diverse approximate solutions of a spin-glass problem. We search for high-quality solutions, meaning their energy is within \emph{approximation ratio} $a_r$ of the best found energy. Formally $Q_{\text{sol}} \leq a_r$. Within those, we chose the largest set of independent solutions. We say that configurations $\bm{s}$ and $\bm{s}'$ are independent if their Hamming distance $d$ is above a chosen threshold $R$ relative to their size $N$. Formally:

\begin{equation}
    \mathcal{D} = \vert \max  \{\bm{s} ; \, \forall_{\bm{s}} \, Q_{\text{sol}} \leq a_r \,\wedge \, \forall_{\bm{s}, \bm{s}'} \, d(\bm{s}, \bm{s}') \geq RN  \} \vert
\end{equation}

To compare the performance of the considered solvers, we first estimate the reference (ground-truth) diversity, $\mathcal{D}_{\mathrm{total}}$, for each problem instance. Specifically, we pool all candidate solutions from all solvers and identify the largest subset of independent, low-energy solutions that lie within the prescribed approximation ratio $a_r$. This selection task can be cast as a maximum-clique problem by constructing a graph whose vertices represent eligible low-energy solutions and whose edges connect pairs of solutions whose Hamming distance exceeds a chosen threshold (so that every pair in the selected subset is mutually ``independent''). The associated decision version of the clique problem is NP-complete~\cite{Karp1972}. We approximate its solution using a randomized heuristic approach. We iterate over the list of states in a random order, building, in the process, a set of independent states. At each iteration step, a configuration is added to the list of independent states if it is independent of all configurations already in the list. We select the largest set obtained across all the restarts of that procedure. Empirically, the clique size typically saturates after about $10^2$ restarts. Finally, we set $\mathcal{D}_{\mathrm{total}}$ to the size of the resulting clique. Our final metric is $\mathcal{D}_{mathrm{approx}}$ defined as:

\begin{equation}
    \mathcal{D}_{\mathrm{approx}} = \frac{\mathcal{D}_{\mathrm{solver}}}{\mathcal{D}_{\mathrm{total}}},
\end{equation}

\noindent where $\mathcal{D}_{\mathrm{solver}}$ is the diversity of solutions found by each solver for a given problem instance.

Both metrics are explored within the time-to-approximation-ratio framework. We are interested in the full time-quality trade-off curve, \textit{i.e.}, comparing the achieved approximation ratio with the computation time required to obtain it.

\subsection{Solvers}

The main method we used for benchmarking is \texttt{SpinGlassPEPS.jl} implementation of the tensor-network (TN) based algorithm described in the previous chapter. Parameter selection was performed separately for each graph type. For example, the $\beta$ parameter was set $\beta = 1/2$ for Pegasus geometries while $\beta=1$ was used for Zephyr topologies. In addition to full-space TN simulations, we additionally conduct simulations with local dimensional reduction of cluster degrees of freedom, as described in~\cite{peps}. Here, for Pegasus-native instances we reduce the $2^{24}$ cluster states to $2^{20}$ and $2^{16}$. For Zephyr-native instances we reduce from $2^{16}$ states to $2^{14}$ and $2^{12}$ states. Due to memory limitations, for the Pegasus instances with $5376$ spins, we only performed calculations with local dimensional reduction to $2^{16}$ states. 

When performing quantum annealing, we have chosen two different total annealing times: $\tau = 200 \, \mu$s and $\tau = 2000 \, \mu$s. Due to time limitations, not all instances were solved within $200 \, \mu$s of total annealing time. Details and physical characteristics of used devices can be found in Appendix~\ref{apx:dwave}. 

Additionally, we employed several different classical and physics-inspired solvers as baselines for benchmarking the quantum annealing devices. In particular, we used simulated annealing (SA), simulated bifurcation (SBM), and parallel annealing (PA). Most of these solvers were implemented by the author in a unified code base~\cite{code}, ensuring consistent instance handling, parameterization, and performance evaluation across the different approaches. Simulated bifurcation is an exception, as we have used an external implementation. All classical computations were performed on the hardware reported in Table~\ref{tab:benchmark-hardware}.

\section{Results}

Unless stated otherwise, we report \emph{median} performance over $20$ randomly generated instances per class and size, \textit{i.e.}, at each runtime we take the median value of the metric across the instance set. We focus on \emph{native} problem graphs to avoid confounding effects from minor embedding. Figs.~\ref{fig:pegasus-bench-E},~\ref{fig:zephyr-bench-E} and~\ref{fig:lattice-bench-E} summarize optimization performance via the approximation ratio $E_{\mathrm{approx}}$, where $E_{\mathrm{approx}}=0$ corresponds to the best energy observed across \emph{all} solvers on that instance set. For Pegasus, we consider $N=216,1176,5376$ spins, for Zephyr $N=332,563$ spins, and for lattice instances $N=400,900,1600$ spins. The largest size for each topology uses the full working graph, and for lattices, it is the biggest King's graph embeddable on the available QPU.

\subsection{Solution quality and time-to-approximation}
\label{sec:results-energy}

\begin{figure}[t]
    \centering
    \includegraphics[width=\textwidth]{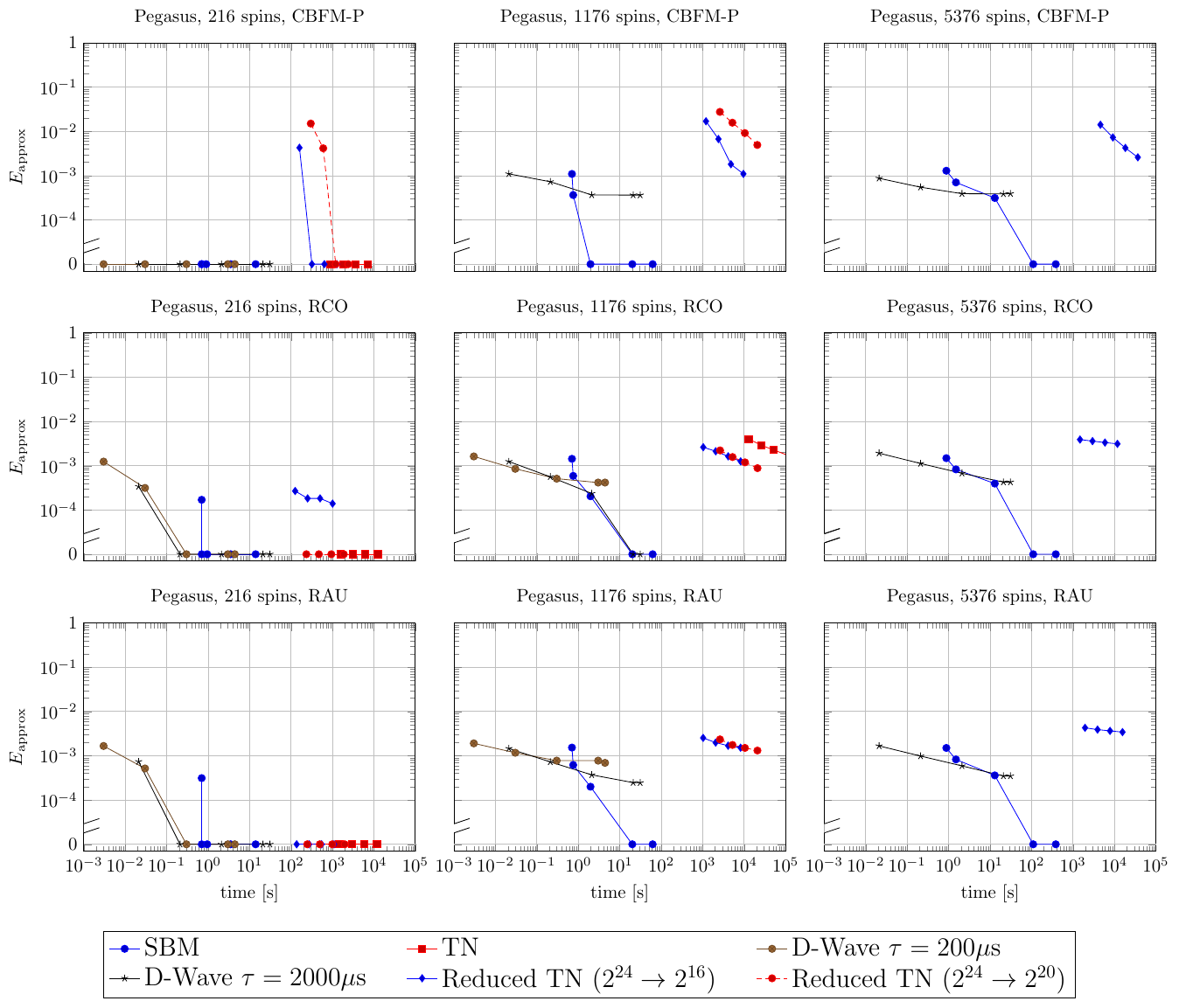}
    \caption{Time-to-approximation-ratio on Pegasus-native Ising instances. Each panel shows the approximation ratio $E_{\mathrm{approx}}=(E-E_{\mathrm{best}})/(2|E_{\mathrm{best}}|)$ versus runtime for three instance classes: CBFM-P (top row), RCO (middle row), and RAU (bottom row), and system sizes $N=216,1176,5376$ spins. We compare simulated bifurcation (SBM), the PEPS-based tensor-network solver (TN), and D-Wave quantum annealing with anneal times $\tau=200\,\mu\mathrm{s}$ and $2000\,\mu\mathrm{s}$. “Reduced TN” denotes TN runs preceded by local dimensional reduction of each 24-spin Pegasus cluster from $2^{24}$ to $2^{16}$ or $2^{20}$ states. Lower values indicate better solutions, with $E_{\mathrm{approx}}=0$ corresponding to the best energy observed across all solvers.}

    \label{fig:pegasus-bench-E}
\end{figure}

\paragraph{Pegasus}
For the smallest Pegasus instances ($N=216$, left column of Fig.~\ref{fig:pegasus-bench-E}), most methods reach $E_{\mathrm{approx}} \approx 0$ within the displayed budgets, but the required time differs by orders of magnitude. Simulated bifurcation (SBM) and quantum annealing (QA) reduce $E_{\mathrm{approx}}$ rapidly and typically attain the best-reference energies in the sub-second to few-second regime. In contrast, the tensor-network (TN) variants reach comparable accuracy only at substantially longer times, indicating that at this scale, TN can match solution quality but at markedly higher computational cost.

At $N=1176$ (middle column), solver behavior separates clearly across all instance classes. SBM is the only method that consistently reaches $E_{\mathrm{approx}}=0$ within moderate runtimes (typically $\mathcal{O}(1)$--$\mathcal{O}(10)$~s for the median instance). QA improves quickly at short times but often saturates at a small nonzero error ($\sim 10^{-4}$--$10^{-3}$), suggesting near-optimal solutions that do not coincide with the best-reference energy within the explored budget. The reduced-TN variants appear only at long runtimes ($\sim 10^{3}$--$10^{4}$~s) and plateau at higher errors (often $10^{-3}$--$10^{-2}$), consistent with an explicit accuracy--tractability trade-off introduced by local-state truncation.

For the largest Pegasus instances ($N=5376$, right column), these trends are even stronger. SBM still reaches $E_{\mathrm{approx}}=0$ in the median curves, albeit at larger time budgets (typically $\sim 10^{2}$~s). QA continues to deliver high-quality solutions quickly but does not reach the best-reference energy within the shown budgets, instead flattening at a small yet nonzero $E_{\mathrm{approx}}$. TN-based approaches remain confined to long runtimes and show clear saturation at higher approximation errors, with no indication (within feasible execution parameters) of closing the gap to SBM at this scale.

\begin{figure}[t]
    \centering
    \includegraphics[width=\textwidth]{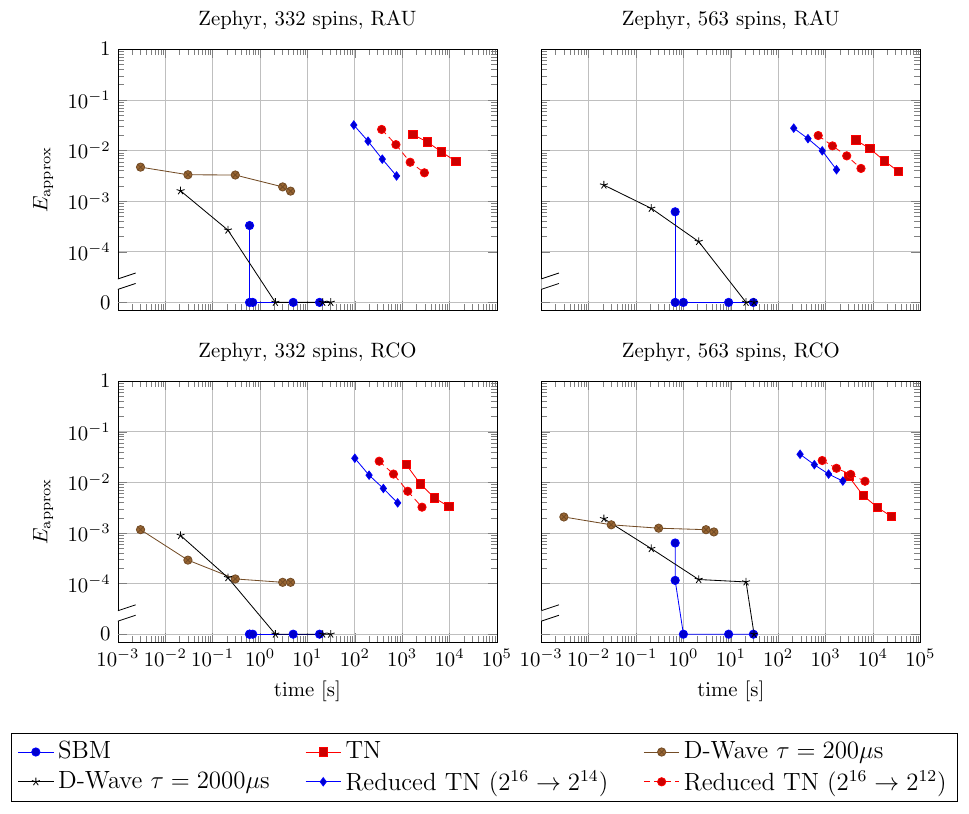}
    \caption{Time-to-approximation-ratio on Zephyr-native Ising instances. Each panel shows the approximation ratio $E_{\mathrm{approx}}=(E-E_{\mathrm{best}})/(2|E_{\mathrm{best}}|)$ versus runtime for two instance classes, RAU (top row), and RCO (bottom row), and system sizes $N=332, 563$ spins. We compare simulated bifurcation (SBM), the PEPS-based tensor-network solver (TN), and D-Wave quantum annealing with anneal times $\tau=200\,\mu\mathrm{s}$ and $2000\,\mu\mathrm{s}$. “Reduced TN” denotes TN runs preceded by local dimensional reduction of each 16-spin Zephyr cluster from $2^{16}$ to $2^{14}$ or $2^{12}$ states. Lower values indicate better solutions, with $E_{\mathrm{approx}}=0$ corresponding to the best energy observed across all solvers.}

    \label{fig:zephyr-bench-E}
\end{figure}

\paragraph{Zephyr}
Figure~\ref{fig:zephyr-bench-E} shows analogous results on native Zephyr instances. Across both sizes and both classes, SBM exhibits the fastest convergence to $E_{\mathrm{approx}}=0$. QA shows a strong dependence on the anneal time: the longer schedule ($\tau=2000\,\mu\mathrm{s}$) approaches SBM performance and reaches $E_{\mathrm{approx}}=0$ for the median instance, while the shorter schedule ($\tau=200\,\mu\mathrm{s}$) improves only modestly with runtime and typically plateaus at a nonzero error (around $10^{-4}$--$10^{-3}$). TN and reduced-TN curves appear predominantly in the long-time regime and remain above the best-reference energy throughout the shown budgets. Both truncation levels improve over full TN, but neither closes the gap to SBM nor to long-anneal QA on these Zephyr sizes.

\paragraph{Lattices}

\begin{figure}
    \centering
    \includegraphics[width=\textwidth]{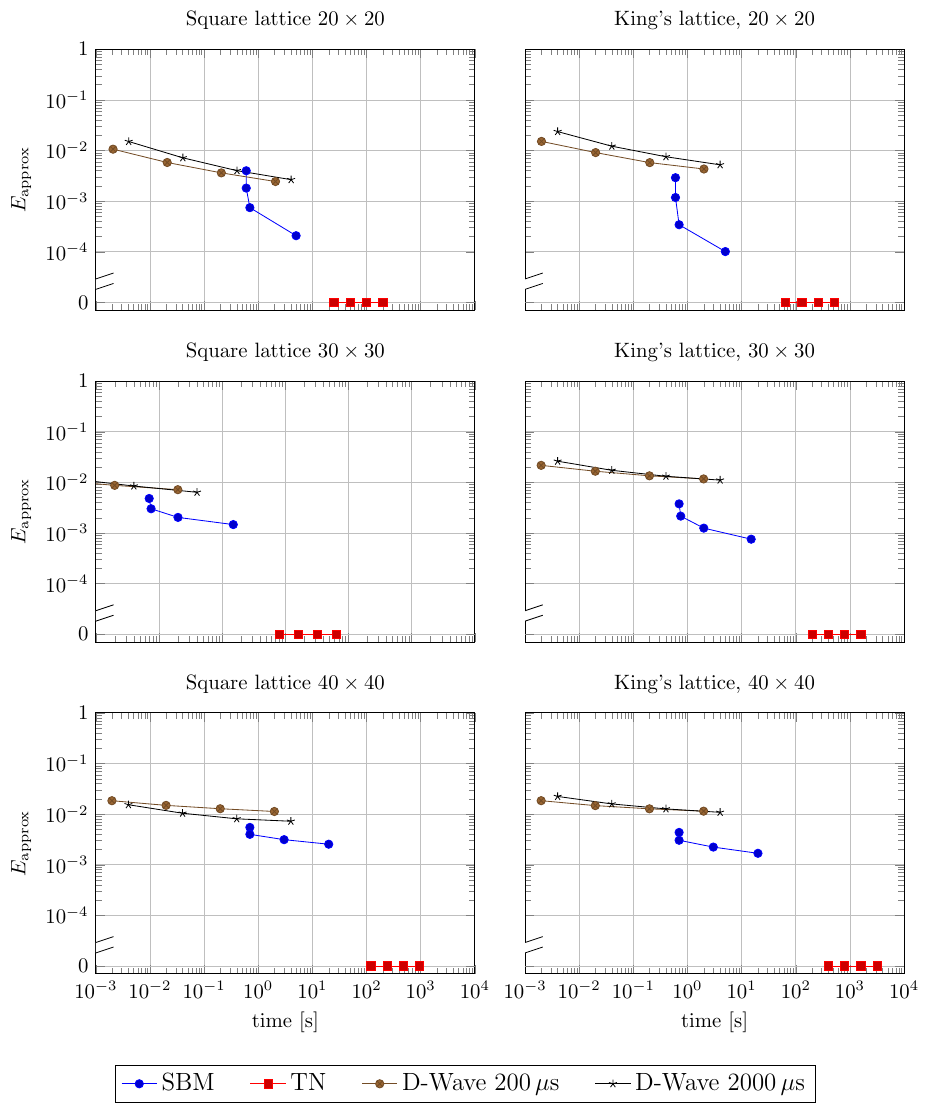}
    \caption{Time-to-approximation-ratio on lattice instances. Each panel shows the approximation ratio $E_{\mathrm{approx}}=(E-E_{\mathrm{best}})/(2|E_{\mathrm{best}}|)$ versus runtime for two instance classes, without diagonal connections (left column) and with them (right column), and system sizes $N=400, 900, 1600$ spins ($20 \times 20$, $30 \times 30$ and $40 \times 40$). We compare simulated bifurcation (SBM), the PEPS-based tensor-network solver (TN), and D-Wave quantum annealing with anneal times $\tau=200\,\mu\mathrm{s}$ and $2000\,\mu\mathrm{s}$. The D-Wave quantum annealers required embedding, as tested instances are not-native. Lower values indicate better solutions, with $E_{\mathrm{approx}}=0$ corresponding to the best energy observed across all solvers.}
    \label{fig:lattice-bench-E}
\end{figure}

Figures~\ref{fig:lattice-bench-E} summarize the time-to-approximation performance on structured 2D lattices (square and king’s graphs) for system sizes $20\times 20$, $30\times 30$, and $40\times 40$. In all six panels, the PEPS-based tensor-network (TN) solver attains $E_{\mathrm{approx}}=0$ for the median instance, i.e., it identifies the best-reference energy among the compared methods, but only in the long-time regime (typically $\sim 10^{1}$--$10^{3}\,$s). By contrast, the simulated bifurcation machine (SBM) reaches near-optimal energies much faster (order-unity to $\sim 10\,\mathrm{s}$), yet it systematically plateaus at a small residual error that grows mildly with size (roughly $E_{\mathrm{approx}}\sim 10^{-4}$--$10^{-3}$). The quantum annealer operates at the shortest times but stabilizes at noticeably higher approximation errors (typically $E_{\mathrm{approx}}\sim 10^{-2}$), with the longer anneal schedule $\tau=2000\,\mu\mathrm{s}$ providing only a modest improvement over $\tau=200\,\mu\mathrm{s}$.


\subsection{Diversity of solutions}
\label{sec:results-diversity}

\begin{figure}
    \centering
    \includegraphics[width=\textwidth]{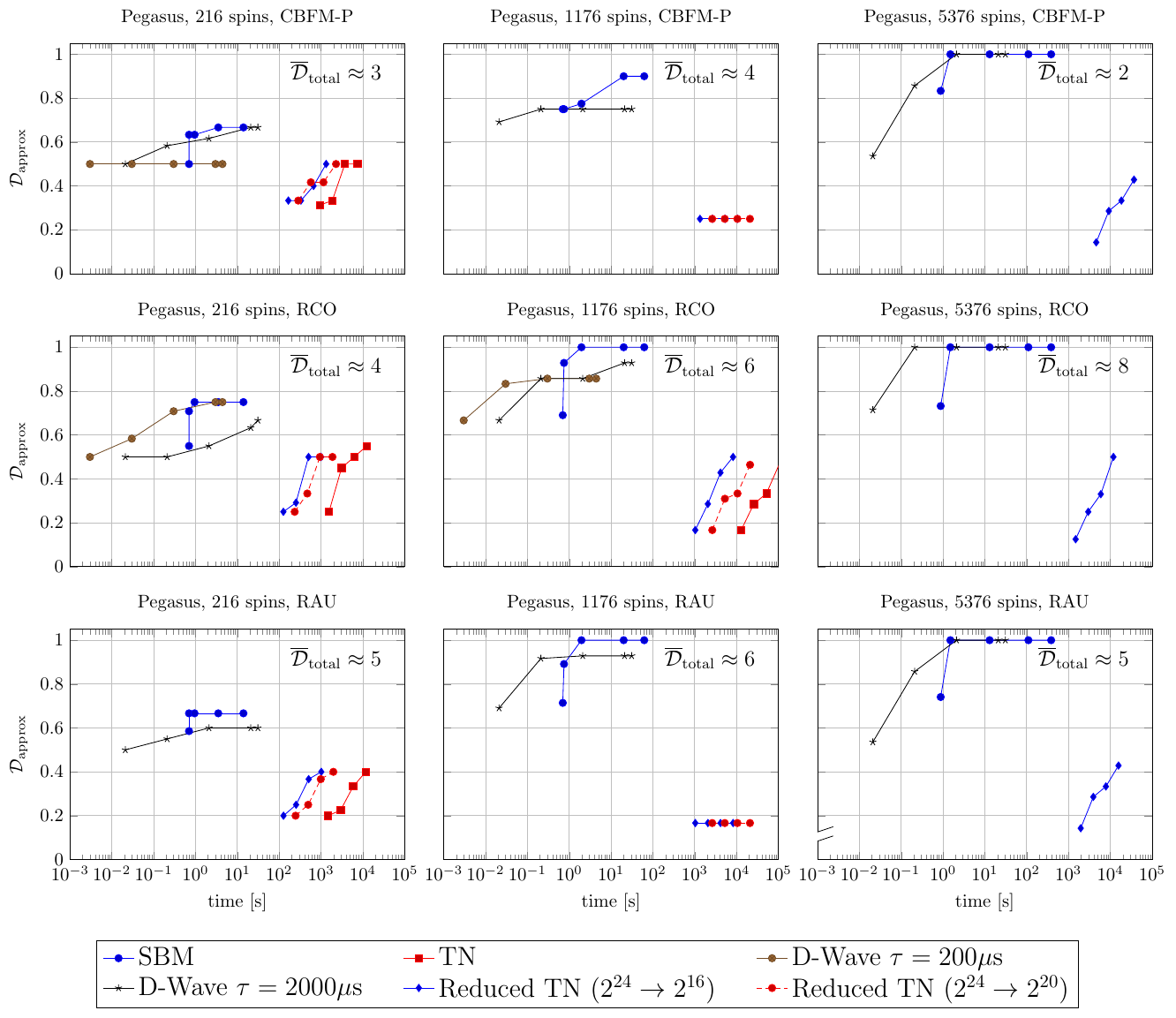}
    \caption{Time-to-diversity-ratio for native Pegasus instances. We set $R=1/2$ and $a_r=0.01$. Each panel shows the time needed to reach the given median fraction of diverse solutions $\mathcal{D}_{\mathrm{approx}}$. The total diversity of each instance is obtained by combining the outputs of all considered solvers. The median among 20 instances is denoted as $\overline{\mathcal{D}}_{\mathrm{total}}$ and annotated in each panel. We considered instance classes: CBFM-P (top row), RCO (middle row), and RAU (bottom row), and system sizes $N=216,1176,5376$ spins. We compare simulated bifurcation (SBM), the PEPS-based tensor-network solver (TN), and D-Wave quantum annealing with anneal times $\tau=200\,\mu\mathrm{s}$ and $2000\,\mu\mathrm{s}$. “Reduced TN” denotes TN runs preceded by local dimensional reduction of each 24-spin Pegasus cluster from $2^{24}$ to $2^{16}$ or $2^{20}$ states.}
    \label{fig:pegasus-diversity}
\end{figure}

\begin{figure}
    \centering
    \includegraphics[width=\textwidth]{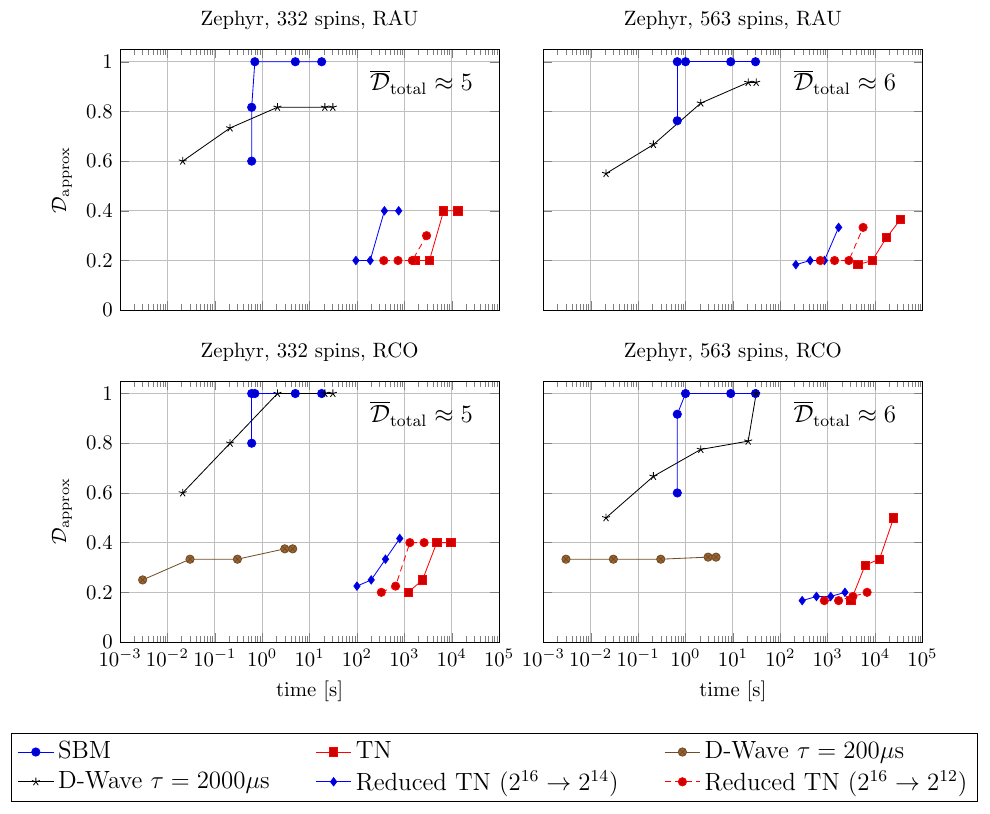}
    \caption{Time-to-diversity-ratio for native Zephyr instances. We set $R=1/2$ and $a_r=0.01$. Each panel shows the time needed to reach the given median fraction of diverse solutions $\mathcal{D}_{\mathrm{approx}}$. The total diversity of each instance is obtained by combining the outputs of all considered solvers. The median among 20 instances is denoted as $\overline{\mathcal{D}}_{\mathrm{total}}$ and annotated in each panel. We considered instance classes: RAU (top row), and RCO (bottom row), and system sizes $N=332,563$ spins. We compare simulated bifurcation (SBM), the PEPS-based tensor-network solver (TN), and D-Wave quantum annealing with anneal times $\tau=200\,\mu\mathrm{s}$ and $2000\,\mu\mathrm{s}$. “Reduced TN” denotes TN runs preceded by local dimensional reduction of each 16-spin Zephyr cluster from $2^{16}$ to $2^{14}$ or $2^{12}$ states.}
    \label{fig:zephyr-diversity}
\end{figure}

Figs.~\ref{fig:pegasus-diversity} and~\ref{fig:zephyr-diversity} report the normalized diversity fraction $D_{\mathrm{approx}}\in[0,1]$ for the same instance sets as above. We use an independence threshold $R=1/2$, \textit{i.e.}, two solutions are considered distinct only if they differ on at least half of the spins, and we restrict attention to near-optimal states within approximation ratio $a_r=0.01$. The total reference diversity is obtained by pooling solutions from all solvers, as described in section~\ref{sec:metrics}. The median pooled value $\overline{D}_{\mathrm{total}}$ is annotated in each panel.

\paragraph{Pegasus.}
Across essentially all Pegasus panels, SBM and QA reach large diversity fractions at substantially shorter times than TN-based approaches. For the larger instances ($N=1176$ and $N=5376$), SBM typically reaches $D_{\mathrm{approx}}\simeq 1$ rapidly, indicating near-complete coverage of the pooled independent low-energy set under the chosen $(R,a_r)$ criteria. QA with the longer anneal ($\tau=2000\,\mu\mathrm{s}$) often follows a similar trend and, on several panels, also approaches $D_{\mathrm{approx}}\approx 1$, whereas the shorter anneal tends to plateau at noticeably smaller diversity fractions. In contrast, TN and reduced-TN variants appear predominantly in the long-time regime (hundreds to tens of thousands of seconds) and achieve only intermediate diversity ($D_{\mathrm{approx}}\sim 0.2$--$0.55$), consistent with a bias toward producing a small subset of near-optimal states rather than broadly exploring the low-energy manifold.

\paragraph{Zephyr.}
On Zephyr (Fig.~\ref{fig:zephyr-diversity}), the pooled reference diversity is modest ($\overline{D}_{\mathrm{total}}\approx 5$--$6$), yet clear method-dependent differences persist. SBM reaches $D_{\mathrm{approx}}\approx 1$ in the short-time regime for both sizes and both classes. QA again depends strongly on anneal time: $\tau=2000\,\mu\mathrm{s}$ approaches full diversity, while $\tau=200\,\mu\mathrm{s}$ plateaus at substantially reduced diversity ($\sim 0.3$--$0.4$). TN-based methods remain far less competitive in terms of time-to-diversity and saturate well below unity even at much longer runtimes.

\section{Conclusion}

In this chapter, we benchmarked D-Wave quantum annealing against competitive classical baselines on instance families that are directly relevant to near-term analog quantum optimization hardware. We focused on (i) \emph{native} Ising problems on Pegasus and Zephyr graphs, thereby avoiding the confounding effects of minor embedding, and (ii) structured quasi-two-dimensional lattice instances that interpolate between regular grids and denser ``king's-graph'' geometries. To make comparisons meaningful across fundamentally different computational paradigms, we evaluated complementary aspects of performance: energy quality via the time-to-approximation-ratio, as well as the ability to generate \emph{diverse} near-optimal configurations.

On native Pegasus/Zephyr instances, the clearest outcome is that simulated bifurcation (SBM) provides the strongest time-to-best-energy performance for the median instance across the tested sizes. Quantum annealing produces high-quality solutions very rapidly, but its performance depends strongly on schedule choice: longer anneals improve both energy and coverage of the low-energy manifold, whereas short anneals tend to plateau at small but nonzero approximation error and reduced diversity. Interpreted operationally, QA is highly competitive as a fast heuristic and sampler under suitable settings, but it does not consistently reach the best-reference energies within the explored budgets at the largest scales.

\texttt{SpinGlassPEPS.jl} serves as a transparent, topology-aware tensor-network baseline, and these experiments quantify its current strengths and limitations. While TN-based methods can achieve good energies and provide a principled classical point of reference aligned with the hardware topology, on large QPU-native random instances, they are dominated in both time-to-approximation and time-to-diversity under feasible contraction and truncation settings. The observed sensitivity to contraction order/boundary choice indicates that the practical bottleneck is not expressivity of the PEPS ansatz per se, but the stability and accuracy of approximate contraction at scale; addressing this is the primary route to improving competitiveness on the largest native instances.

A further lesson is that energy-only benchmarking is incomplete. For several instance families, the ranking of solvers depends on whether the task is \emph{rapidly finding one} near-optimal configuration or \emph{efficiently exploring} a diverse set of low-energy configurations. In this sense, solution diversity is not merely an auxiliary metric but a task-defining criterion that changes how ``performance'' should be interpreted for quantum annealers and other Ising machines.

Overall, the chapter establishes an experimental protocol and a set of metrics that support controlled, apples-to-apples comparisons between quantum annealers and modern classical solvers on hardware-relevant problems. The quantitative gaps and plateaus identified here motivate the subsequent analysis of physical resource costs and post-processing strategies, which aim to connect algorithmic performance to thermodynamic efficiency and to assess how much additional improvement can be extracted beyond raw device sampling.

%% file: chapters/thermo.tex
\chapter{Thermodynamic Properties of Quantum Annealers}
\chaptermark{Thermodynamic Properties of QA}
\label{chap:thermo}

In the previous chapter, we benchmarked quantum annealers against strong classical and quantum-inspired heuristics. While quantum annealers can generate low-energy solutions very rapidly, their performance often saturates at a non-zero approximation error, and they do not reliably reach the global optima achieved by the best classical solvers (such as SBM) on the largest instances. This indicates that benchmarking based solely on algorithmic performance captures only part of the relevant behavior of these devices.

If a quantum annealer does not provide a clear advantage in speed or solution quality, it becomes crucial to ask a more fundamental questions: how does it compute? Is it operating as an efficient thermal machine? Quantum annealers are physical systems operating far from equilibrium, exchanging energy with external controls and their environment. From this perspective, optimization performance must be analyzed together with the thermodynamic cost of computation. In particular, this allows us to analyze whether the observed "saturation" reflects a trade-off between success probability, thermalization, and energy dissipation rather than a purely algorithmic or fundamental limitation.

In this chapter, we analyze quantum annealers as thermodynamic machines and shift the focus of benchmarking from computational metrics to physical ones. Using fluctuation relations and thermodynamic uncertainty relations, we extract hardware-certified bounds on work, heat, entropy production, and power directly from experimentally accessible energy statistics. This framework enables energy-aware benchmarking of quantum annealers and provides a complementary perspective on their performance, independent of time-to-solution and solution quality alone.

\paragraph{Author contributions} 
I designed the experimental protocol (including the selection of schedules and control parameters) and collected data on D-Wave hardware. I then assisted in the analysis and interpretation of results by processing the experimental outputs, extracting the relevant performance figures of merit, and quantifying the associated energetic/thermodynamic implications. 

\subsection*{Relation to published work}
This chapter builds on the author's published work on thermodynamic characterization of quantum annealers~\cite{thermo} and represents the underlying framework (fluctuation relations, pseudo-likelihood thermometry, and TUR-certified bounds on work/heat/entropy production) in a unified notation consistent with the thesis. While parts of the methodology and selected baseline experiments correspond to the published study, the results reported here substantially extend it: all experiments performed on native Pegasus connectivity, as well as the complete set of two-dimensional instance studies (including the scans over turning point $s$ and cycle time $\tau$ and the associated operating-mode/efficiency analysis), are new.

\section{Quantum annealer as a thermal machine}
\label{sec:thermo-theory}

In this chapter, we model quantum annealing as a thermodynamic process. The algorithmic principles of quantum annealing and their implementation in D-Wave devices were introduced in Chapter~\ref{chap:AQC-QA}. Here, we adopt a complementary perspective and focus on the device's physical operation, treating the annealing protocol as a driven, non-equilibrium process in which energy is exchanged with both external control fields and the environment.

\begin{figure}
    \centering
    \includegraphics[width=\textwidth]{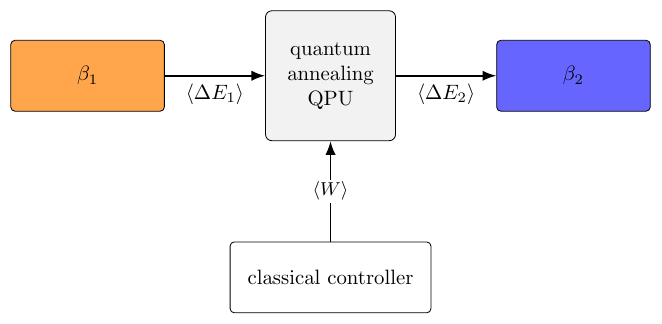}
    \caption{Schematic representation of a D-Wave processor operating under a reverse annealing protocol and viewed as a thermal machine. The system is initialized in a thermal state at inverse temperature $\beta_1$ and exchanges work with external controls and heat with an environment at inverse temperature $\beta_2$.}
    \label{fig:dwave-thermal-machine}
\end{figure}

Experimental and theoretical studies indicate that D-Wave quantum annealers operate as driven open quantum systems~\cite{Gardas2018,Morrell2023}. During the annealing cycle, the processor is subject to time-dependent electromagnetic control fields that perform work on the system while simultaneously interacting with a low-temperature environment that enables heat exchange~\cite{Buffoni2020}. From this perspective, the quantum annealer functions as a thermodynamic machine rather than a closed computational device. A schematic representation of this viewpoint, illustrated for the reverse annealing protocol, is shown in Fig.~\ref{fig:dwave-thermal-machine}.

We assume that, at the beginning of the annealing cycle, the combined state of the processor (system $S$) and its environment ($E$) is given by a product of thermal states,

\begin{equation}
    \rho = \frac{e^{-\beta_1 H_S}}{Z_S} \otimes \frac{e^{-\beta_2 H_E}}{Z_E},
\end{equation}

\noindent where $H_S$ and $H_E$ denote the Hamiltonians of the system and the environment, respectively, and $Z_S$ and $Z_E$ are the corresponding partition functions. The inverse temperatures $\beta_1$ and $\beta_2$ characterize the initial thermal states of the processor and the environment.

To describe energy exchange during the annealing process, we employ the quantum exchange fluctuation theorem~\cite{Jarzynski2004,Sone2023}, which relates the probabilities of forward and backward energy exchange events,

\begin{equation}
    \frac{P(\Delta E_1, \Delta E_2)}{P(-\Delta E_1, -\Delta E_2)} 
    = e^{\beta_1 \Delta E_1 + \beta_2 \Delta E_2}.
\end{equation}

\noindent Here, $\Delta E_1$ and $\Delta E_2$ denote the energy changes of the processor and its environment during a single annealing run, and $P(\Delta E_1, \Delta E_2)$ is the joint probability distribution of observing these energy changes. This relation provides the theoretical foundation for extracting thermodynamic information—such as entropy production, heat, and work—from experimentally accessible energy statistics and serves as the basis for the analysis presented in the remainder of this chapter.

We focus on three thermodynamic quantities that characterize the operation of a quantum annealer viewed as a thermal machine: the average entropy production $\langle \Sigma \rangle$, the average work performed by the external driving $\langle W \rangle$, and the average heat exchanged with the environment $\langle Q \rangle$. In the context of quantum annealing, these quantities can be defined as~\cite{Buffoni2020}:

\begin{equation}
\langle \Sigma \rangle \equiv \beta_1 \langle \Delta E_1 \rangle + \beta_2 \langle \Delta E_2 \rangle,
\end{equation}

\begin{equation}
\langle W \rangle \equiv \langle \Delta E_1 \rangle + \langle \Delta E_2 \rangle,
\end{equation}

\begin{equation}
\langle Q \rangle \equiv -\langle \Delta E_2 \rangle,
\end{equation}

\noindent where $\Delta E_1$ and $\Delta E_2$ denote, respectively, the energy changes of the processor and its environment during an annealing cycle. Entropy production quantifies irreversibility, work captures the energetic cost of external control, and heat measures energy dissipated into the environment.

A key practical limitation is that, in experiments on D-Wave quantum annealers, only the processor's energy change, $\Delta E_1$, is directly accessible. The environmental energy change $\Delta E_2$ is not measurable, and consequently, none of the above thermodynamic quantities can be obtained directly. Instead, they can be \emph{rigorously lower bounded} using thermodynamic uncertainty relations (TURs)~\cite{Uffink1999}, which relate dissipation to fluctuations of experimentally observable quantities.

In particular, the entropy production satisfies:

\begin{equation}
\label{eq:entropy-tur}
\langle \Sigma \rangle \geq 2 g\left(\frac{\langle \Delta E_1 \rangle}{\sqrt{\langle \Delta E_1^2 \rangle}}\right),
\end{equation}

\noindent while the average heat and work are bounded as

\begin{equation}
\label{eq:heat-tur}
-\langle Q \rangle \geq \frac{2}{\beta_2} g\left(\frac{\langle \Delta E_1 \rangle}{\sqrt{\langle \Delta E_1^2 \rangle}}\right) -
 \frac{\beta_1}{\beta_2} \langle \Delta E_1 \rangle,
  \end{equation}

  \begin{equation}
  \label{eq:work-tur}
  \langle W \rangle \geq \frac{2}{\beta_2} g\left(\frac{\langle \Delta E_1 \rangle}{\sqrt{\langle \Delta E_1^2 \rangle}}\right)
+ \left(1 - \frac{\beta_1}{\beta_2}\right) \langle \Delta E_1 \rangle,
  \end{equation}

\noindent where $g(x) = x \tanh^{-1}(x)$. These inequalities depend only on the first two moments of the processor energy change and therefore provide *hardware-certified bounds* on dissipation, work, and heat using experimentally accessible data alone.

To evaluate these bounds, the inverse temperature of the environment $\beta_2$ must be estimated independently. To achieve this, we employ the pseudo-likelihood method~\cite{Benedetti2016}, which infers an effective bath temperature from the statistics of the measured spin configurations. Given a dataset $\mathcal{D} = {\bm{s}^1, \ldots, \bm{s}^D}$ of $D$ samples produced by the quantum annealer, where $\bm{s}^d = (s_1^d, \ldots, s_N^d)$, $d=1,\ldots,D$, the environmental inverse temperature is defined as

\begin{equation}
\label{eq:beta-2}
\beta_2 = \arg\max_{\beta} \Lambda(\beta),
\end{equation}

\noindent with the pseudo-likelihood function

\begin{equation}
\label{eq:lambda-beta}
\Lambda(\beta) = -\frac{1}{ND} \sum_{i=1}^{N} \sum_{d=1}^{D}
\ln\left[1 + \exp\left(-2\beta s_i^d \Bigl(h_i + \sum_{j \in \mathcal{N}(i)} J_{ij} s_j^d\Bigr)\right)\right].
\end{equation}

\noindent Here, $\mathcal{N}(i)$ denotes the set of spins coupled to $s_i$. This procedure yields an effective temperature $T_2 = \beta_2^{-1}$ of the environment and completes the set of inputs required to evaluate the TUR bounds used throughout the remainder of this chapter.

Combining $\langle \Sigma \rangle \ge 0$ with $\langle W \rangle = \langle \Delta E_1 \rangle + \langle \Delta E_2 \rangle$ and $0<\beta_1<\beta_2$ admits only four consistent sign patterns for $\bigl(\langle \Delta E_1\rangle,\langle \Delta E_2\rangle,\langle W\rangle\bigr)$ and consequently modes of operation~\cite{Buffoni2020}:

\begin{center}
\begin{tabular}{lccc}
\toprule
Operation & $\langle \Delta E_1\rangle$ & $\langle \Delta E_2\rangle$ & $\langle W\rangle$\\
\midrule
Refrigerator [R] & $\ge 0$ & $\le 0$ & $\ge 0$\\
Engine [E] & $\le 0$ & $\ge 0$ & $\le 0$\\
Accelerator [A] & $\le 0$ & $\ge 0$ & $\ge 0$\\
Heater [H] & $\ge 0$ & $\ge 0$ & $\ge 0$\\
\bottomrule
\label{tab:thermal-modes}
\end{tabular}
\end{center}

\noindent In practice, we determine the regime by evaluating $\langle \Delta E_1\rangle$ from data, using \eqref{eq:heat-tur}--\eqref{eq:work-tur} to infer the compatible signs of $\langle Q\rangle$ and $\langle W\rangle$. The power bound is conservatively estimated by:

\begin{equation}
    P_{\mathrm{bound}} \ge \frac{W}{\tau},
\end{equation}

\noindent with $W$ defined as lower bound on $\langle W\rangle$.

\section{Methodology}
\label{sec:thermo-exp}

All experiments were performed across multiple generations of D-Wave quantum annealers to probe the robustness of the thermodynamic analysis across hardware platforms. Specifically, we used a D-Wave 2000Q processor based on the Chimera topology, an Advantage\_system6.4 processor implementing the Pegasus topology, and an Advantage2\_system4.3 processor implementing the Zephyr topology. In the remainder of this chapter, these devices are referred to as \emph{Chimera}, \emph{Pegasus}, and \emph{Zephyr}, respectively. A summary of their physical characteristics is provided in Appendix~\ref{apx:dwave}.

All measurements were performed using reverse annealing protocols, with and without a mid-anneal pause. In both cases, the annealing parameter $s(t)$ starts at $s=1$, corresponding to a classical Ising Hamiltonian, decreases to a programmable minimum value $s_a$, and subsequently returns to $s=1$ over a total annealing time $\tau$. For the reverse annealing schedule without a pause, we use

\begin{equation}
    s(t) =
    \begin{cases}
        1 - 2(1 - s_a)\dfrac{t}{\tau}, & t \in \left[0, \dfrac{\tau}{2}\right], \\[6pt]
        -1 + 2s_a + 2(1 - s_a)\dfrac{t}{\tau}, & t \in \left[\dfrac{\tau}{2}, \tau\right].
    \end{cases}
\end{equation}

\noindent When a mid-anneal pause is employed, the schedule takes the form

\begin{equation}
    s(t) =
    \begin{cases}
        1 - 3(1 - s_a)\dfrac{t}{\tau}, & t \in \left[0, \dfrac{\tau}{3}\right], \\[6pt]
        s_a, & t \in \left[\dfrac{\tau}{3}, \dfrac{2\tau}{3}\right], \\[6pt]
        -1 + 3s_a + 3(1 - s_a)\dfrac{t}{\tau}, & t \in \left[\dfrac{2\tau}{3}, \tau\right].
    \end{cases}
\end{equation}

\noindent In both cases, each annealing run is initialized in a classical spin configuration $\bm{s}^{(0)}$ sampled from the Boltzmann distribution of the problem Hamiltonian $H_{\mathrm{problem}}$ at an effective inverse temperature $\beta_1$. The sampling is done approximately by Markov Chain Monte Carlo (MCMC)~\cite{Mossel2013}.

Our analysis focuses on both computational and thermodynamic performance metrics. As a computational figure of merit, we introduce a notion of \emph{computational efficiency} $\eta_{\mathrm{comp}}$, defined as the ratio between the probability of successfully sampling the ground state and the energetic cost of the annealing process,

\begin{equation}
    \label{eq:comp-eff}
    \eta_{\mathrm{comp}} \leq \frac{\mathcal{P}_{\mathrm{GS}}}{\langle W \rangle},
\end{equation}

\noindent where $\langle W \rangle$ denotes the average work performed during an annealing cycle. The ground-state success probability is defined as

\begin{equation}
    \label{eq:succ-prob}
    \mathcal{P}_{\mathrm{GS}} = \mathbb{P}\!\left(s^\star \in \bm{s}\right),
\end{equation}

\noindent with $s^\star$ denoting a ground-state configuration and $\bm{s}$ the output spin configuration of a single annealing run.

We emphasize that $\eta_{\mathrm{comp}}$ is not intended as a universal measure of algorithmic performance. Rather, it is chosen to capture a physically motivated trade-off between solution quality and energetic cost, which is particularly relevant when quantum annealers operate in a regime resembling that of a thermal accelerator~\cite{Buffoni2020}.

To complement this computational perspective, we consider the \emph{thermodynamic efficiency}

\begin{equation}
    \label{eq:th-eff}
    \eta_{\mathrm{th}} \leq -\frac{\langle W \rangle}{\langle Q \rangle},
\end{equation}

\noindent which quantifies how efficiently work input is converted into heat exchanged with the environment. Unlike the efficiency of a heat engine, $\eta_{\mathrm{th}}$ is not constrained by the Carnot bound~\cite{Callen2006}, as the annealer operates as a driven non-equilibrium system rather than as a cyclic engine between two equilibrium reservoirs.

For problem instances where the ground-state energy $E^\star$ can be reliably identified, we additionally evaluate the quality of the obtained solutions via

\begin{equation}
    \label{eq:sol-quality}
    Q_{\mathrm{GS}} = \left\langle \frac{E_{\mathrm{exp}}}{E^\star} \right\rangle,
\end{equation}

\noindent where $E_{\mathrm{exp}}$ is the energy of the sampled configuration and the average is taken over all collected samples corresponding to a given data point.

Finally, we stress that the quantities $\eta_{\mathrm{comp}}$ and $\eta_{\mathrm{th}}$ obtained in this work are not direct measurements, but thermodynamic lower bounds derived from thermodynamic uncertainty relations. Their evaluation relies on accurate statistics of the processor energy change, $\Delta E_1$, and a reliable estimate of the effective environmental inverse temperature, $\beta_2$. While this introduces systematic uncertainty in absolute values, the comparative conclusions drawn throughout this chapter remain robust, as all results are obtained under closely matched experimental conditions and analyzed consistently across devices and schedules.

\subsection{Experimental procedure}

All experiments concerning thermodynamic and computational efficiency were performed on the D-Wave 2000Q quantum annealer.
The quantum processing unit (QPU) implements a Chimera graph and comprises $2041$ physical qubits connected by $5974$ couplers. 

For each parameter pair $(s_a,\tau)$ (and, when applicable, each field strength), we collect $M$ independent readouts. Depending on resource availability, we use $M \in [10^2,10^4]$.

Each run begins from a classical spin configuration $\bm{s}$ sampled from the Boltzmann distribution (see Eq.~\eqref{eq:gibbs}) at an effective inverse temperature $\beta_1=1$ using a Markov chain Monte Carlo (MCMC) procedure~\cite{Mossel2013}.

In these experiments, we consider (among others) length-$N=300$ chains with disordered couplings and fields, and we scan the turning point $\bar{s}$ for multiple cycle times $\tau$.

\section{Results}
\label{sec:thermo-results}

In this section, we present the experimental results of the thermodynamic analysis of quantum annealers. We begin by examining standard computational metrics—ground-state success probability and solution quality—to characterize the device's algorithmic behavior under different annealing protocols. We then interpret these observations through the lens of thermodynamic efficiency, using the bounds introduced in Section~\ref{sec:thermo-theory}.

\begin{figure}[t]
    \centering
    \includegraphics[width=\textwidth]{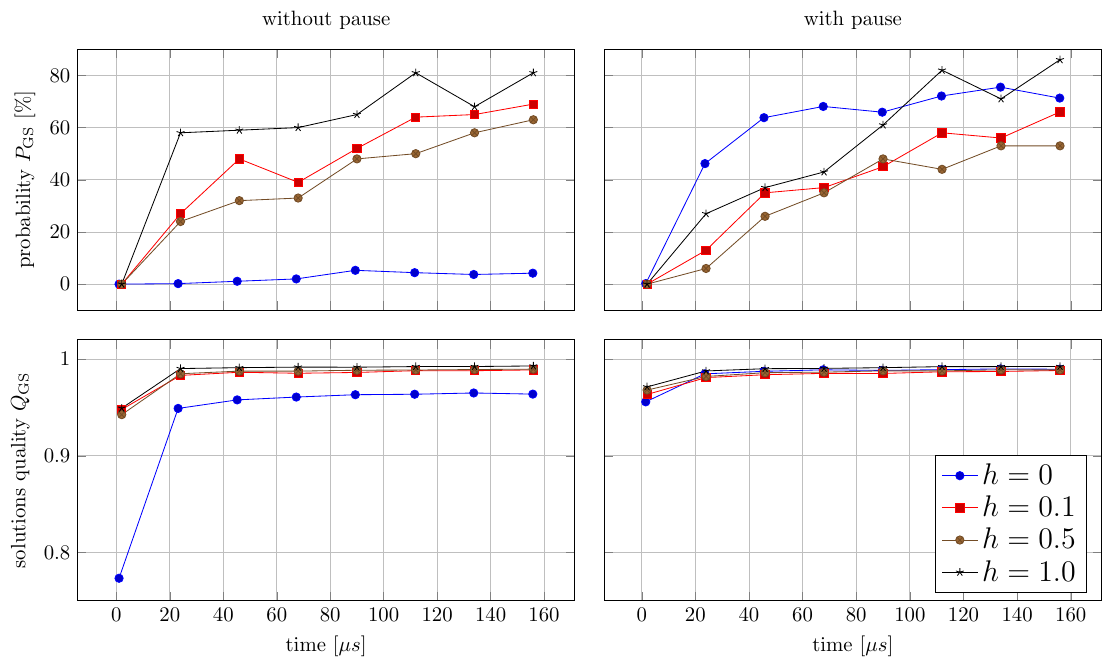}
    \caption{Ground-state success probability $\mathcal{P}_{\mathrm{GS}}$ and average solution quality $Q_{\mathrm{GS}}$ for an antiferromagnetic Ising chain with $N=300$ spins, obtained using reverse annealing schedules with and without a mid-anneal pause and for different values of the external magnetic field $h$. Each data point is averaged over $100$ annealing runs with $10$ samples per run.}
    \label{fig:prob-quality}
\end{figure}

Figure~\ref{fig:prob-quality} summarizes the success probability $\mathcal{P}_{\mathrm{GS}}$ and the average solution quality $Q_{\mathrm{GS}}$ for reverse annealing schedules with and without a mid-anneal pause. For unbiased instances ($h=0$), the minimum annealing parameter was set to $s_a = 0.5$, while for biased instances ($h \neq 0$) we used $s_a = 0.41$. These values were determined empirically as those yielding the highest success probabilities under the respective conditions.

\subsection{Computational performance}

For unbiased instances without a pause, the success probability remains below approximately $10\%$ across the full range of annealing times. In practical terms, this implies that in a batch of $10^3$ samples, fewer than one hundred configurations correspond to the true ground state. At the same time, the solution quality $Q_{\mathrm{GS}}$ is relatively high, exceeding $0.75$ already for short anneals and saturating near $0.95$. This indicates that the device frequently samples low-energy local minima that are close in energy to the ground state but rarely reaches the true optimum.

Introducing a uniform external magnetic field substantially alters this behavior. For $h=1$, the success probability increases dramatically, reaching values of order $80\%$, while $Q_{\mathrm{GS}}$ approaches unity. Since the external field is applied uniformly to all spins, it does not change the identity of the ground state. Instead, it reshapes the energy landscape by lifting degeneracies and suppressing competing low-energy configurations, thereby facilitating relaxation into the true ground state. This observation highlights that the saturation seen in the unbiased case is not purely a dynamical limitation but is strongly influenced by the structure of the energy landscape.

The effect of introducing a mid-anneal pause is particularly pronounced in the unbiased case. Pausing the anneal at the minimum value $s_a$ results in a substantial increase in the ground-state success probability while leaving the average solution quality largely unchanged. This suggests that the pause does not significantly alter the set of low-energy configurations explored by the system, but rather redistributes their populations. A natural interpretation is that the pause allows for passive thermal relaxation, during which diabatic and thermally excited states decay toward the ground state under environmental coupling. In contrast, when an external magnetic field is present and the energy landscape is already strongly biased toward the optimum, the same pause provides little benefit and can even slightly reduce the success probability, indicating that additional thermalization may promote transitions away from the ground state.

\subsection{Thermodynamic efficiency}

These algorithmic observations are reflected in the thermodynamic efficiency bounds derived from thermodynamic uncertainty relations. In the absence of an external magnetic field and without a pause, the bound on computational efficiency $\eta_{\mathrm{comp}}$ increases only slowly with annealing time and never exceeds a few percent. In contrast, the bound on thermodynamic efficiency $\eta_{\mathrm{th}}$ decreases monotonically with annealing time, while remaining close to unity. This behavior is consistent with a regime in which the device expends work primarily to maintain near-equilibrium sampling around low-energy local minima, rather than to reliably drive the system into the global ground state.

The introduction of a mid-anneal pause leads to a qualitatively different regime. For unbiased instances, inserting a pause results in a sharp increase of the ground-state success probability, from single-digit percentages to values approaching $80\%$. This behavior is consistent with the interpretation that the pause allows the system to relax quasi-passively under environmental coupling, enabling diabatic and thermal excitations accumulated during the reverse anneal to decay into the ground state. Notably, in this regime the solution quality remains high throughout, indicating that the pause primarily affects the population of the ground state rather than the overall energy distribution.

Interestingly, when an external magnetic field is present, the insertion of a pause no longer improves—and in some cases slightly degrades—the ground-state success probability, despite maintaining high $Q_{\mathrm{GS}}$. This suggests that when the energy landscape is already strongly biased toward the ground state, additional thermal relaxation during the pause can promote transitions away from the optimal configuration. From a thermodynamic perspective, this highlights that pauses do not universally improve performance but instead shift the balance between work, heat dissipation, and relaxation in a manner that depends sensitively on the underlying energy landscape.

Together, these results illustrate that quantum annealers can operate in distinct thermodynamic regimes depending on schedule design and problem structure. High-quality solutions can be obtained with relatively low dissipation even when ground-state success probabilities remain modest, while improved success probabilities often come at the cost of increased reliance on thermal relaxation. This reinforces the central message of this chapter: algorithmic performance and thermodynamic efficiency provide complementary, and sometimes competing, perspectives on the operation of quantum annealers.

\begin{figure}[t]
        \centering
    \includegraphics[width=\textwidth]{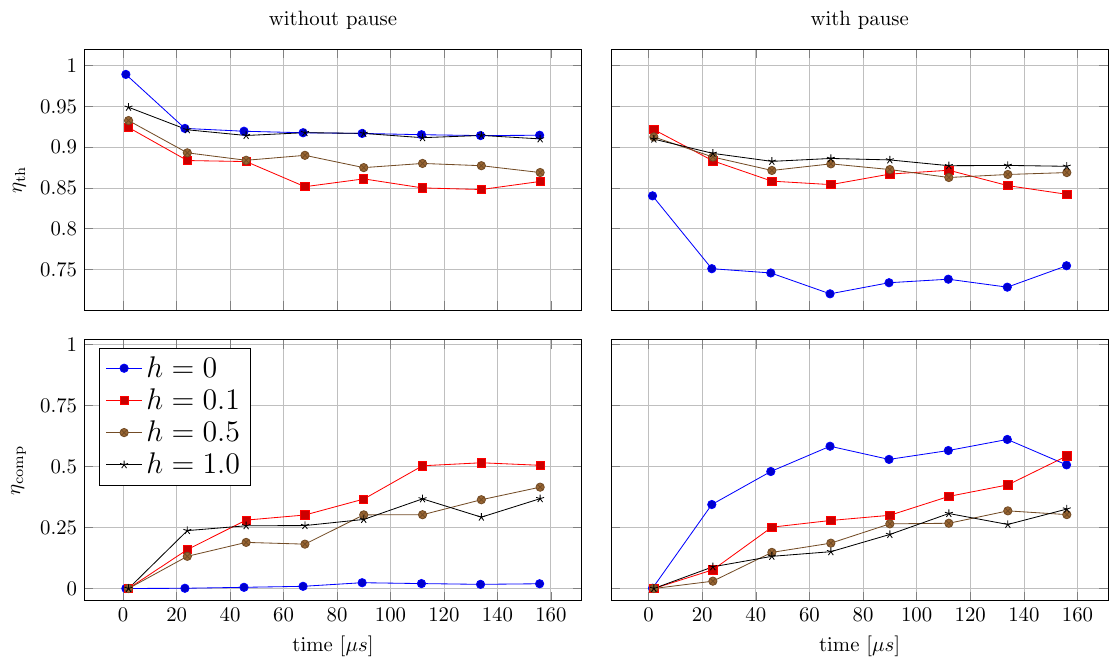}
    \caption{Thermodynamic and computational efficiency for an antiferromagnetic Ising chain with $N=300$ spins, obtained using reverse annealing schedules with and without a mid-anneal pause and for different values of the external magnetic field $h$. Each data point is averaged over $100$ annealing runs with $10$ samples per run.}
    \label{fig:thermo-eff}
\end{figure}

\subsection{Thermodynamic properties}

We begin by characterizing the thermodynamic properties of a simple, well-controlled system: a one-dimensional Ising chain with $N=300$ spins. The couplings are drawn uniformly at random from $[-1,1]$, and longitudinal fields from $[-h,h]$. The instance is embedded on the D-Wave processor and initialized in a Gibbs state of the problem Hamiltonian at inverse temperature $\beta_1=1$. This setting allows us to isolate the intrinsic thermodynamic behavior of the annealer with minimal embedding overhead and a well-understood equilibrium reference.

\begin{figure}[t]
    \centering
    \includegraphics[width=\linewidth]{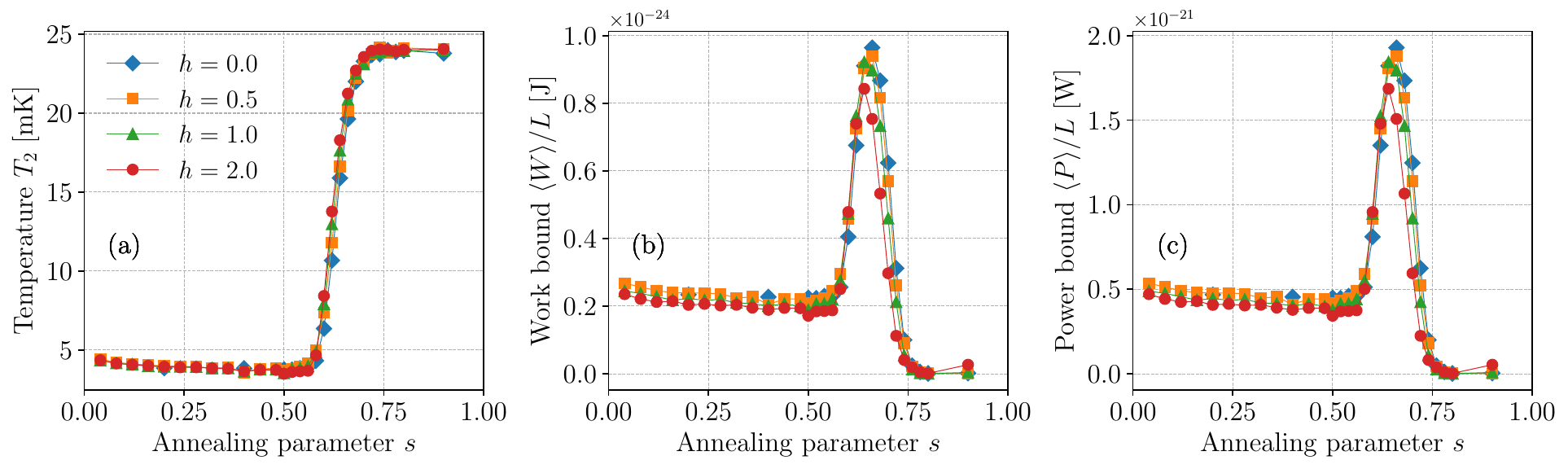}
    \caption[Thermodynamic properties Ising chain]{Thermodynamic response of a one-dimensional Ising chain ($N=300$) programmed on a Pegasus QPU. From the statistics of the processor energy change, we infer (i) an \emph{effective} environment temperature $T_2(s)$ via pseudo-likelihood thermometry and (ii) thermodynamically certified \emph{lower bounds} on the work and power per spin using fluctuation relations and thermodynamic uncertainty relations.
Results are shown as a function of the reverse-annealing turning point $s$ for several values of the longitudinal field $h$.}
    \label{fig:thermo-1D}
\end{figure}

The observed behavior in Fig.~\ref{fig:thermo-1D} is consistent with both device specifications and equilibrium theory. According to manufacturer data, the QPU operates at temperatures below approximately $20\,\mathrm{mK}$~\cite{dwave}, and the inferred effective bath temperatures $T_2(s)$ remain within this envelope, up to calibration uncertainty and the fact that $T_2$ is extracted from qubit statistics rather than measured directly. From a theoretical perspective, the one-dimensional transverse-field Ising model exhibits a quantum critical point at $\Gamma/J=1$~\cite{Pfeuty1970}. Expressing the annealing Hamiltonian as $H(s)=A(s)H_x+B(s)H_p$ with $J$ normalized to unity, this condition corresponds to $A(s)/B(s)\approx 1$, which on D-Wave processors occurs near the mid-anneal crossing of the schedule functions $A(s)$ and $B(s)$.

In this critical region, enhanced susceptibility and energy fluctuations are expected. Correspondingly, we observe a sharp crossover near $s\approx 0.6$, where the inferred $T_2(s)$ rises rapidly and the TUR-certified lower bounds on work and power exhibit pronounced peaks. Small changes in $s$ within this interval produce large changes in relaxation behavior and dissipation, indicating that the system is most thermodynamically active in this narrow window. This phenomenology aligns with previous experimental studies showing that dynamics, thermal repopulation, and the effectiveness of pauses are strongest near and just after the minimum gap.

Taken together, the data support the following picture: (i) across the annealing parameter $s$ the environment behaves as an effectively cold bath in the millikelvin regime consistent with QPU specifications; (ii) near $s \approx 0.6$ the chain traverses its critical region (defined by $A(s)/B(s) \approx 1$), which amplifies energy fluctuations and irreversibility; and (iii) this amplification manifests simultaneously in T2(s) and in the TUR-certified bounds on work and power. The co-occurrence of these signatures at the same s is the expected thermodynamic hallmark of the quantum critical point for this instance.

\paragraph{Two-dimensional instances}

We now extend the analysis to more realistic two-dimensional Ising instances embedded on the Pegasus architecture. We consider three representative problem classes: a random uniform (RAU), only random couplings (RCO), and corrupted biased ferromagnet for Pegasus (CBFM-P). Their detailed description can be found in Chapter~\ref{chap:bench}. For each instance, we scan the reverse-annealing turning point $s$ and repeat the experiment for several cycle times $\tau$.

\begin{figure}[t]
    \centering
    \includegraphics[width=\textwidth]{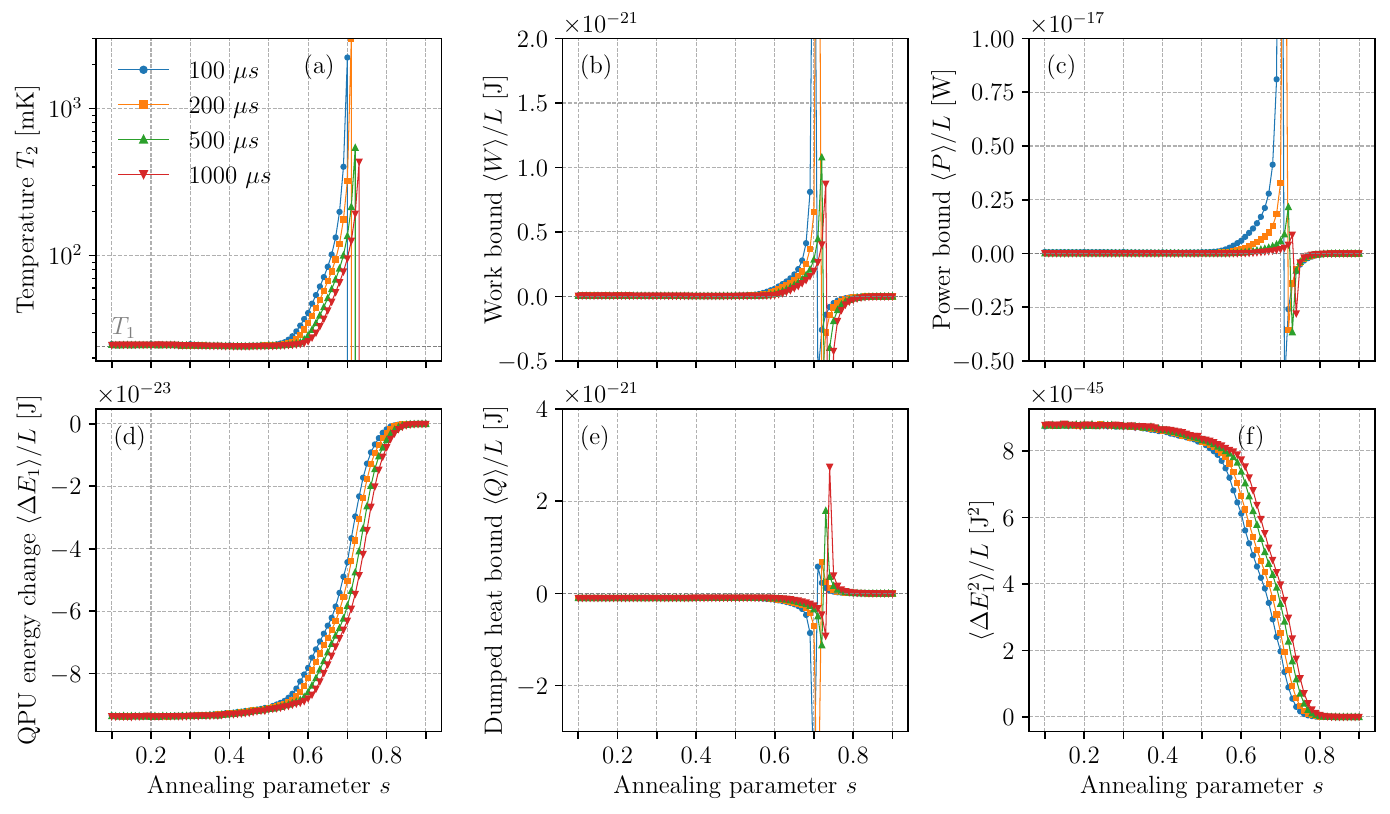}
    \caption{\textbf{Two-dimensional uniform instance on Pegasus.}
Scan of the reverse-annealing turning point $s$ for cycle times $\tau=100,200,500,1000\,\mu$s.
Panels show (a) pseudo-likelihood effective bath temperature $T_2(s)$ (with $T_1$ as reference),
(b) TUR-certified lower bound on the average work per spin $\langle W\rangle/L$,
(c) corresponding power bound $\langle P\rangle/L$,
(d) mean processor energy change $\langle\Delta E_1\rangle/L$,
(e) TUR lower bound on dumped heat $-\langle\Delta E_2\rangle/L$, and
(f) variance $\langle\Delta E_1^2\rangle/L$.
All observables exhibit a pronounced crossover in a narrow ``active window'' of $s$ (vertical guides), where susceptibility and dissipation are maximal.}
    \label{fig:2d-uniform}
\end{figure}

\begin{figure}[t]
    \centering
    \includegraphics[width=\textwidth]{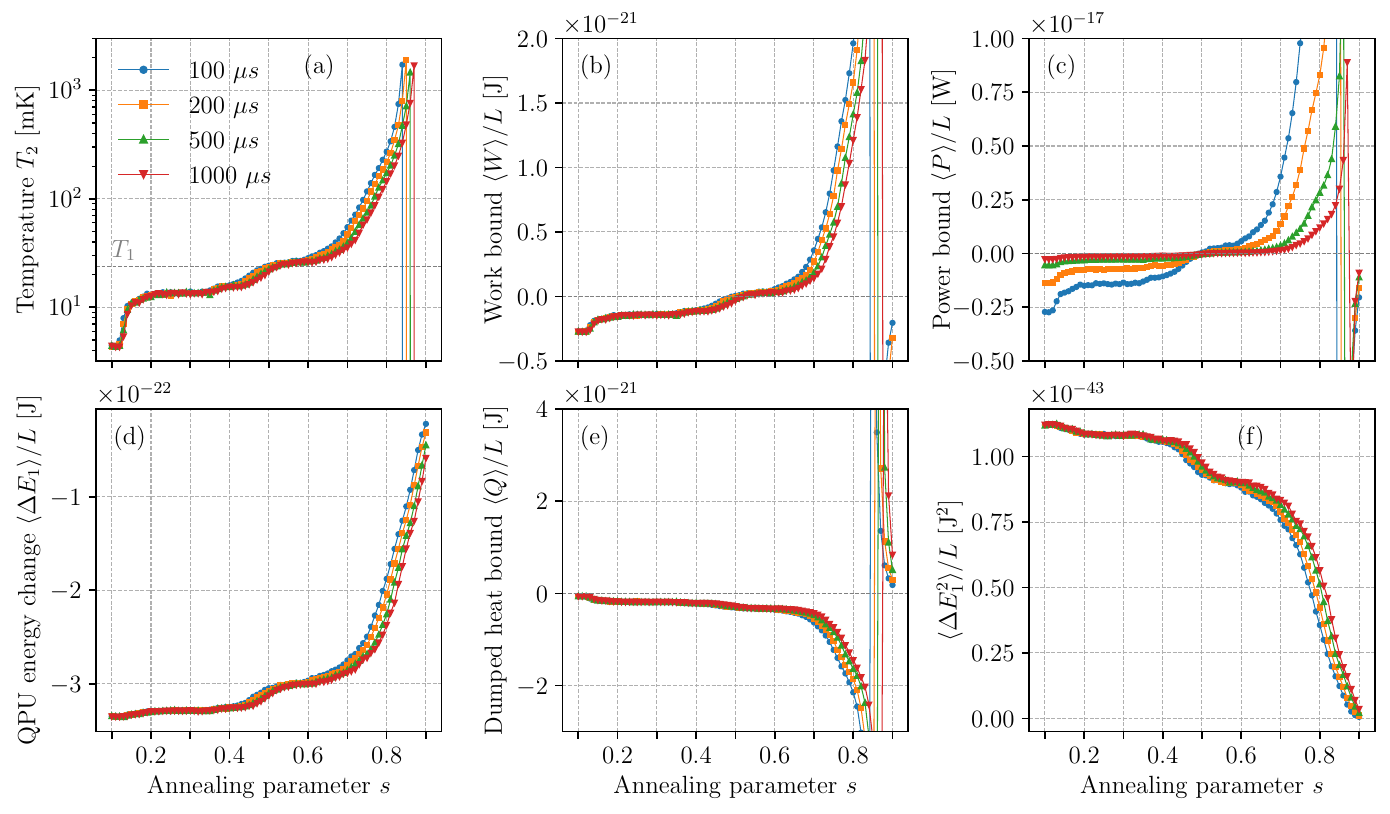}
    \caption{\textbf{Two-dimensional constant instance on Pegasus.}
Same analysis as Fig.~\ref{fig:2d-uniform}.
Instance ruggedness broadens the active window and raises the off-window baseline, while longer cycle times reduce peak magnitudes, consistent with suppressed finite-time irreversibility.}
    \label{fig:2d-constant}
\end{figure}

\begin{figure}[t]
    \centering
    \includegraphics[width=\textwidth]{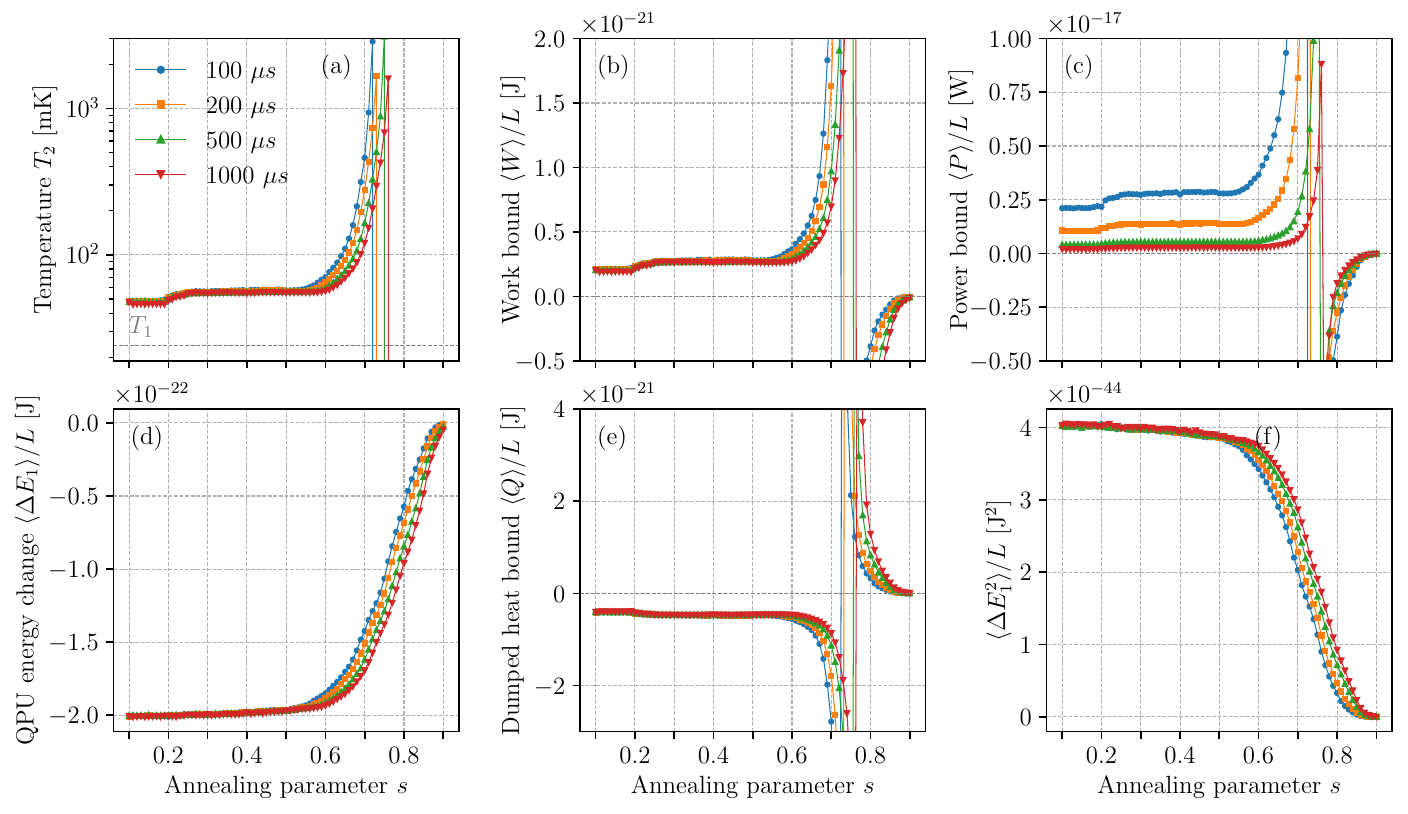}
    \caption{\textbf{Two-dimensional CBFM instance on Pegasus.}
Same analysis as Fig.~\ref{fig:2d-uniform}.
The effects of ruggedness are most pronounced here, with elevated dissipation outside the active window and strong sensitivity to cycle time.}
    \label{fig:2d-cbfm}
\end{figure}

Across all three two-dimensional instances shown in Figs.~\ref{fig:2d-uniform}-\ref{fig:2d-cbfm}, the thermodynamic response is sharply localized in the annealing coordinate. For most values of $s$, the inferred bath temperature remains close to the baseline scale set by $T_1$, and the TUR-certified bounds on work and power are small. As $s$ approaches a narrow ``active window'', all observables change rapidly and in concert: the effective temperature rises, the bounds on work and power peak, and both the mean $\langle\Delta E_1\rangle/L$ and the variance $\langle\Delta E_1^2\rangle/L$ exhibit a pronounced crossover.

Because TUR bounds depend simultaneously on the drift and fluctuations of $\Delta E_1$, the co-occurrence of large mean energy changes and enhanced variance provides direct evidence that this window corresponds to the regime of maximal thermodynamic activity. Importantly, this behavior cannot be attributed to a single statistic alone; it reflects the system's collective response to the competition between driver and problem terms.

\subsection{Thermal operating modes of quantum annealers}

\begin{figure}[t]
    \centering
    \begin{subfigure}[b]{0.48\textwidth}
        \centering
        \includegraphics[width=\textwidth]{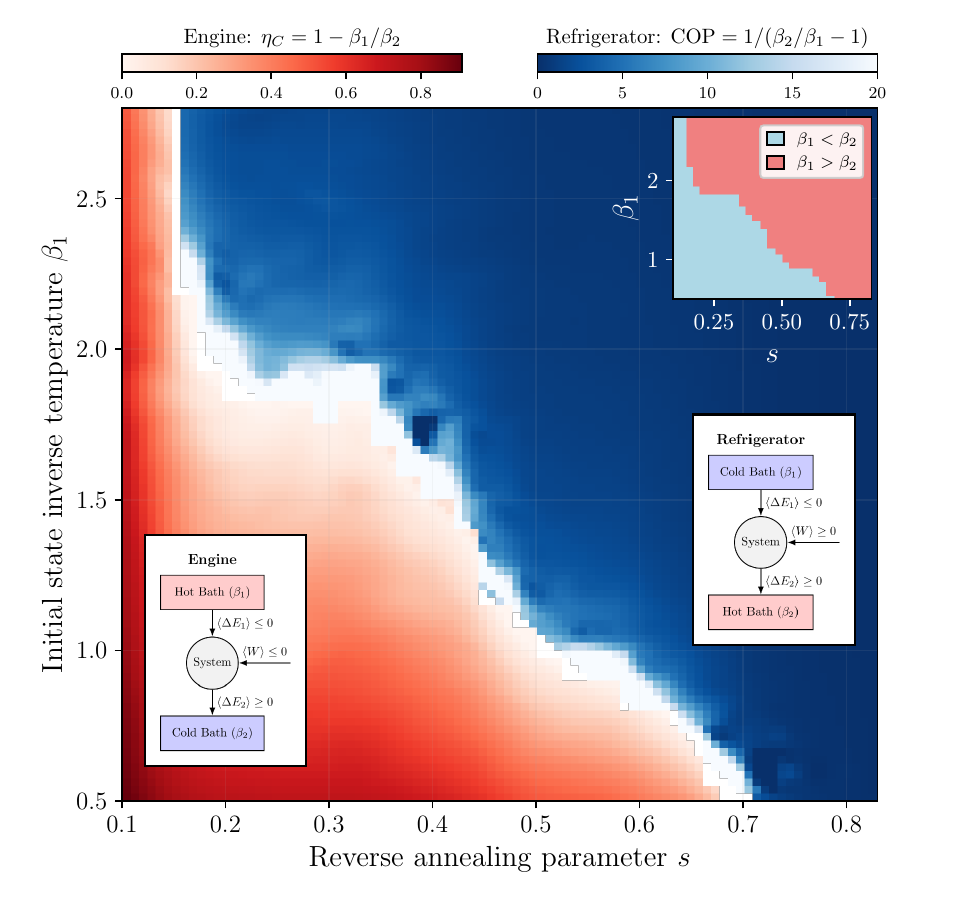}
        \caption{Pegasus}
        \label{fig:phase-diagram-p}
    \end{subfigure}\hfill
    \begin{subfigure}[b]{0.48\textwidth}
        \centering
        \includegraphics[width=\textwidth]{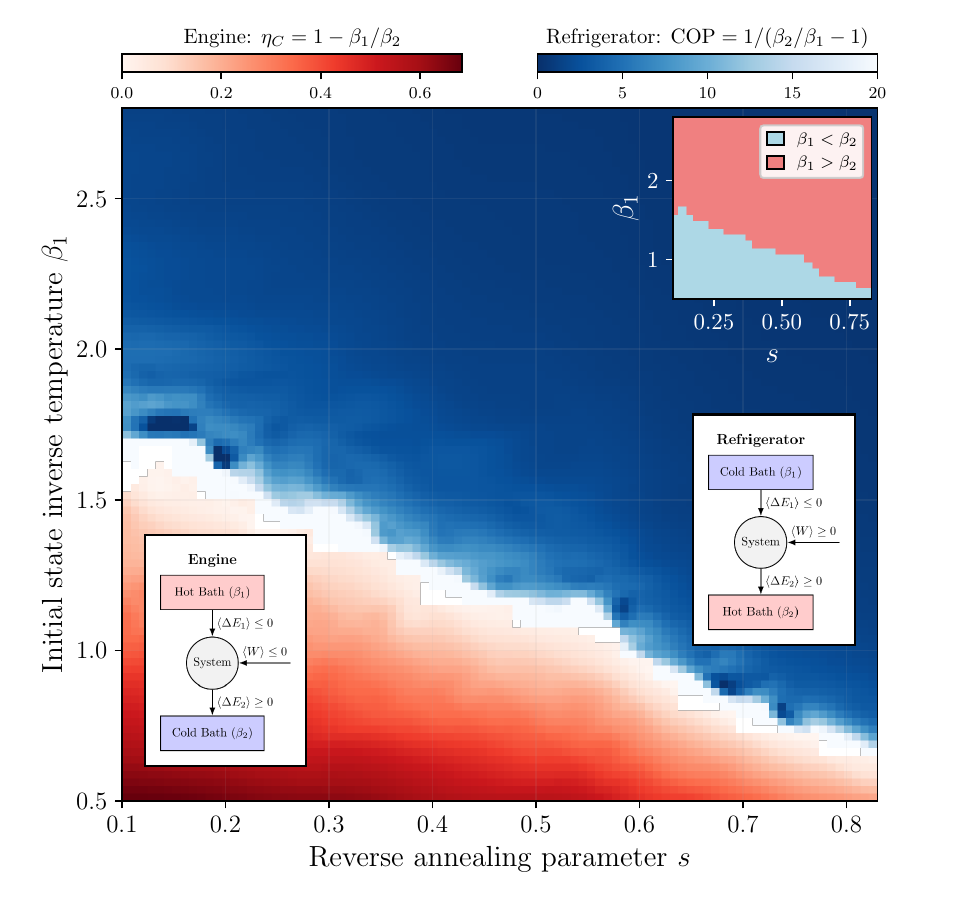}
        \caption{Zephyr}
        \label{fig:phase-diagram-z}
    \end{subfigure}
    \caption{\textbf{Operating-mode phase diagrams of Pegasus and Zephyr.}
Operating regimes in the $(\beta_1,s)$ plane for (a) Pegasus and (b) Zephyr.
Colors denote engine (E), refrigerator (R), heater (H), and accelerator (A) modes.
Color intensity indicates the ideal Carnot efficiency or coefficient of performance for orientation. Finite-time operation remains below these ideal limits.
Engine operation appears only when the prepared state is sufficiently hot relative to the bath and when the schedule traverses the susceptible mid-anneal window.}
    \label{fig:phase-diagram}
\end{figure}

Figure~\ref{fig:phase-diagram} summarizes the operating mode of the quantum annealer as a function of two experimentally accessible control parameters: the preparation inverse temperature $\beta_1$ and the reverse-annealing turning point $s$. Each point in the $(\beta_1,s)$ plane is classified into one of four thermodynamic modes: engine, refrigerator, heater, or accelerator, based solely on the sign structure of the cycle-averaged energy exchanges (see~\ref{tab:thermal-modes}).

A key observation is that the separation between operating modes bends sharply across a relatively narrow interval of $s$. Outside this interval, the annealing cycle acts as a weak perturbation, and the operating mode is governed primarily by the thermodynamic bias set by $\beta_1$ relative to the effective bath temperature $\beta_2$. Inside the window, however, small changes in $s$ substantially modify excitation production and relaxation pathways, allowing the inferred work and heat to change sign.

The geometry of Fig.~\ref{fig:phase-diagram} therefore encodes the interplay of two independent controls: the thermal bias determined by the preparation temperature and the dynamical activation controlled by the annealing schedule. This demonstrates that the operational mode of a quantum annealer is not fixed but can be continuously tuned through scheduling design, reinforcing the view of quantum annealers as programmable non-equilibrium thermal machines rather than static optimization devices.

\section{Conclusion}

In this chapter, we examined quantum annealers from a thermodynamic perspective, linking computational performance to physical cost. By modeling reverse annealing on D-Wave processors as a driven open-system process, we complemented standard benchmarking metrics with thermodynamically certified bounds on work, heat, entropy production, and power derived from experimentally accessible energy statistics.

Using the fluctuation theorem and the thermodynamic uncertainty relations, we showed that dissipation and energetic activity are sharply localized in a narrow mid-anneal window, where the driver and problem Hamiltonians compete most strongly. In both one- and two-dimensional instances, this window is characterized by enhanced energy fluctuations, elevated effective bath temperatures, and peaks in the certified bounds on work and power, while outside it the system operates close to equilibrium at millikelvin temperatures. These observations are consistent with equilibrium theory and prior studies of critical and near-critical annealing dynamics.

From a computational standpoint, the thermodynamic analysis clarifies the role of schedule design. Mid-anneal pauses can significantly increase ground-state success probabilities by enabling thermal relaxation, but this effect depends sensitively on the energy landscape's structure. In particular, when the landscape is already strongly biased by an external field, pausing may become counterproductive. This demonstrates that improvements in success probability often reflect a trade-off between thermalization and dissipation rather than purely coherent dynamics.

Finally, by classifying reverse-annealing cycles into distinct operating modes—accelerator, engine, refrigerator, and heater—we showed that quantum annealers function as programmable non-equilibrium thermal machines. The resulting operating-mode diagrams highlight the joint role of thermodynamic bias and schedule-induced dynamical activation. Overall, this chapter establishes thermodynamic efficiency as an independent and complementary axis for benchmarking quantum annealers, providing insight into how performance gains are achieved and at what physical cost.

%% file: chapters/error-correction.tex
\chapter{Error Mitigation in Quantum Annealers Using Reinforcement Learning}
\chaptermark{RL for QA}
\label{chap:rl}

In the realm of quantum annealers, the pursuit of error-handling strategies is paramount to harnessing the full power of these devices. In this chapter, we present a new error-mitigation algorithm based on deep reinforcement learning. First, we will describe the general framework of the proposed method. Next, we will delve into the details of the employed model. Lastly, we will present the obtained results.

The introduction to reinforcement learning, the language it uses, and its standard mathematical framework are in Appendix~\ref{apx:rl}. 

\paragraph{Author contributions}
I  conceived and designed the reinforcement-learning based error mitigation framework presented in this chapter, implemented the computational model and training procedure, performed all numerical experiments on quantum annealing data, and carried out the analysis and interpretation of the results.

\paragraph{Relation to published work}
The material presented in this chapter is based on previously published work~\cite{rl}. No new conceptual or experimental contributions beyond that publication are introduced in this chapter; rather, the content is integrated here to provide a coherent and self-contained account within the broader benchmarking framework of the thesis.

\section{Reinforcement Learning for Error Mitigation}

\begin{figure}[t]
    \centering
    \includegraphics[width =\textwidth]{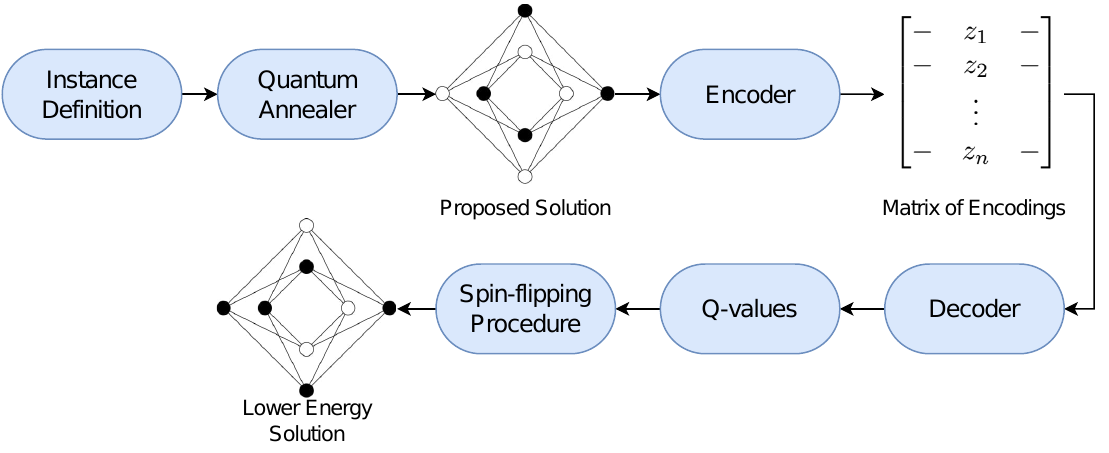}
    \caption{Overview of our method. Arrows represent consecutive steps. First, we define the Ising instance. Next, we obtain a candidate solution from a quantum annealer, where white (black) nodes represent spins with $\sigma_i = 1$ ($-1$), and edges represent couplings. In the following step, we encode this instance using a graph neural network into a matrix of embeddings, where each row $z_i$ corresponds to the embedding of vertex $i$. This matrix is then passed through a decoder to obtain the Q-values of actions associated with each vertex. Finally, the spin-flipping procedure is carried out by flipping spins one by one according to the Q-values and recording the energy after each step. The solution is the lowest energy state found during this procedure.
    }
    \label{fig:rl-schema}
\end{figure}

We employ a standard reinforcement-learning formulation based on a Markov Decision Process (MDP)~\cite{Sutton2018}, in which an agent interacts with an environment over discrete time steps $t = 0, 1, \ldots, T$. At each step, the agent observes a state $s_t$ and selects an action $a_t$ from the action set $\mathcal{A}$ according to its policy $\pi$. The environment then returns a scalar reward, $r_t$, and transitions the agent to the next state, $s_{t+1}$. This interaction proceeds until a terminal state $s_T$ is reached, which concludes the episode. The procedure is then repeated for subsequent episodes.

\subsection{Ising game}

Inspired by ``the spin-flipping procedure'' from \cite{Fan2023}, we propose to treat the error correction process as a one-player game. In the so-called ``Ising Game'', the board is defined as a given Ising state, that is, spin configuration $s= [s_1, \ldots, s_n]$, graph $\mathcal{G}$, and parameters ($(J_{ij})_{(i,j)}$, $(h_i)_i$ ). In each move, the player can choose one vertex $i$ and ``flip'' its spin, switching $\uparrow$ to $\downarrow$ or vice versa. The goal of the game is to obtain the lowest possible value of $H_{\text{Ising}}$  within a given number of moves (denoted $k$). The limitation is that each spin can be ``flipped'' only once. This reduces the action space size from $n^k$ to $n(n-1)\ldots (n-k+1)$. 

The score, obtained after the game ends, is defined as the difference between the lowest energy found during play and the starting energy.

\begin{equation*}
	v = E_{start} - E_{best}
\end{equation*}

This game can be presented formally in the language of the Markov Decision Process~\cite{Puterman1990}.

Starting at $t=0$, the agent flips a single spin at each time step, thereby transitioning to a new state corresponding to a different spin configuration. The terminal state $s_{T}$ is reached once every spin has been flipped exactly once. The solution returned by the procedure is the spin configuration $\sigma$ with the lowest energy encountered during the episode.

\section{Model}

Our model architecture is inspired by DIRAC (\textbf{D}eep reinforcement learning for sp\textbf{I}n-glass g\textbf{R}ound-st\textbf{A}te \textbf{C}alculation), an encoder-decoder design introduced in~\cite{Fan2023}. It exploits the fact that an Ising model instance is fully specified by its underlying graph: the couplings $J_{ij}$ are treated as edge weights, while the external magnetic field $h_i$ and spin variables $\sigma_i$ are encoded as node features. The model uses a two-stage process. First, the encoder maps the entire spin-glass instance to a low-dimensional embedding, assigning each node a vector representation. Next, the decoder uses these node embeddings to compute the $Q$-value for each admissible action, and the agent selects the action with the highest $Q$-value. In the following, we will describe these components in detail.

\subsection{Using graph neural networks to exploit QPU structure}

As described above, the Ising spin-glass instance can be formulated in terms of graph theory. It allows us to employ graph neural networks~\cite{Scarselli2009,Zhou2020}, neural networks designed to take graphs as input. We use a modified SGNN (Spin Glass Neural Network)~\cite{Fan2023} to obtain node embedding. To capture the coupling strengths and external field strengths ($J_{ij}$ and $h_i$), which are crucial for determining spin glass ground states, SGNN performs two updates at each layer: the edge-centric update and the node-centric update.

Let $z_{(i,j)}$ denote the embedding of edge $(i,j)$ and $z_i$ the embedding of node $i$. In the edge-centric update, the embedding vectors of its incident nodes are first aggregated (i.e., for edge $(i,j)$ the embeddings $z_i$ and $z_j$), and the result is concatenated with the current edge embedding $z_{(i,j)}$. The resulting vector is then passed through a nonlinear transformation, for example, the rectified linear unit $\text{ReLU}(x) = \max(0, x)$. Mathematically, this update can be expressed as:

\begin{equation}
\label{eq:rl-edge}
    z_{(i,j)}^{k+1} = \text{ReLU}(\gamma_\theta(z_{(i,j)}^{k}) \oplus \phi_\theta(z_{i}^{k} + z_{j}^{k})),
\end{equation}

\noindent where $z_{(i,j)}^{k}$ denotes encoding of edge $(i,j)$ after $k$ layers. Correspondingly, $z_i^k$ expresses an analog encoding of node $i$. $\gamma_\theta$ and $\phi_\theta$ are differentiable functions of parameter set $\theta$. Symbol $\oplus$ is used to denote the concatenation operation.

The node-centric update is defined in an analogous fashion. It aggregates embeddings of incident edges and then concatenates them with the self-embedding $z_i$. Using notation from EQ.~\eqref{eq:rl-edge}, the formal description is:

\begin{equation}
	\label{eq:rl-node}
    z_i^{k+1} = \text{ReLU}(\phi_{\theta}(z_i^k) \oplus \gamma_\theta (E_i^k)), \quad E_i^k = \sum_j z_{(i,j)}^k.
\end{equation}

Edge features are initialized with the edge weights ${J_{ij}}$. Defining suitable node features is less straightforward, since the node weights ${h_i}$ and spins ${\sigma_i}$ alone may not be sufficiently informative for the task at hand. Therefore, we augment the node features with each node's graph position.

We also include pooling layers that are not present in the original design. After repeated concatenations, the embedding vectors become high-dimensional, so pooling is employed to reduce the model's dimensionality while retaining the most informative components of each vector. Since every node is a potential candidate for an action, we refer to the final encoding of node $i$ as its action embedding, denoted $Z_i$. To represent the entire Chimera graph, we use a state embedding $Z_s$ defined as the sum of all node embedding vectors. This simple aggregation scheme has been shown to be an effective approach to graph-level encoding~\cite{Khalil2017}.

\subsection{Model training}

We trained our model on randomly generated Chimera instances. We found that the minimal viable size of a training instance is $C_3$. Smaller instances lack inter-cell couplings, which are essential in the full Chimera topology, resulting in poor performance. The couplings $J_{ij}$ and local fields $h_i$ were sampled from a normal distribution $\mathcal{N}(0,1)$, and the initial spin configuration was chosen uniformly at random. To introduce low-energy configurations into the training set, we applied the following preprocessing step: for each generated instance, with probability $p = 10\%$, we performed simulated annealing before passing it through the SGNN.

Our goal is to learn an approximation of the optimal action-value function
$Q(a, s; \Theta)$. To this end, we employ standard $n$-step deep
Q-learning~\cite{Mnih2016} with an experience replay
buffer~\cite{Lin1992,Mnih2015}.

During each episode, the agent generates a trajectory of states, actions, and rewards
\begin{equation}
    \tau = (s_0, a_0, r_0, \ldots, s_{T-1}, a_{T-1}, r_{T-1}, s_T),
\end{equation}
\noindent where $s_T$ is the terminal state. From this trajectory, we construct
$n$-step transitions
\begin{equation}
    \tau_t^{n} = (s_t, a_t, r_{t,t+n}, s_{t+n}),
\end{equation}
\noindent which are stored in the replay buffer $\mathcal{B}$. The $n$-step return is defined as
\begin{equation}
    r_{t,t+n} = \sum_{k=0}^{n} \gamma^k r_{t+k},
\end{equation}
\noindent where $\gamma \in (0,1]$ denotes the discount factor.


\section{Results and discussion}

For our experiments, we combined reinforcement learning with simulated annealing. We evaluate a reinforcement-learning–augmented simulated annealing post-processing method (SAwR) against standard simulated annealing (SA). Low-energy initial configurations were obtained from the D-Wave $2000$Q quantum annealer using default parameters. For each Chimera size  $C_4$, $C_8$, $C_{12}$, and $C_{16}$ (corresponding to $128$, $512$, $1152$, and $2048$ spins), 500 random Ising instances were generated from the same distribution used during training of the reinforcement-learning agent. Both classical methods were applied to the same initial quantum annealing samples.

\begin{figure}[t]
\centering
\begin{subfigure}{0.49\textwidth}
    \includegraphics[width=\textwidth]{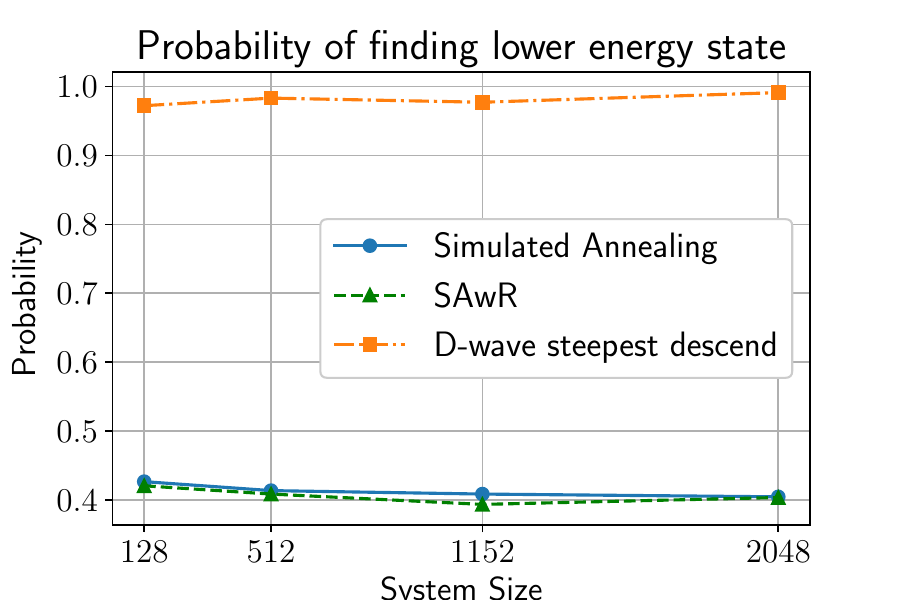}

\end{subfigure}
\hfill
\begin{subfigure}{0.49\textwidth}
    \includegraphics[width=\textwidth]{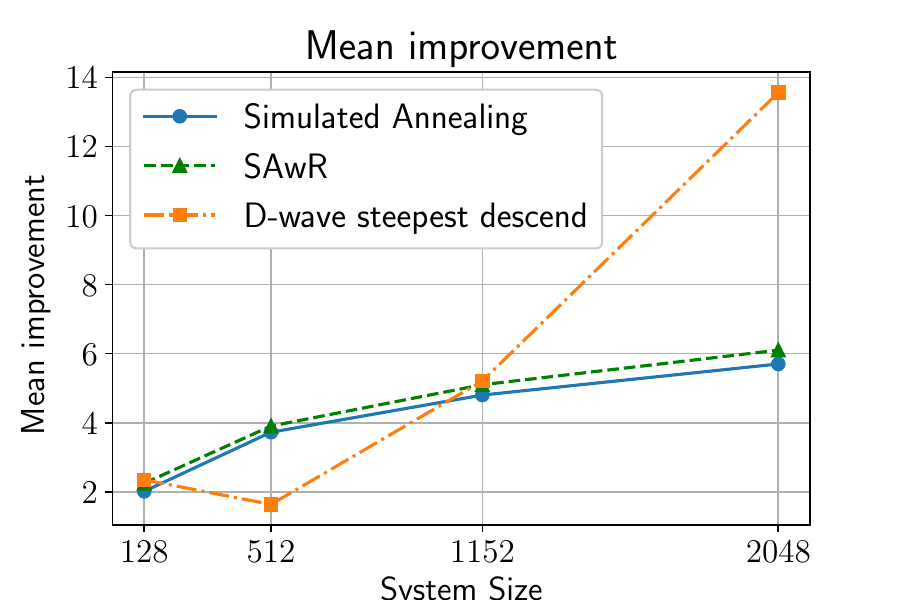}
\end{subfigure}
\caption{Performance of simulated annealing with reinforcement (SAwR) compared to standard simulated annealing and greedy search (D-Wave steepest descend). Results are averaged over $500$ random Ising instances for each Chimera size. Left: probability of finding a configuration with energy lower than the initial quantum-annealing sample. Right: mean energy improvement, defined as the difference between the initial energy and the lowest energy found during post-processing.}
\label{fig:rl-results}
\end{figure}

Performance is assessed using two metrics:  the probability of improvement, defined as the fraction of instances for which a lower-energy configuration than the initial state is found, and the energy improvement, defined as the difference between the initial energy and the lowest energy obtained during post-processing. As shown in Fig.~\ref{fig:rl-results}, SAwR achieves a consistently higher probability of improvement than standard SA across all problem sizes. While the absolute differences are small, their persistence across sizes indicates a systematic effect rather than statistical noise. In contrast, the mean energy improvement shows only marginal differences between the two methods and remains comparable within statistical fluctuations.

To assess whether reinforcement learning is necessary, we performed preliminary comparisons with a greedy single-spin-flip hill-climbing heuristic. These tests indicate that greedy descent frequently becomes trapped in nearby local minima and performs worse than both SA and SAwR. The reinforcement-learning agent occasionally accepts locally unfavorable moves that enable escape from such minima, suggesting that it captures non-trivial structural information about typical error patterns or local landscape features induced by the hardware embedding. Overall, the results indicate that reinforcement learning provides a modest but consistent robustness advantage in post-processing quantum annealing outputs, without yielding a significant increase in the depth of energy improvements.

\section{Conclusion}

In this chapter, we investigated a reinforcement-learning approach designed to enhance the quality of solutions returned by a quantum annealing processor through post-error correction. The method operates directly on the samples generated by the annealer and learns to identify and reverse local errors that frequently arise from noise, imperfect embeddings, or thermal excitations during annealing. By iteratively refining raw outputs, the agent increases the probability of recovering true low-energy configurations without requiring modifications to the annealing schedule or additional hardware calls.

The empirical results indicate that such an agent can, in controlled settings, improve success probability and reduce the energy of returned configurations. However, this study should be regarded as preliminary. The experiments were conducted on selected instance classes and moderate system sizes, and the training procedure was tailored to these specific conditions. Questions of scalability, robustness across diverse problem families, transferability between devices or embeddings, and stability under different noise regimes remain open. Moreover, the design of a practically deployable agent would require systematic investigation of training cost, generalization performance, and integration with other mitigation or schedule-level optimization techniques. Thus, while the results suggest that reinforcement learning can be a viable post-processing tool for quantum annealers, substantial further research is necessary before such methods can be considered mature or broadly applicable.

%% file: chapters/sim.tex
\chapter{Exact Simulation of Adiabatic Quantum Dynamics for Small Systems}
\chaptermark{Simulation of quantum dynamics}
\label{chap:sim}

On D-Wave quantum annealers, the standard forward-anneal time is specified in microseconds, but newer systems also expose a fast-anneal protocol that enables much shorter, effectively nanosecond-scale schedules, intended to probe “coherent” dynamics rather than optimize best-possible solution quality~\cite{fast_anneal}. In this ultrafast limit, the evolution is generally diabatic (non-adiabatic), and the environmental interaction is minimal~\cite{King2022,King2023}. 

We assess the fidelity of the exact simulation of the D-Wave quantum system at ultra-short annealing times. We are especially interested in the effect of device errors.

\paragraph{Author contributions}
The author implemented and ran exact closed-system simulations of the annealing dynamics, augmented the simulator with an effective control-error (coupling-disorder) model, and conducted corresponding QPU experiments in the ultrafast annealing regime on native, embedding-free instances. The author then performed end-to-end data processing and conducted a distribution-level comparison between simulation and hardware, quantifying how the effective error model alters agreement and deriving the chapter's main empirical conclusions and figures.

\paragraph{Relation to published work}
This chapter presents original, previously unpublished work. No results from the author's publications are reused here. All results and figures were produced for this dissertation.

\section{Simulation of quantum dynamics}

Numerical simulations were performed using \texttt{QuantumAnnealing.jl}~\cite{Morrell2024}. It solves ordinary differential equations arising in dynamic quantum systems. Specifically, it solves the Schrödinger equation with time-dependent Hamiltonian $H(t)$, acting over a set of $n$ qubits in natural units:

\begin{equation}
  i  \frac{d}{dt} \ket{\Psi(t)} = H(t) \ket{\Psi(t)},
  \label{eq:schrodinger}
\end{equation}

\noindent where $H(t)$ is a quantum annealing Hamiltonian with transverse field Ising model (see Eq.~\ref{eq:ising-dwave}), $t\in [0, \tau]$ is the annealing time, and the initial condition $\ket{\Psi(0)}$ is given. As we assume natural units, we take $\hbar = 1$.  The used initial state is:

\begin{equation}
  \ket{\Psi(0)} = \ket{\Psi_0} = \bigotimes^n \frac{1}{\sqrt{2}} (\ket{\uparrow} + \ket{\downarrow}),
  \label{fig:psi-0}
\end{equation}

\noindent which corresponds to the state of minimum energy at the beginning of anneal.

When solving equation~\eqref{eq:schrodinger} it is useful to write the solution in terms of the time-evolution operator $U(t, t_0)$:

\begin{equation}
  \ket{\Psi (t)} = U(t, t_0) \ket{\Psi (t_0)},
  \label{eq:time-evolution-operator}
\end{equation}

\noindent where $\ket{\Psi (t_0)} = \ket{\Psi_0}$ provides the initial condition for time evolution. The operator $U(t, t_0)$ must satisfy:

\begin{equation}
  i \frac{d}{dt} U(t, t_0) = H(t)U(t, t_0), \quad U(t_0, t_0) = \mathbb{I}.
  \label{eq:operator-requirement}
\end{equation}

As the system described here is time-dependent, the naive solution:

\begin{equation}
  U(t, t_0) = \exp \left( -i \int_{t_0}^{t} H(t)dt \right) ,
\end{equation}

\noindent doesn't hold. It can be shown that Eq.~\eqref{eq:operator-requirement} is not satisfied when $H(t)$, $H(t')$ don't commute, as is the case here.

The standard approach to solving the time-dependent Schrödinger equation is to use \emph{time-ordered exponential}:

\begin{equation}
  U(t, t_0) = \mathcal{T} \exp \left( -i \int_{t_0}^{t} H(t)dt \right)
  \label{eq:time-ordered-solution}
\end{equation}

\noindent where $\mathcal{T}$ is the \emph{time-ordering operator} and is defined as ordering~\cite{Kosovtsov2009}:

\begin{align}
  \mathrm{T} \{\mathrm{L}(t_1), \mathrm{L}(t_2), \ldots, \mathrm{L}(t_n)\} = \underset{t_{\alpha_1} \geq t_{\alpha_2} \geq \ldots \geq t_{\alpha_n}}{\mathrm{L}(t_{\alpha_1})\mathrm{L}(t_{\alpha_2})\ldots\mathrm{L}(t_{\alpha_n})},
\end{align}

\noindent where $\mathrm{L}(t)$ is a linear operator. The formula in Eq.~\eqref{eq:time-ordered-solution} is computationally expensive.

Another option is to simulate the evolution of the quantum state $\ket{\Psi(t)}$ by Magnus expansion~\cite{Magnus1954}. It is a well-established numerical method for simulating quantum dynamics~\cite{Wall2012}. The Magnus expansion solves the system of ODEs:

\begin{equation}
  \frac{d}{dt} \ket{\Psi(t)} = \mathcal{A}(t) \ket{\Psi(t)}, \quad \ket{\Psi(0)} = \ket{\Psi_0}.
  \label{eq:magnus-ode}
\end{equation}

\noindent Here we set $\mathcal{A}(t) = -i H(t)$. In general, $\mathcal{A}(t)$ can be any linear operator. The solution is given by:

\begin{align}
  \begin{split}
  \ket{\Psi(t)} &= \exp(\Omega(t))\ket{\Psi(0)} \\
  \Omega(t) &= \sum_{k=1}^{\infty} \Omega_{k}(t).
  \end{split}
\end{align}

\noindent The full recursive formula for $\Omega_{k}(t)$ can be found in~\cite{Arnal2018}. For our simulations, we will use the fourth-order Magnus expansion. There are several reasons for this choice. First, as a single step of the Magnus expansion truncated to $\Omega_k$ has error rate $\epsilon = \mathcal{O}(t^{k+1})$~\cite{Sanchez2011}, $\Omega_4$ is accurate enough. Additionally, it captures meaningful time-ordering effects with good precision~\cite{Iserles2001}. As such, here we will provide the first four terms. It is worth mentioning that \texttt{QuantumAnnealing.jl} uses by default fourth-order Magnus expansion, and those functions were independently optimized~\cite{Morrell2024}. 


\begin{align}
\Omega_1(t) &= \int_{0}^{t} \mathcal{A}(t_1)\, dt_1 \notag\\
\Omega_2(t) &= \frac{1}{2} \int_{0}^{t} dt_1 \int_{0}^{t_1} dt_2
  \Bigl[\mathcal{A}(t_1), \mathcal{A}(t_2)\Bigr] \notag\\
\Omega_3(t) &= \frac{1}{6} \int_{0}^{t} dt_1 \int_{0}^{t_1} dt_2 \int_{0}^{t_2} dt_3
\begin{aligned}[t]
\Bigl(
&\bigl[\mathcal{A}(t_1), [ \mathcal{A}(t_2), \mathcal{A}(t_3)] \bigr] \\
&+ \bigl[\mathcal{A}(t_3), [ \mathcal{A}(t_2), \mathcal{A}(t_1)] \bigr]
\Bigr)
\end{aligned}
\label{eq:magnus} \\
\Omega_4(t) &= \frac{1}{12} \int_{0}^{t} dt_1 \!\int_{0}^{t_1} dt_2 \!\int_{0}^{t_2} dt_3 \!\int_{0}^{t_3} dt_4
\begin{aligned}[t]
\Bigl(
&\Bigl[ \bigl[[\mathcal{A}(t_1), \mathcal{A}(t_2)], \mathcal{A}(t_3) \bigr], \mathcal{A}(t_4) \Bigr]\\
&+ \Bigl[ \mathcal{A}(t_1), \bigl[ [\mathcal{A}(t_2), \mathcal{A}(t_3)], \mathcal{A}(t_4) \bigr]  \Bigr]\\
&+ \Bigl[ \mathcal{A}(t_1), \bigl[ \mathcal{A}(t_2), [\mathcal{A}(t_3), \mathcal{A}(t_4)] \bigr]  \Bigr]\\
&+ \Bigl[ \mathcal{A}(t_2), \bigl[ \mathcal{A}(t_3), [\mathcal{A}(t_4), \mathcal{A}(t_1)] \bigr]  \Bigr]
\Bigr),
\end{aligned}
\notag
\end{align}

\noindent Here $[\cdot, \cdot]$ is the matrix commutator, \textit{i.e.} $[A, B] = AB - BA$.

When applied to simulating a quantum system, the operator $\mathcal{A}(t)$ is anti-Hermitian. This means that even when the higher-order terms of $\Omega(t)$ are omitted to provide an approximation for the full series, this approximation is still unitary~\cite{Blane2009}. Therefore, these truncated series are valid quantum operators and the quantum state obtained by the solver is properly normalized. The unitary matrices constructed by the Magnus expansion are then applied to the density operator $\rho(t)$, using the fact that $\rho(t) = \ket{\Psi(t)} \bra{\Psi(t)}$. Finally, the obtained solution is:

\begin{equation}
  \rho(t) = \exp (\Omega(t)) \rho_0 \exp(\Omega(t))^{\dagger}, \quad \rho_0 = \rho(0).
\end{equation}

We introduced errors into the simulated system using the error model presented in Eq.~\eqref{eq:ising-ice}. For each coupling $J_{ij}$, we independently generated a perturbation $\delta J_{ij} \sim \mathcal{N}(0,\sigma)$ from a normal distribution centered at zero, and added it to the original value, $J_{ij} \rightarrow J_{ij} + \delta J_{ij}$. The standard deviation $\sigma$ serves as a control parameter for the noise strength, allowing us to systematically tune the magnitude of the effective control errors. Each time we simulated the instance with errors, we drew $4000$ samples and then averaged the results.

This is a Closed System Simulation (CSS), meaning it ignores interactions with the environment, including thermal ones. As we showed in an earlier chapter, there are interactions between the QPU and its environment. This implies that some inaccuracy in the simulation is expected.



\subsection{QPU data collection}

\begin{figure}
  \begin{center}
    \includegraphics[width=0.48\textwidth]{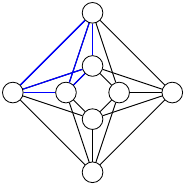}
  \end{center}
  \caption{Example of a natural embedding of complete graph $K_4$ into a Pegasus unit cell}
  \label{fig:k4-embedding}
\end{figure}

For the experiments, we generated $40$ random $K_4$ Ising instances with couplings only, \textit{i.e.}, fully connected four-spin graphs with $h_i=0$. These instances embed natively into a Pegasus topology (example embedding shown in Fig.~\ref{fig:k4-embedding}), eliminating embedding overhead and keeping the study focused on intrinsic device dynamics.

Our goal was to select problems that exhibit nontrivial diabatic behavior in the fast-anneal regime, rather than instances whose output distribution effectively equilibrates too quickly. To this end, we drew couplings $J_{ij}$ independently from a uniform distribution on $[-1,1]$ and performed a screening simulation for each candidate instance using a closed-system Schrödinger evolution implemented in \texttt{QuantumAnnealing.jl}. We computed the ground-state probability as a function of the annealing time $\tau$, and retained only instances for which this dependence indicated sufficiently slow relaxation (\textit{i.e.}, ground-state probability did not trivially saturate at short $\tau$).

During data collection, we tiled the QPU with many disjoint copies of each $K_4$ instance, so that nearly all available qubits were active while ensuring that no two copies overlapped. This approach leverages the full chip in each run and averages over local fabrication-induced inhomogeneities, reducing the sensitivity of the results to imperfections in any particular region of the device.

\subsection{Performance metric}

In our considerations, we are interested in the distributions of states returned by the simulation and the quantum annealer. Due to the small size of the instance, we can track the evolution of all possible classical states. We will focus on the ground state of the instance, which is easy to find via brute force search. The main performance metric will be the probability of finding the ground state.

In addition to the ground-state probability, we compare the full output distributions using two additional metrics. The first is the total variation distance (TVD) between the simulated distribution $P_{\mathrm{sim}}$ and the empirical QPU distribution $P_{\mathrm{dwave}}$, defined as

\begin{equation}
  \mathrm{TVD} = \frac{1}{2} \sum_{\bm{s} \in \mathcal{S}} \left| P_{\mathrm{sim}}(\bm{s}) - P_{\mathrm{dwave}}(\bm{s}) \right|,
\end{equation}

\noindent where \(\mathcal{S}\) is the set of all computational-basis spin configurations. In our case, for a $K_4$ problem $|\mathcal{S}|=2^4$. The simulated probabilities are obtained from the final density operator as $P_{\mathrm{sim}}(\bm{s})=\bra{\bm{s}}\rho(\tau)\ket{\bm{s}}$. The QPU probabilities are estimated from measurement outcomes by the relative frequencies,
$P_{\mathrm{dwave}}(\bm{s}) \approx n(\bm{s})/N$, where $n(\bm{s})$ is the number of times state $\bm{s}$ was observed in $N$ annealing runs.

The second distribution-level metric is the \emph{classical} fidelity between $P_{\mathrm{sim}}$ and $P_{\mathrm{dwave}}$, defined as:

\begin{equation}
  \mathcal{F} = \left( \sum_{\bm{s} \in \mathcal{S}} \sqrt{P_{\mathrm{sim}}(\bm{s})\, P_{\mathrm{dwave}}(\bm{s})} \right)^2,
\end{equation}

To model control errors, for each coupler $(i,j)\in\mathcal{E}$ we draw an independent Gaussian perturbation $\delta J_{i,j}\sim\mathcal{N}(0,\sigma)$ and construct a perturbed instance by modifying the original couplings (e.g., $\tilde{J}_{i,j}=J_{i,j}+\delta J_{i,j}$). We then run the simulation for this modified model. The entire procedure is repeated $M=4000$ times, and all reported quantities are averaged over the resulting ensemble.

\section{Results}

\begin{figure}[h]
  \centering
  \includegraphics[width=\textwidth]{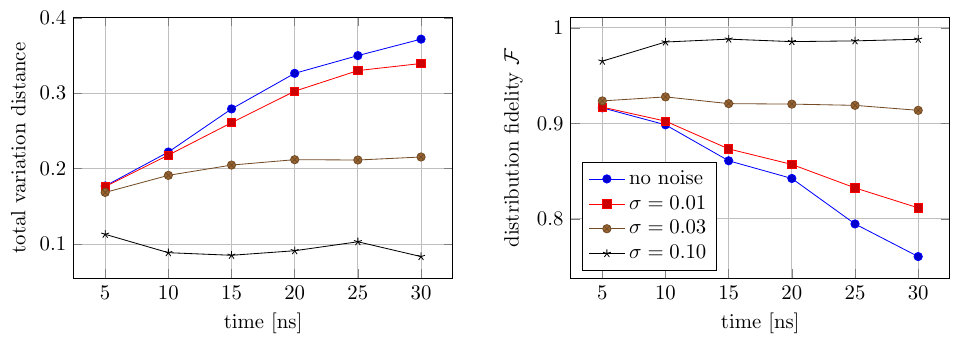}
  \caption[Median TVD and classical fidelity vs annealing time.]{Median total variation distance (left) and classical fidelity $\mathcal{F}$ (right) between the simulated output distribution $P_{\mathrm{sim}}$ and the empirical QPU distribution $P_{\mathrm{dwave}}$ as a function of the fast-anneal duration $\tau$. Each curve corresponds to a coupling-noise strength $\sigma$ in the ICE model. For each instance, simulated quantities are averaged over $M=4000$ independent noise realizations and the plotted points show the median over $40$ random $K_4$ instances. Lower TVD and higher $\mathcal{F}$ indicate better agreement.}
  \label{fig:tvd-fidelity}
\end{figure}

\begin{figure}[h]
  \centering
  \includegraphics[width=\textwidth]{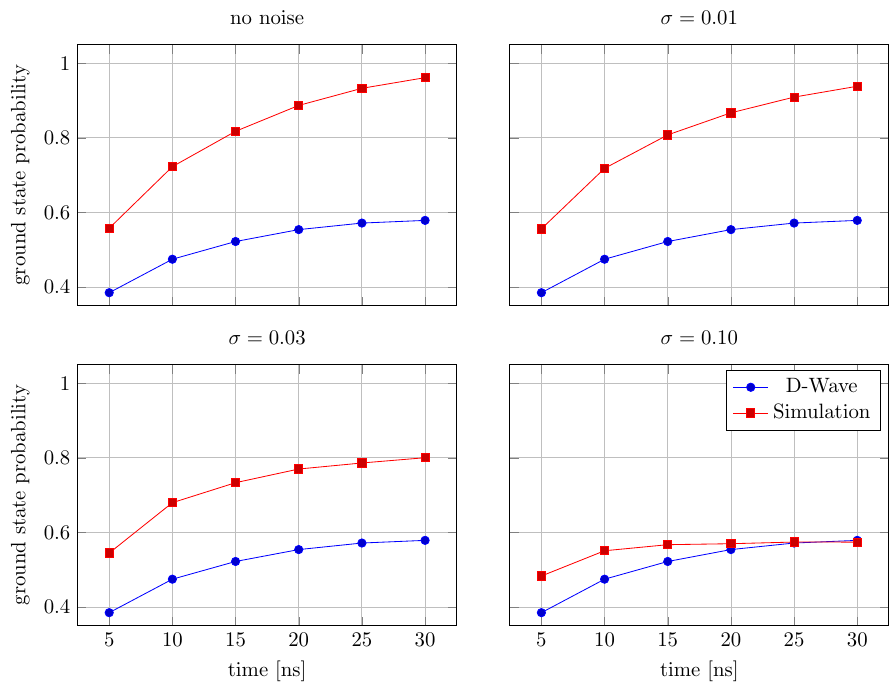}
  \caption[Median ground-state probability vs annealing time.]{Median ground-state probability as a function of the fast-anneal duration $\tau$, comparing QPU measurements (blue) with closed-system simulations (red). Each panel corresponds to a different coupling-noise strength $\sigma$ in the ICE perturbation model. Simulated values are averaged over $M=4000$ noise realizations per instance and the median is taken over the same set of $40$ random $K_4$ instances.}
  \label{fig:gs-probability}
\end{figure}

Figure~\ref{fig:tvd-fidelity} summarizes distribution-level agreement between the closed-system simulation and the QPU as a function of the fast-anneal duration $\tau$. We report the median of the total variation distance (TVD) and the classical fidelity $\mathcal{F}$ between $P_{\mathrm{sim}}$ and $P_{\mathrm{dwave}}$ for several strengths of the coupling-noise model parameterized by $\sigma$. Figure~\ref{fig:gs-probability} shows the corresponding median ground-state probability, which provides an easily interpretable marginal statistic of the same distributions.

In the absence of control errors ($\sigma=0$), the simulation systematically overestimates the ground-state probability (Fig.~\ref{fig:gs-probability}) and the disagreement with the measured output distribution increases with $\tau$: TVD rises from $\approx 0.18$ at $5\,\mathrm{ns}$ to $\approx 0.37$ at $30\,\mathrm{ns}$, while the fidelity drops from $\approx 0.92$ to $\approx 0.76$ (Fig.~\ref{fig:tvd-fidelity}). Introducing coupling disorder improves the agreement monotonically. The best overall match across all metrics is obtained for $\sigma\approx 0.10$, where TVD is reduced to $\approx 0.08$-$0.11$ and the fidelity increases to $\approx 0.97$-$0.99$ over the full $\tau$ range (Fig.~\ref{fig:tvd-fidelity}). In the same setting, the simulated and measured ground-state probabilities nearly coincide for $\tau\gtrsim 15\,\mathrm{ns}$ (Fig.~\ref{fig:gs-probability}).

The value $\sigma\approx 0.1$ is substantially larger than typical expectations for intrinsic ICE on these devices~\cite{dwave,Pearson2019,Zaborniak2021}. Importantly, in the present closed-system model $\sigma$ should therefore be interpreted as an \emph{effective} disorder strength that compensates for missing physical effects rather than as a direct estimate of the hardware calibration error. Plausible contributors include residual open-system relaxation during the fast anneal, correlated and/or biased control errors not captured by independent Gaussian coupler noise, and global miscalibration (e.g., an effective scale factor in the implemented Hamiltonian). Disentangling these mechanisms would require extending the model beyond CSS and/or fitting an additional global rescaling parameter alongside the ICE-like perturbations.

\section{Conclusion}

In this chapter we evaluated how accurately an exact closed-system simulation reproduces the output statistics of D-Wave's ultrafast (nanosecond-scale) annealing protocol. Using fourth-order Magnus integration in \texttt{QuantumAnnealing.jl}, we simulated the full unitary dynamics of small native $K_4$ instances and compared the resulting measurement distributions with QPU data using ground-state probability, total variation distance, and classical fidelity.

Across the tested annealing times, the best agreement between simulated and experimental distributions was obtained only after introducing relatively strong effective coupling disorder, with an optimal match around $\sigma\approx 0.1$. Because this value exceeds typical expectations for intrinsic control errors, it should be interpreted as an \emph{effective} parameter that likely absorbs additional discrepancies beyond the independent Gaussian ICE model, including residual open-system effects in the fast-anneal regime and possible global calibration or scale-factor mismatches. Overall, the results indicate that exact closed-system simulation captures qualitative features of the diabatic dynamics on small instances, but quantitative agreement with hardware distributions requires extending the model to include more realistic error channels and, potentially, dissipative dynamics.

%% file: chapters/last.tex
\chapter{Summary}
\label{chap:last}

The work presented in this thesis aims to assess quantum annealers as computational devices for hard Ising and QUBO optimization, in direct comparison with advanced classical and quantum-inspired methods, while maintaining explicit tracking of the underlying physics and errors. Within this broad goal, the thesis combines algorithmic development, systematic benchmarking, thermodynamic analysis, and machine-learning-based error mitigation into a coherent narrative that spans from abstract models to concrete devices.

A central methodological contribution of this thesis is the development of \texttt{SpinGlassPEPS.jl}, a tensor-network-based solver that operates directly on the native geometries of contemporary quantum annealers. Building on earlier work~\cite{Rams2021}, the solver implements a branch-and-bound search in probability space guided by approximate PEPS contractions. By exploiting sparse tensor representations, clustering, and GPU acceleration, it can handle Pegasus- and Zephyr-type graphs without embedding overhead. Within the thesis, \texttt{SpinGlassPEPS.jl} serves as a topology-aware classical baseline, enabling structurally aligned comparisons with quantum annealers and providing access to low-energy spectra, localized excitations, and droplet structures.

Using this solver, the thesis presents a systematic benchmark of D-Wave quantum annealers against selected classical baselines, including simulated bifurcation machines. The benchmarks focus on random spin-glass instances defined directly on Pegasus and Zephyr graphs, with system sizes reaching several thousand spins. Performance is assessed using multiple complementary metrics, including the diversity of low-energy states, to capture both optimization performance and sampling behavior.

The results demonstrate that \texttt{SpinGlassPEPS.jl} provides a strong and physically interpretable classical reference, but also exposes its practical limitations. For the largest random instances, the approximate tensor-network contractions required for scalability introduce errors that prevent the solver from matching the solution quality of highly optimized GPU-accelerated heuristics within comparable runtimes. Consequently, tensor-network methods currently play a more effective role as topology-aware benchmarking and analysis tools for small-to-medium scale problems, rather than as wall-clock-competitive solvers for the largest instances. These findings clarify the regime in which tensor-network approaches are most informative for evaluating quantum annealing hardware.

A third pillar of the thesis is a physical, thermodynamic view of quantum annealing. By treating the annealer as a thermodynamic machine driven along a programmable schedule, the work relates success probability and solution quality to quantities such as energy dissipation, entropy production, and effective temperature.  The results demonstrate that carefully chosen pauses can simultaneously enhance computational performance and reduce thermodynamic cost relative to simple reverse annealing, while also revealing regimes in which the applied longitudinal field becomes detrimental once pausing is introduced.


This thermodynamic lens reveals that annealing time and schedule shape are not merely tuning parameters for optimization, but fundamental control knobs that balance coherence, thermalization, and energetic overhead. It suggests that future benchmarking of quantum hardware should routinely incorporate physical efficiency metrics alongside purely algorithmic ones.

The final main component of the thesis is practical: a reinforcement-learning-based post-processing strategy for mitigating errors in quantum annealers. Here, the outputs of a D-Wave device are treated as inputs to an RL agent that learns how to navigate the local energy landscape by flipping spins in a way that systematically corrects typical error patterns. Once trained, the agent improves both success probability and energy accuracy across test instances.

The method integrates naturally as an extra step in a standard annealing workflow and is orthogonal to hardware-level improvements. It exploits structure in observed errors without requiring modifications to the underlying QPU and can be iterated multiple times or combined with other mitigation techniques. This illustrates how modern learning techniques can be used to “wrap” noisy quantum devices in adaptive classical controllers.


The technical appendices complement the main narrative by collecting background material that underpins these results. The introduction to tensor networks, including notation, matrix-product states, and PEPS, is presented in Appendix~\ref{apx:tn}. It provides the conceptual and mathematical tools required to follow the construction of \texttt{SpinGlassPEPS.jl}. Likewise, Appendix~\ref{apx:rl} focuses on deep reinforcement learning, including countable Markov decision processes and value-based methods. It supplies the formal framework behind the RL-driven error-mitigation pipeline. Finally, Appendix~\ref{apx:dwave} presents the physical properties of D-Wave's quantum annealing devices used in this work.

\section{Future work}

Several concrete directions for future research follow from this synthesis. On the methodological front, further development of \texttt{SpinGlassPEPS.jl} can proceed along three complementary directions that strengthen its role as a QPU-aligned classical baseline. First, the feature set can be broadened by incorporating alternative PEPS contraction schemes, such as Corner Transfer Matrix approaches~\cite{Lukin2024,Mangazeev2024}, extending the range of supported quasi-2D geometries of the Potts Hamiltonian (for example, Lechner-Hauke-Zoller architecture~\cite{Lechner2015}), and adding routines for estimating thermodynamic quantities like approximate free energies~\cite{Nishimori2025}. Second, enabling multi-GPU execution in the workflow of the software will require a careful re-examination of the underlying tensor-network algorithm~\cite{peps} to identify the most efficient use of hardware, in particular, whether parallelism is best achieved by contracting different parts of the network concurrently or by splitting large tensors across several devices and processing them sequentially. Finally, automating the embedding of Ising and QUBO instances into the Potts representation relevant for Pegasus and Zephyr graphs~\cite{Gomez-Tejedor2025,Pelofske2024b} would remove a manual step currently required from the user.

Thermodynamically, the efficiency framework developed here could be applied to newer Pegasus and Zephyr-based devices. These architectures offer higher connectivity, improved noise characteristics, and more complex structures, making them well-suited for probing the trade-off between energetic performance and computational complexity in a more realistic setting. This would enable systematic tests of how the energetic “cost of computation” scales with connectivity, noise properties, problem structure, and system size, and would help assess the scalability of our findings to larger-scale quantum systems.

Finally, the reinforcement-learning-based mitigation could be enhanced by integrating it with run-time controls of the annealing schedule.

%% file: appendices/appendix-quantum.tex
\chapter{Quantum Computing Conventions}
\label{apx:qconventions}

This appendix fixes the basic notation used throughout the thesis for qubits and
Hamiltonians relevant to quantum annealing. We set $\hbar=1$.

\section{Qubits and computational basis}

An $N$-qubit register is described by the Hilbert space
\begin{equation}
\mathcal{H} = \left(\mathbb{C}^{2}\right)^{\otimes N}.
\end{equation}

We use Dirac notation $\ket{\psi}\in\mathcal{H}$ and $\bra{\psi}=\ket{\psi}^\dagger$.
The \emph{computational basis} is the common eigenbasis of all $\sigma_i^{z}$.
For a single qubit we use the $\sigma^z$ eigenstates

\begin{equation}
\ket{\uparrow}=
\begin{bmatrix}
1\\
0
\end{bmatrix},
\qquad
\ket{\downarrow}=
\begin{bmatrix}
0\\
1
\end{bmatrix},
\end{equation}
with eigenvalue equations
\begin{equation}
\sigma^{z}\ket{\uparrow}=+1 \cdot \ket{\uparrow},\qquad
\sigma^{z}\ket{\downarrow}=-1 \cdot \ket{\downarrow}.
\end{equation}
A computational-basis configuration of $N$ qubits is a tensor product
\begin{equation}
\ket{\mathbf{s}}=\ket{s_1}\otimes\ket{s_2}\otimes\cdots\otimes\ket{s_N},
\qquad s_i\in\{\uparrow,\downarrow\},
\end{equation}
equivalently labeled by classical Ising spins $s_i\in\{+1,-1\}$ via
$+1\equiv\ket{\uparrow}$ and $-1\equiv\ket{\downarrow}$.

\section{Pauli operators and locality}

The Pauli matrices are
\begin{equation}
\sigma^{z}=
\begin{bmatrix}
1 & 0\\
0 & -1
\end{bmatrix},
\qquad
\sigma^{x}=
\begin{bmatrix}
0 & 1\\
1 & 0
\end{bmatrix}.
\end{equation}
For an operator acting on qubit $i$ we write
\begin{equation}
\sigma_i^\alpha
= I^{\otimes (i-1)}\otimes \sigma^\alpha \otimes I^{\otimes (N-i)},
\qquad \alpha\in\{x,z\},
\end{equation}
i.e., $\sigma^\alpha$ on site $i$ and the identity $I$ on all other sites.

\section{Annealing Hamiltonians and the classical energy}

Quantum annealing implements a time-dependent Hamiltonian interpolating between
a \emph{driver} (transverse-field) term and a \emph{problem} (Ising) term. We write
\begin{equation}
H(t)=H\!\left(s(t)\right),\qquad s(t)\in[0,1],
\end{equation}
and use the standard Ising problem Hamiltonian (diagonal in the computational basis)
\begin{equation}
H_P = \sum_{(i,j)\in E} J_{ij}\,\sigma_i^{z}\sigma_j^{z} + \sum_{i=1}^N h_i\,\sigma_i^{z},
\label{eq:HP_app}
\end{equation}
together with a transverse-field driver, typically of the form
\begin{equation}
H_X = -\sum_{i=1}^N \sigma_i^{x}.
\label{eq:HX_app}
\end{equation}
A common schedule parametrization is $H(s)=A(s)H_X + B(s)H_P$ (up to
conventional prefactors), where $A(s)$ decreases and $B(s)$ increases with $s$.

Because $H_P$ is diagonal in the computational basis, each measurement outcome
$\ket{\mathbf{s}}$ corresponds to a classical Ising configuration $\mathbf{s}\in\{\pm1\}^N$
with energy
\begin{equation}
E(\mathbf{s})=\bra{\mathbf{s}}H_P\ket{\mathbf{s}}
= \sum_{(i,j)\in E} J_{ij}\,s_i s_j + \sum_{i=1}^N h_i\,s_i,
\end{equation}
where $s_i=+1$ for $\ket{\uparrow}$ and $s_i=-1$ for $\ket{\downarrow}$.
Thus, sampling in the computational basis returns classical spin configurations.

For completeness, if binary variables $x_i\in\{0,1\}$ are used (QUBO notation),
we employ the standard mapping
\begin{equation}
s_i = 2x_i-1,
\qquad
x_i=\frac{s_i+1}{2},
\end{equation}
which relates bitstrings to $\sigma^z$ outcomes.

\section{Dynamics and expectation values}

Closed-system unitary dynamics are governed by the Schr\"odinger equation
\begin{equation}
i\,\frac{d}{dt}\ket{\psi(t)} = H(t)\ket{\psi(t)}.
\end{equation}
For any observable $O$, its expectation value in state $\ket{\psi}$ is
$\langle O\rangle = \bra{\psi}O\ket{\psi}$.
In practice, quantum annealers are open systems; nevertheless, the conventions
above define the operators, basis, and the classical energy function used to interpret
measured samples.




%% file: appendices/appendix-tn.tex
\chapter{Introduction to Tensor Networks}
\label{apx:tn}

Tackling complicated structures of modern QPUs requires employing advanced tensor-network techniques. To help better understand the inner workings of the presented algorithm, we dedicate part of this chapter to introducing the basics of tensor networks. We will describe the notation used, tensor diagrams, and common tensor structures. We want to stress that this section is not designed as a comprehensive introduction. The goal is to give context to the presented algorithm. For interested readers, we recommend the following works~\cite{Bridgeman2017,Cichocki2014,Evenbly2022,Orus2014,Orus2019,Ran2020}, which are the basis for this section. 

\section{Notation and Terminology}

Tensors are mathematical structures that describe multilinear relationships between objects~\cite{Roman2005}. They can be commonly thought of as multidimensional arrays of numbers. For our needs, a tensor is defined as a set of numbers labeled by $N$ indices, where $N$ is called the \emph{order}\footnote{In the literature, one may find the term \emph{rank} used instead} of the tensor~\cite{Ran2020}. For example, a scalar ($x$) is labeled as a zero index and called a $0$th-order tensor. Similarly, a vector is a $1$st-order tensor, a matrix is a $2$nd-order tensor, etc.

\begin{definition}[Index Contraction~\cite{Orus2014}]
    An \emph{index contraction} is the sum over all the possible values of the repeated indices of a set of tensors. 
\end{definition}

For example, the matrix product of $A$ and $B$ can be expressed as:

\begin{equation}
    \label{eq:tn-ex1}
    C_{ik} = \sum_{j=1}^{D} A_{ij} B_{jk}.
\end{equation}

\noindent This is the contraction of index $j$, which amounts to the sum over its $D$ possible values. Thus, matrix multiplication is a contraction between two $2$nd-order tensors at one shared index. It is possible to have a more complicated case:

\begin{equation}
    \label{eq:tn-ex2}
    D_{ijk} = \sum_{l=1}^{D_1}\sum_{m=1}^{D_2}\sum_{n=1}^{D_3} A_{ljm} B_{iln} C_{nmk}.
\end{equation}

\noindent Here, we are contracting multiple indices with a different number of possible values $D_i$. Indices that are left after contraction (in equation~\eqref{eq:tn-ex2} these are indices $i$, $j$, and $k$) are called \emph{open}, and contracted ones are called \emph{bond} or \emph{auxiliary} indices. The number $D$ of possible values that an index can have is called the \emph{bond dimension}. Additionally, in cases where tensors are describing some physical system, the open indices are often called \emph{physical} (as they correspond to some physical quantity or observable). Similarly, bond indices are called \emph{virtual} indices. Additionally, when describing index contractions, the sum symbols are often omitted as they are obvious from the context.

\begin{definition}[Tensor Network~\cite{Orus2014}]
    A \emph{Tensor Network} (TN) is a set of tensors where some, or all, of its indices are contracted according to some pattern. Contracting the indices of a TN is called, for simplicity, \emph{contracting the TN}.
\end{definition}

It is convenient to use diagrammatic notation when dealing with tensor networks. They are represented by a diagram where tensors are denoted by shapes and indices are represented by lines emerging from the shapes. Thus, a TN is expressed as a collection of shapes connected by lines. If two tensors are connected by a line, then their common index is contracted. If a line connects to only one tensor, it represents an open index. Sometimes used shapes and direction of legs may carry additional meaning~\cite{Biamonte2020,Bridgeman2017}, however, in general, neither carries any special significance.
 
\begin{figure}[!h]
    \includegraphics[width=\textwidth]{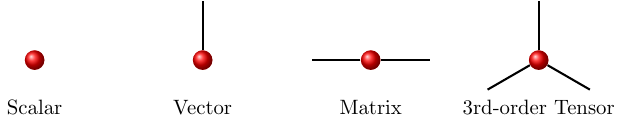}
    \caption{The graphical representation of scalar, vector, matrix, and $3$rd-order tensor in diagrammatic notation}
\end{figure}

It is very important to note that the total number of operations that must be done in order to obtain the final result of a TN contraction depends heavily on the order in which indices in the TN are contracted~\cite{Evenbly2022}. In general, the computational cost of pairwise tensor contraction between tensors $A$ and $B$ can be expressed as:

\begin{equation}
    \text{cost} \sim \frac{\lvert \dim(A)\rvert \times \lvert \dim(B) \lvert}{\lvert \dim(A \cap B) \rvert},
\end{equation}

\noindent where $\lvert \dim(A)\rvert$ denotes total dimension of $A$ (\textit{i.e.}, the product of its index dimensions) and $\lvert \dim(A \cap B) \rvert$ the total dimension of the shared indices. For example, the cost of matrix multiplication in~\eqref{eq:tn-ex1}, assuming that each index dimension has size $D$, is:

\begin{equation*}
    \mathcal{O}\left(\frac{(D \times D)  \times (D \times D)}{D}\right) = \mathcal{O}(D^3),
\end{equation*}

\begin{figure}
    \centering
    \begin{subfigure}[l]{\textwidth}
        \centering
        \captionsetup{labelformat=parens}
        \caption{}
        \includegraphics[width=\textwidth]{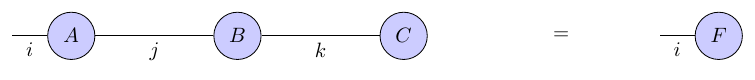}
        \label{fig:tn-contraction-example-a}
    \end{subfigure}
    
    \hfill
    
    \begin{subfigure}[l]{\textwidth}
        \centering
        \captionsetup{labelformat=parens}
        \caption{}
        \includegraphics[width=\textwidth]{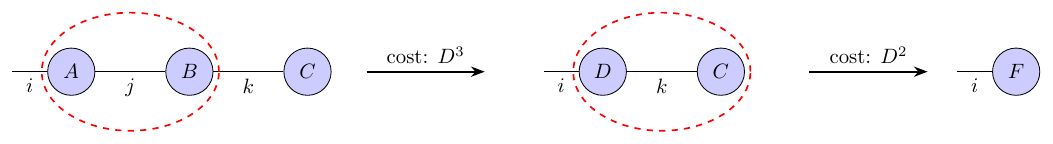}
        \label{fig:tn-contraction-example-b}
    \end{subfigure}
    
    \hfill
    
    \begin{subfigure}[l]{\textwidth}
        \centering
        \captionsetup{labelformat=parens}
        \caption{}
        \includegraphics[width=\textwidth]{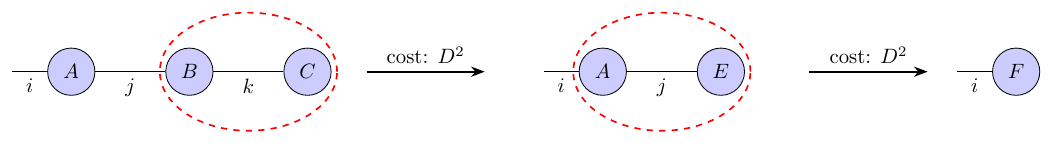}
        \label{fig:tn-contraction-example-c}
    \end{subfigure}
    
    \caption{Example of how the order of index contraction influences the computational cost of tensor network contraction. \ref{fig:tn-contraction-example-a} Contraction of a small tensor network consisting of three tensors $\{A, B, C\}$ into a tensor $F$. The tensor $A$ and $B$ are $D\times D$ matrices and $C$ is a $D$-element vector. \ref{fig:tn-contraction-example-b} The cost of the following sequence in (b) is $O(D^3 + D^2) = O(D^3)$. \ref{fig:tn-contraction-example-c} The cost from (c) is $O(2D^2) = O(D^2)$.}
    \label{fig:tn-contraction-example}
\end{figure}

\noindent Here, we use standard asymptotic $\mathcal{O}$ notation. The example of how the order of index contraction influences the computational cost of tensor network contraction is presented in Fig.~\ref{fig:tn-contraction-example}.

\section{Matrix Product States (MPS)}

The Matrix Product States (MPS)~\cite{Fannes1992}, also called ``Tensor Train'' in literature~\cite{Oseledets2011}, is a simple, but powerful one-dimensional TN state~\cite{Ran2020}. This is because it is a fundamental tool in many methods for simulating 1D quantum many-body systems, such as the Density Matrix Renormalization Group (DMRG)~\cite{White1992,Schollwock2011}, Time-Evolving Block Decimation (TEBD)~\cite{Vidal2003}, and Product Wave Function Renormalization Group (PWFRG)~\cite{Nishino1995}.

\begin{figure}[!h]
 
    \centering
    \includegraphics[width=\textwidth]{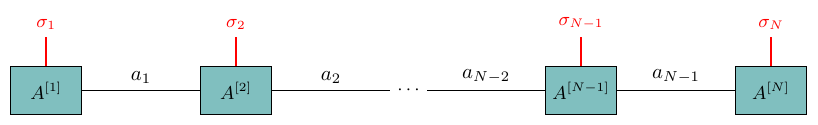}
    \caption{The graphical representation of 4-site MPS with open boundary conditions.}
    \label{fig:mps}
\end{figure}

\begin{definition}[Matrix Product State (MPS)~\cite{Orus2014}]
    The MPS is a TN formed by the contraction of $N$ tensors arranged into a one-dimensional chain (see Fig.~\ref{fig:mps}). In an MPS, there is one tensor per site in the many-body system. The connecting bond indices that glue the tensors together can take up to $D$ values, and the open indices correspond to the physical degrees of freedom of the local Hilbert spaces, which can take up to $p$ values.
\end{definition}

Given a general pure quantum state on a 1D lattice with $N$ sites:

\begin{equation}
    \ket{\psi} = \sum_{\sigma_1, \ldots, \sigma_N} C_{\sigma_1, \ldots, \sigma_N} \ket{\sigma_1, \ldots, \sigma_N},
\end{equation}

\noindent the MPS representation of its coefficients is~\cite{Ran2020}:

 \begin{equation}
 \label{eq:mps-basic}
     C_{\sigma_1, \ldots, \sigma_N} = \sum_{a_1,\ldots,a_{N-1}} A^{[1]}_{\sigma_1,a_1} A^{[2]}_{\sigma_2,a_1 a_2} \cdots A^{[N-1]}_{\sigma_{N-1}, a_{N-2} a_{N-1}} A^{[N]}_{\sigma_N, a_{N}},
 \end{equation}

 where $A^{i}_{\sigma_{i}, a_{i-1}, a_{i}}$ are tensors obtained by repeatedly applying SVD (or QR) decomposition to the coefficient matrix $C_{\sigma_1, \ldots, \sigma_N}$~\cite{Schollwock2011}. In fact, the MPS given by equation \eqref{eq:mps-basic} has an open boundary condition, and can be generalized to a periodic boundary condition as:

 \begin{equation}
 \label{eq:mps-basic-periodic}
     C_{\sigma_1, \ldots, \sigma_N} = \sum_{a_1,\ldots,a_{N-1}} A^{[1]}_{\sigma_1,a_Na_1} A^{[2]}_{\sigma_2,a_1 a_2} \cdots A^{[N-1]}_{\sigma_{N-1}, a_{N-2} a_{N-1}} A^{[N]}_{\sigma_N, a_{N}},
 \end{equation}

where all tensors are third order. The graphical representation of MPS is presented in Fig.~\ref{fig:mps}.

\section{Projected Entangled Pair States (PEPS)}
The PEPS~\cite{Verstraete2004} tensor network is the natural generalization of MPS to higher spatial dimensions. In general, it can be defined on any $N$-dimensional regular lattice~\cite{Ran2020}, but here we will focus on the two-dimensional case. PEPS is at the core of several methods for simulating two-dimensional quantum lattice systems, such as Tensor Renormalization Group (TRG)~\cite{Levin2007} or Corner Transfer Matrices (CTM) and Corner Tensors methods~\cite{Orus2009,Orus2012}. In this work, we will consider only two-dimensional rectangular lattices. Of course, one can also define PEPS on other types of lattices, such as honeycomb, triangular, kagome, etc.~\cite{Orus2014} However, the \texttt{SpinGlassPEPS} algorithm and software employ rectangular lattices, and they will be the focus.

\begin{definition}[Projected Entangled Pair States (PEPS)~\cite{Orus2014}]
    The PEPS is a TN formed by the contraction of $K$ tensors arranged into an $N$-dimensional regular lattice (\textit{e.g.} square grid, honeycomb, triangular, etc.). In a PEPS, there is one tensor per site in the many-body system. The connecting bond indices that glue the tensors together can take up to $D$ values, and the open indices correspond to the physical degrees of freedom of the local Hilbert spaces, which can take up to $p$ values.
\end{definition}

\begin{figure}
    \centering
    \includegraphics[width=\textwidth]{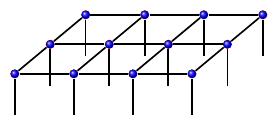}
    \caption{Graphical representation of PEPS tensor network on a $3 \times 4$ rectangular grid. It is worth noting that tensors can have different orders depending on their position on the grid. The boundary ones are $3$rd-order (corners) or $4$th-order, while those "inside" are $5$th-order tensors.}
    \label{fig:peps}
\end{figure}

%% file: appendices/appendix-rl.tex
\chapter{Basics of Deep Reinforcement Learning}
\label{apx:rl}

Reinforcement learning (RL) is one of the three major machine-learning paradigms, alongside supervised and unsupervised learning.
In RL, an agent improves its decision-making by interacting with an environment and receiving feedback in the form of scalar rewards.
In contrast to supervised learning, RL does not assume access to labeled input--output pairs; instead, the agent must discover which actions lead to high long-term reward through trial and error, while balancing exploration and exploitation \cite{Sutton2018,Szepesvari2022}.

\begin{figure}
	\centering
	\includegraphics[width=0.8\textwidth]{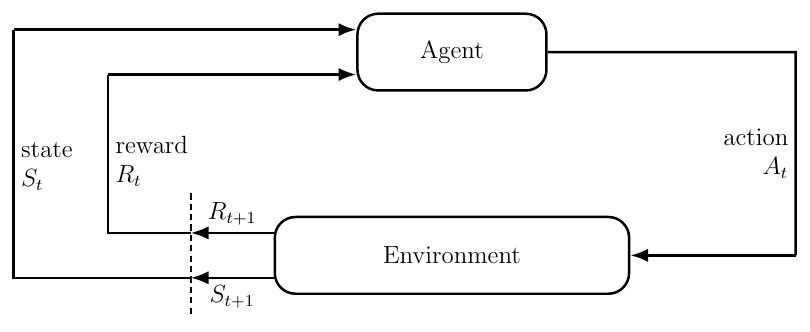}
	\caption{Schematic RL loop. At each discrete time step $t$, the agent observes a state $S_t$, selects an action $A_t$ according to a policy $\pi$, receives a reward $R_t$, and transitions to a next state $S_{t+1}$.}
\end{figure}

\section{Markov decision processes}

Throughout this dissertation we consider the standard RL setting formalized as a Markov decision process (MDP) \cite{Sutton2018,Szepesvari2022}.
Time is discrete, $t = 0,1,\ldots$, and the environment evolves in response to the agent's actions.

\begin{definition}[Markov decision process \cite{Sutton2018,Szepesvari2022}]
An (episodic) MDP is a tuple
\begin{equation}
	\mathcal{M} = (\mathcal{S}, \mathcal{A}, P, r, \gamma),
\end{equation}
where $\mathcal{S}$ is a (typically finite or countable) state space, $\mathcal{A}$ is an action space, $P(\cdot \mid s,a)$ is a transition kernel over $\mathcal{S}$, $r:\mathcal{S}\times\mathcal{A}\times\mathcal{S}\to\mathbb{R}$ is a (possibly stochastic) reward mechanism, and $\gamma\in(0,1]$ is the discount factor.
The Markov property states that the conditional distribution of $(s_{t+1},r_t)$ depends on the past only through the current pair $(s_t,a_t)$.
\end{definition}

In many texts, the reward is modeled as part of the transition kernel.
This is particularly convenient for countable state spaces.
Following \cite{Szepesvari2022}, one can equivalently define a \emph{transition probability kernel} $\mathcal{P}_0(\cdot\mid s,a)$ over $\mathcal{S}\times\mathbb{R}$, which jointly specifies the next-state and immediate-reward distribution.
From $\mathcal{P}_0$, the \emph{state transition kernel} and the \emph{expected immediate reward} are obtained as
\begin{align}
	P(s' \mid s,a) &= \int_{\mathbb{R}} \mathcal{P}_0\big((s',\rho)\mid s,a\big)\, d\rho,\\
	\bar r(s,a) &= \mathbb{E}[r_t\mid s_t=s,a_t=a]
	= \sum_{s'\in\mathcal{S}} \int_{\mathbb{R}} \rho\; \mathcal{P}_0\big((s',\rho)\mid s,a\big)\, d\rho.
\end{align}

\paragraph{Episodes and return.}
An episode is a trajectory $(s_0,a_0,r_0,s_1,a_1,r_1,\ldots,s_T)$ that terminates at a terminal state $s_T$.
The (discounted) return from time $t$ is
\begin{equation}
	R_t \;=\; \sum_{k=0}^{T-t-1} \gamma^k r_{t+k}.
\end{equation}
The discount factor $\gamma$ controls how strongly the agent prefers immediate rewards over delayed rewards; $\gamma<1$ also ensures that the infinite-horizon return is well-defined for continuing tasks \cite{Sutton2018}.

\section{Policies and value functions}

A \emph{policy} is a rule for selecting actions.
We allow stochastic policies $\pi(a\mid s)$, which define a probability distribution over actions in each state.
A deterministic policy is a special case where $\pi(a\mid s)$ is a point mass.

\paragraph{State-value and action-value functions.}
For a policy $\pi$, the \emph{state-value function} and \emph{action-value function} are defined as \cite{Sutton2018}
\begin{align}
	V^\pi(s) &= \mathbb{E}_\pi\!\left[R_t \mid s_t=s\right],\\
	Q^\pi(s,a) &= \mathbb{E}_\pi\!\left[R_t \mid s_t=s,\; a_t=a\right],
\end{align}
where the expectation is taken over trajectories generated by following policy $\pi$ and the environment dynamics.
The \emph{optimal} value functions are
\begin{equation}
	V^*(s) = \max_\pi V^\pi(s),\qquad Q^*(s,a) = \max_\pi Q^\pi(s,a).
\end{equation}
Any policy that is greedy with respect to $Q^*$ is optimal.

\paragraph{Bellman equations.}
Value functions satisfy recursive (Bellman) relations \cite{Bellman57a,Sutton2018}.
For any fixed policy $\pi$,
\begin{align}
	V^\pi(s) &= \sum_{a\in\mathcal{A}} \pi(a\mid s)\sum_{s'\in\mathcal{S}} P(s'\mid s,a)\Big(\bar r(s,a,s') + \gamma V^\pi(s')\Big),\\
	Q^\pi(s,a) &= \sum_{s'\in\mathcal{S}} P(s'\mid s,a)\Big(\bar r(s,a,s') + \gamma \sum_{a'\in\mathcal{A}} \pi(a'\mid s')Q^\pi(s',a')\Big).
\end{align}
Optimal values satisfy the Bellman \emph{optimality} equations,
\begin{align}
	V^*(s) &= \max_{a\in\mathcal{A}} \sum_{s'\in\mathcal{S}} P(s'\mid s,a)\Big(\bar r(s,a,s') + \gamma V^*(s')\Big),\\
	Q^*(s,a) &= \sum_{s'\in\mathcal{S}} P(s'\mid s,a)\Big(\bar r(s,a,s') + \gamma \max_{a'\in\mathcal{A}} Q^*(s',a')\Big).
\end{align}

\section{Model-based vs.\ model-free learning, and on-policy vs.\ off-policy}

A central distinction in RL is whether the agent has (or learns) an explicit model of the environment dynamics \cite{Sutton2018,Szepesvari2022}.
\begin{itemize}
	\item \textbf{Model-based RL} uses (known or learned) $P$ and $r$ to plan, e.g., by dynamic programming, tree search, or trajectory rollouts.
	\item \textbf{Model-free RL} does not explicitly build $P$; instead it learns value functions and/or policies directly from experience.
\end{itemize}
Another key distinction concerns which policy generates data:
\begin{itemize}
	\item \textbf{On-policy} methods learn about the same policy that is used to act (e.g., SARSA).
	\item \textbf{Off-policy} methods learn a target policy from data collected by a different behavior policy (e.g., Q-learning), enabling experience reuse and more flexible exploration.
\end{itemize}

\section{Value-based methods and temporal-difference learning}

In many applications, directly computing $Q^*$ is infeasible due to large state spaces.
Value-based RL therefore relies on sample-based updates.
Temporal-difference (TD) learning combines bootstrapping (using current value estimates) with Monte Carlo sampling \cite{Sutton2018}.

\paragraph{TD error.}
Given an estimate $Q(s,a)$, the one-step TD error is
\begin{equation}
	\delta_t = r_t + \gamma \max_{a'} Q(s_{t+1},a') - Q(s_t,a_t),
\end{equation}
where the maximization corresponds to greedy control (off-policy).
In tabular settings, Q-learning performs the update
\begin{equation}
	Q(s_t,a_t) \leftarrow Q(s_t,a_t) + \alpha\, \delta_t,
\end{equation}
with stepsize $\alpha>0$.

\paragraph{Deep Q-learning.}
In deep RL, $Q$ is represented by a neural network $Q(s,a;\Theta)$ with parameters $\Theta$.
A standard training objective is the squared TD error (a regression loss) \cite{Sutton2018}:
\begin{equation}
	\mathcal{L}(\Theta) = \mathbb{E}\Big[\big(y_t - Q(s_t,a_t;\Theta)\big)^2\Big],
	\qquad
	y_t = r_t + \gamma \max_{a'} Q(s_{t+1},a';\Theta^-),
\end{equation}
where $\Theta^-$ denotes parameters of a (slowly updated) target network.
In practice, stability is improved by (i) \emph{experience replay}, which trains on minibatches sampled from a buffer of past transitions, and (ii) the target network, which reduces harmful feedback loops during bootstrapping.

\paragraph{Exploration.}
Value-based agents typically use stochastic exploration strategies, such as $\varepsilon$-greedy action selection:
with probability $\varepsilon$ choose a random action, otherwise choose $a=\arg\max_{a'}Q(s,a')$.
Other approaches include Boltzmann/softmax policies and explicit exploration bonuses \cite{Sutton2018}.

\section{Policy gradients and actor--critic methods}

An alternative to value-based control is to directly optimize a parameterized policy $\pi_\Theta(a\mid s)$.
Let $J(\Theta)$ denote the expected return from an initial-state distribution.
Policy gradient methods estimate $\nabla_\Theta J(\Theta)$ from sampled trajectories and apply stochastic gradient ascent \cite{Sutton2018,Szepesvari2022}.
A classical estimator uses
\begin{equation}
	\nabla_\Theta J(\Theta) \approx \sum_{t=0}^{T-1} \nabla_\Theta \log \pi_\Theta(a_t\mid s_t)\, (R_t - b(s_t)),
\end{equation}
where $b(s)$ is a baseline (often $V^\pi(s)$) that reduces variance without biasing the gradient.
Actor--critic methods combine an \emph{actor} (the policy) with a \emph{critic} (a learned value function) to provide low-variance learning signals via TD errors \cite{Sutton2018}.

%% file: appendices/appendix-dwave.tex
\chapter{Physical Properties of Used D-Wave Machines}
\label{apx:dwave}

D-Wave reference results for the Pegasus instances were obtained using the Advantage\_system6.1, while Zephyr instances were solved on the Advantage2\_prototype1.1. The properties of both machines are detailed in Table~\ref{tab:machine_prop}.

\hskip-4.0cm

\begin{sidewaystable}[p]
\centering
\caption{Physical properties of the D-Wave Advantage\_system6.1 machine and Advantage2\_prototype1.1}
\begin{tabular}{ | l|l|l|}
 \hline
\textbf{Parameter} & \textbf{Advantage\_system6.1} & \textbf{Advantage2\_prototype1.1} \\ 
 \hline
 \hline
Qubits & $5616$ & $563$   \\  
\hline
 Couplers & $40135$  & $4790$   \\
 \hline
 Qubit temperature (mK) & $16.0 \pm 0.1$ & $13.9 \pm 1.0$\\
 \hline
$\text{M}_{\text{AFM}}$ (pH)\footnote{Maximum available mutual inductance achievable between pairs of flux qubit bodies.} & $1.554$ & $0.582$  \\
 \hline
 $L_q$ (pH)\footnote{Qubit inductance.} & $382.180$ & $142.920$  \\
 \hline
$C_q$ (fF) \footnote{Qubit capacitance.} &  $118.638$ &  $169.388$\\
 \hline
$I_c$ ($\mu A$)\footnote{Qubit critical current.} & $1.994$ & $4.083$\\
 \hline
 Average single qubit thermal width (Ising units) & $0.221$  & $0.117$\\
 \hline
 FM problem freezeout (scaled time) & $0.073$ & $0.015$ \\
 \hline
 Single qubit freezeout (scaled time) & $0.616$ & $0.619$ \\
 \hline
$\Phi^i_{\text{CCJJ}}$ ($\Phi_0$)\footnote{Initial value of the external flux applied to qubit compound Josephson-junction structures at the start of an anneal ($s=0$).}& $-0.624$ & $-0.686$\\
 \hline
$\Phi^f_{\text{CCJJ}}$ ($\Phi_0$) \footnote{Final value at the end of an anneal ($s=1$).} & $-0.723$ & $-0.766$\\
 \hline
Readout time range ($\mu$s)\footnote{Typical readout times for reading between one qubit and the full QPU.} & $18.0$ to $173.0$  & $15.0$ to $48.0$\\
 \hline
 Programming Time\footnote{Typical for problems run on this QPU. Actual problem programming times may vary slightly depending on the nature of the problem.} ($\mu$s) & $\sim 14200$ & $\sim 5500$  \\
 \hline
 QPU delay time per sample ($\mu$s) & $20.5$ & $21.0$  \\
 \hline
 Readout Error Rate\footnote{Error rate when reading the full system.} & $\leq 0.001$  & $\leq 0.001$ \\
 \hline
\end{tabular}
\label{tab:machine_prop}
\end{sidewaystable}